\pdfoutput=1
\documentclass[11pt,twoside,a4paper,cmspaper,final,collab]{cms-tdr}
\def\svnVersion{8721300}\def\svnDate{2024/06/15}\def\cmsCernNoTag{CERN-EP-2023-279}\def\cmsCernDate{\today}\def\cmsMessage{Published in Physical Review D as \href{http://dx.doi.org/10.1103/PhysRevD.109.112013}{\doi{10.1103/PhysRevD.109.112013}.}}
\begin{document}\cmsNoteHeader{SMP-21-004}

\renewcommand{\cmsCollabName}{The CMS and TOTEM Collaborations}
\renewcommand{\cmsTag}{CMS-\cmsNUMBER\\&TOTEM 2024-001\\}
\renewcommand{\cmsPubBlock}{\begin{tabular}[t]{@{}r@{}l}&CMS \cmsSTYLE~\cmsNUMBER\\&TOTEM 2024-001\\\end{tabular}}
\renewcommand{\cmsCopyright}{\copyright\,\the\year\ CERN for the benefit of the CMS and TOTEM Collaborations.}
\renewcommand{\appMsg}{See Appendices~\ref{app:collab} and~\ref{app:totem} for the lists of collaboration members}
\renewcommand{\cmslogo}{\includegraphics[height=2cm]{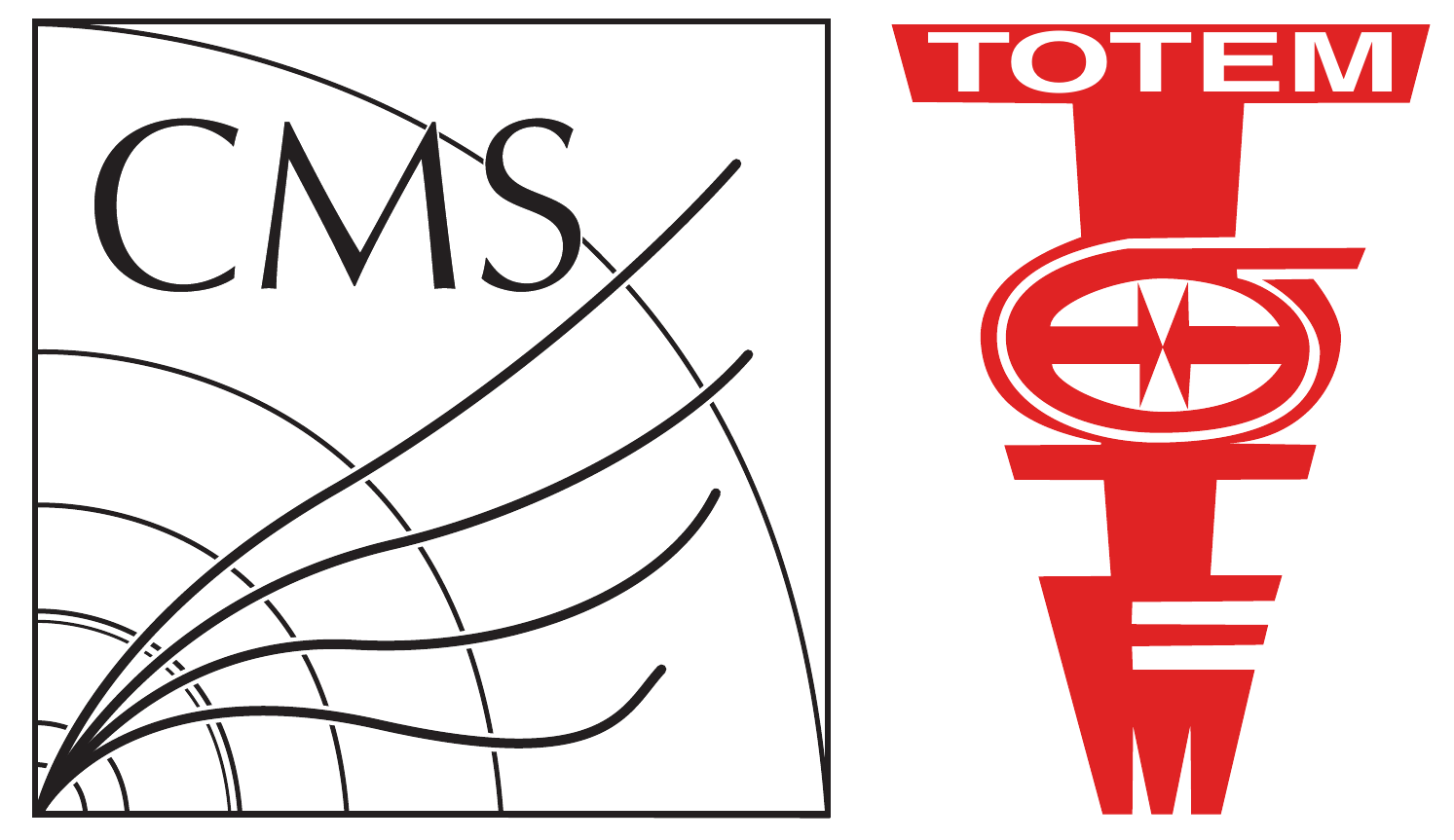}}

\newcommand{\dime}{\textsc{Dime}\xspace}
\newcommand{\supc}{\textsc{SuperChic2}\xspace}
\newcommand{\genex}{\textsc{GenEx}\xspace}
\newcommand{\gran}{\textsc{Graniitti}\xspace}
\newcommand{\prof}{\textsc{Professor}\xspace}

\renewcommand{\P}{\mathrm{I\!P}\xspace}
\DeclareRobustCommand{\PShp}{{\HepParticle{h}{}{+}}\Xspace}
\DeclareRobustCommand{\PShm}{{\HepParticle{h}{}{-}}\Xspace}
\newcommand{\phhp}{\ensuremath{\Pp(\PShp\PShm)\Pp}\xspace}

\newlength\cmsTabSkip\setlength{\cmsTabSkip}{1ex}

\cmsNoteHeader{SMP-21-004}
\title{Nonresonant central exclusive production of charged-hadron pairs
       in proton-proton collisions at \texorpdfstring{$\sqrt{s} = 13\TeV$}{sqrt(s) = 13 TeV}}

\renewcommand{\cmsCollabName}{CMS and TOTEM Collaborations}

\date{\today}

\abstract{The central exclusive production of charged-hadron pairs in pp
collisions at a centre-of-mass energy of 13\TeV is examined, based on data
collected in a special high-$\beta^*$ run of the LHC. The nonresonant continuum
processes are studied with the invariant mass of the centrally produced
two-pion system in the resonance-free region, $m_{\PGpp\PGpm} < 0.7\GeV$ or
$m_{\PGpp\PGpm} > 1.8\GeV$. Differential cross sections as functions of the
azimuthal angle between the surviving protons, squared exchanged four-momenta,
and $m_{\PGpp\PGpm}$ are measured in a wide region of scattered proton
transverse momenta, between $0.2$ and $0.8\GeV$, and for pion rapidities
$\abs{y} < 2$. A rich structure of interactions related to double-pomeron
exchange is observed. A parabolic minimum in the distribution of the two-proton
azimuthal angle is observed for the first time. It can be interpreted as an
effect of additional pomeron exchanges between the protons from the
interference between the bare and the rescattered amplitudes. After model
tuning, various physical quantities are determined that are related to the
pomeron cross section, proton-pomeron and meson-pomeron form factors, pomeron
trajectory and intercept, and coefficients of diffractive eigenstates of the
proton.}

\hypersetup{%
pdfauthor={CMS Collaboration},%
pdftitle={Nonresonant central exclusive production of charged hadron pairs in proton-proton collisions at sqrt(s) = 13 TeV},%
pdfsubject={CMS, TOTEM},%
pdfkeywords={CMS, TOTEM, physics, central exclusive production}}

\maketitle 

\section{Introduction}
\label{sec:intro}

The cross sections of proton-proton ($\Pp\Pp$) and proton-antiproton
($\Pp\PAp$) interactions steadily rise with the centre-of-mass energy, and
approach each other at high energies~\cite{ParticleDataGroup:2022pth}. This
observation was initially explained assuming the exchange of a family of
particles~\cite{Berestetsky:1961zat} with the vacuum quantum numbers of the
pomeron~\cite{Donnachie:2002en}. This pomeron is now interpreted in terms of the
sum of ladder-type diagrams (multiperipheral model~\cite{Amati:1962nv})
composed of gluons. Pomeron physics, its nonperturbative characteristics, and
its relations with the theory of the strong interaction (quantum
chromodynamics, QCD) are a topic of ongoing research both experimentally and
theoretically~\cite{Donnachie:2002en,Meyer:2004jc,TOTEM:2020zzr}.

\begin{figure*}[htb]

 \centering
 \includegraphics[width=\textwidth]{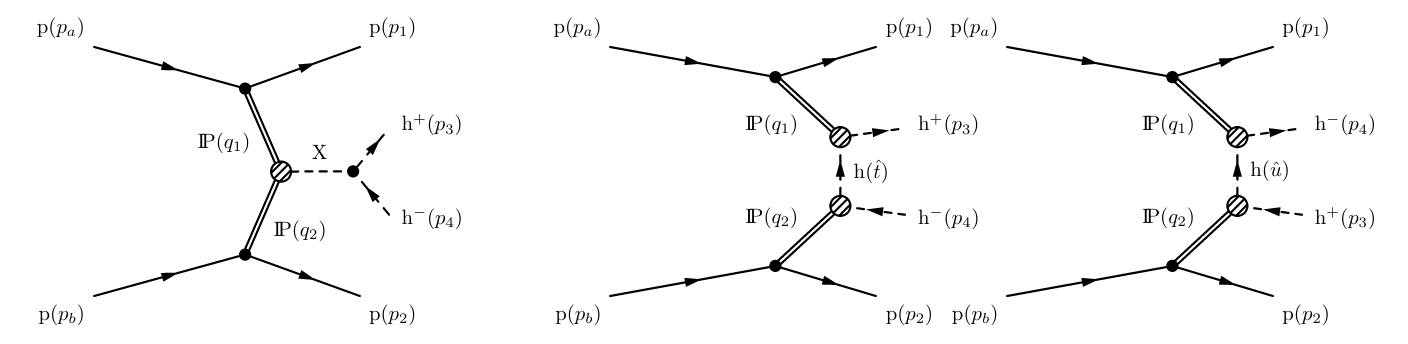}

 \caption{Born-level Feynman diagrams for central exclusive production of
hadron pairs via double-pomeron exchange, depicting resonant (left) and
nonresonant continuum (centre: $t$-channel, right: $u$-channel) contributions.}

 \label{fig:feynman}

\end{figure*}

In $\Pp\Pp$ collisions, the central exclusive production of a few particles
(Fig.~\ref{fig:feynman}) offers a clean laboratory for the study of various
phenomena~\cite{Khoze:2001xm,Albrow:2010yb}, and is often seen as a discovery
channel for new physics coupled to quarks and gluons.
At energies above 20\GeV, the exchange of hadrons is
suppressed~\cite{Lebiedowicz:2009pj} and, for proton transfer momenta above 0.2\GeV,
these central exclusive processes are dominated by double-pomeron exchange.
They provide a gluon-rich environment, which may lead to the creation of
hadrons free of valence quarks, collectively called
glueballs~\cite{Ochs:2013gi}, whose existence is yet to be experimentally
demonstrated.

Double-pomeron exchange processes in $\Pp\Pp$ collisions were intensively
studied at CERN in the 1990s~\cite{Armstrong:1991ch,Antinori:1995wz} at
$\sqrt{s} = 12.7$, $23.8$, and $29\GeV$, with the most significant results
published by the WA102
Collaboration~\cite{Close:1999bi,Barberis:1999zh,Close:2000dx,Kirk:2000ws},
which concluded that the pomeron has a vector-like behaviour. With the advent
of high energy collider experiments, there is a renewed interest in the study
of central exclusive production, especially in double-pomeron exchange
processes. Measurements in $\Pp\PAp$ collisions at $\sqrt{s} = 0.9$ and
$1.96\TeV$ were performed by the CDF Collaboration~\cite{Aaltonen:2015uva} at
the Fermilab Tevatron, and in $\Pp\Pp$ collisions by the STAR
Collaboration~\cite{Adam:2020sap} at $\sqrt{s} = 0.2\TeV$ at the BNL RHIC.
Recent measurements in $\Pp\Pp$ collisions at $\sqrt{s} = 7\TeV$ were reported by the
ATLAS Collaboration~\cite{ATLAS:2022uef} at the LHC.

The CMS Collaboration at the LHC has recently published a study of central
exclusive $\PGpp\PGpm$ production at $\sqrt{s} = 5.02$ and
$13\TeV$~\cite{Sirunyan:2020cmr} exploiting the presence of rapidity ($y$) gaps
between the $\PGpp\PGpm$ system and the outgoing (untagged) protons. The
present study at $\sqrt{s} = 13\TeV$ is based on data with an integrated
luminosity, $4.7\pbinv$, that is more than four orders of
magnitude higher, where the kinematic variables of both the forward-scattered
protons and the centrally produced charged-hadron pair are measured with high
precision.

Central exclusive production of hadron pairs is an active area of theoretical
study, and several groups are investigating the scalar, vector, or tensor nature
of the pomeron~\cite{Lebiedowicz:2009pj,Lebiedowicz:2015eka}, producing
detailed predictions for $\PGpp\PGpm$~\cite{Lebiedowicz:2016ioh,Ryutin:2021ipc},
$\PKp\PKm$~\cite{Lebiedowicz:2018eui}, and $\Pp\PAp$
production~\cite{Lebiedowicz:2018sdt,Ryutin:2023waz}, and studying various absorptive effects
and properties of diffractive
eigenstates~\cite{Ryskin:2012az,HarlandLang:2012qz,Khoze:2013dha}. Monte Carlo
(MC) event generators~\cite{Harland-Lang:2013dia} that describe these processes
are available with empirical
parametrisations~\cite{Grau:2012wy,Fagundes:2013aja}.

In this paper, after a theoretical introduction, details of the CMS and TOTEM
detectors, and the outline of the analysis are presented in
Section~\ref{sec:intro}. The conditions of data-taking are discussed in
Section~\ref{sec:datataking}, with the reconstruction and corrections of
scattered protons in the Roman pots (Section~\ref{sec:protons}) and those of
produced particles in the central region (Section~\ref{sec:particles}).  Event
classification is described in Section~\ref{sec:classification}. Results, model
tuning, and data-model comparisons are shown next (Section~\ref{sec:results}),
and the paper ends with a summary in Section~\ref{sec:summary}.

\subsection{Theoretical background}
\label{sec:theory}

Proton-proton collisions can be either elastic or inelastic. The latter include
double-pomeron exchange, single- and double-diffractive, as well as
nondiffractive and photon fusion processes. The elastic scattering of protons
is essentially understood through the exchange of pomerons and also
odderons~\cite{Lukaszuk:1973nt}. Born-level Feynman diagrams of processes
playing a role in central exclusive production of charged-hadron pairs via
double-pomeron exchange are shown in Fig.~\ref{fig:feynman}. We distinguish two
main processes. The two pomerons may fuse through a short-lived resonance (X),
which in turn decays into a pair of oppositely charged hadrons. Alternatively,
the two pomerons may interact via the exchange of a virtual hadron (h), leading
to the production of a pair of oppositely charged hadrons in a nonresonant
process, the topic of the present study. Both $t$-channel and $u$-channel
graphs must be included. The variables of the outgoing protons are labelled
with subscripts 1 and 2, and the produced hadron ones with subscripts 3 and 4.

Such a simple picture becomes more complicated when the effects of suppression,
absorption, and other corrections are properly included. Here we
follow the notations of Ref.~\cite{Lebiedowicz:2009pj}. The matrix element for
the nonresonant continuum process is

\ifthenelse{\boolean{cms@external}}{
\begin{multline}
\mathcal{M}
  = M_{13}(t_1,s_{13}) \frac{F_m^2(\hat{t})}{\hat{t} - m^2} M_{24}(t_2,s_{24}) \\
  + M_{14}(t_1,s_{14}) \frac{F_m^2(\hat{u})}{\hat{u} - m^2} M_{23}(t_2,s_{23}),
 \label{eq:M}
\end{multline}
}{
\begin{equation}
\mathcal{M} =
    M_{13}(t_1,s_{13}) \frac{F_m^2(\hat{t})}{\hat{t} - m^2} M_{24}(t_2,s_{24}) +
    M_{14}(t_1,s_{14}) \frac{F_m^2(\hat{u})}{\hat{u} - m^2} M_{23}(t_2,s_{23}),
 \label{eq:M}
\end{equation}
}
where $p_\text{a}$ and $p_\text{b}$ are the momenta of the incoming protons,
$t_1 = (p_1 - p_\text{a})^2$, $t_2 = (p_2 - p_\text{b})^2$, $M_{ik}$ denotes
the ``interaction'' between one of the scattered protons and one of the created
hadrons, $s_{ik} = (p_i + p_k)^2$, $\hat{t} = (p_3 - q_1)^2 = (p_4 - q_2)^2$
and $\hat{u} = (p_4 - q_1)^2 = (p_3 - q_2)^2$. The meson-pomeron form factor
$F_m(\hat{t})$ is included twice, once for each vertex connected to the meson.
The propagator $(\hat{t} - m^2)^{-1}$ of the virtual hadron is also present,
where $m$ is the mass of the exchanged meson. At high meson-proton
centre-of-mass energies ($\sqrt{s_{ik}} > 20\GeV$) the double-pomeron exchange
contribution dominates,

\begin{equation}
 M_{ik}(t_i,s_{ik}) = i s_{ik} \sigma_0
  \left(\frac{s_{ik}}{s_0} \right)^{\alpha_\P(t_i) - 1} F_p(t_i),
 \label{eq:pom_only}
\end{equation}
which is the product of the proton-hadron cross section term $s^{\alpha_\P}$
and the exponential proton-pomeron form factor $F_p(t) = \exp(B_\P/2 \, t)$.
From a fit to the energy dependence of the measured cross
sections~\cite{Donnachie:1992ny}, the strength parameter is $\sigma_0 \approx
13.63\unit{mb}$ ($\PGpp\PGpm$ production) or $11.82\unit{mb}$ ($\PKp\PKm$
production), $s_0 = 1\GeV^2$ is a dimensional constant, $B_\P$ is related to
the proton-pomeron vertex (usually $B_\P = 5.5\text{--}6.0\GeV^2$ is assumed),
and the soft-pomeron trajectory is $\alpha_\P(t) = 1.0808 + 0.25\GeV^{-2} t =
\alpha_\P(0) + \alpha_\P'(0) t$. The cross section is proportional to
${\abs{\mathcal{M}}}^2/s^2$. The strength parameter (referred to as $\sigma_0$ in
the following), the intercept $\alpha_\P(0)$, and the slope $\alpha_\P'(0)$ of
the pomeron trajectory are usually taken from the Donnachie--Landshoff
model~\cite{Donnachie:1992ny}, based on a fit to the $\sqrt{s}$-dependence of
the total cross sections. A more complete analysis was performed by the COMPETE
Collaboration~\cite{Cudell:2001pn}.
 
The meson-pomeron form factor takes into account that the virtual meson is
off-shell. There are several
parametrisations~\cite{Lebiedowicz:2009pj,Harland-Lang:2013dia} available for
$F_m(\hat{t})$:

\begin{equation}
  \begin{cases}
   \exp(b_\text{exp} (\hat{t} - m^2))
                                 & \text{(exponential)}, \\
   \exp\Bigg(b_\text{ore} \Big[a_\text{ore} - \sqrt{\smash[b]{a_\text{ore}^2 - (\hat{t} - m^2)}}\Big] \Bigg)
                                 & \text{(Orear~\cite{Orear:1964zz})}, \\
   1/(1 - b_\text{pow} (\hat{t} - m^2))
                                 & \text{(power-law)}.
  \end{cases}
 \label{eq:meson-pomeron}
\end{equation}
The Orear-type parametrisation is given in Ref.~\cite{Lebiedowicz:2009pj}. They are equal to unity if the hadron is on-shell, $F_m(m^2) = 1$. The
parameters $b_\text{exp}$, $b_\text{pow}$, $a_\text{ore}$, and $b_\text{ore}$
are determined from measurement. To account for the low-mass diffractive
dissociation of the proton, a multi-channel framework is needed, where the
incoming proton is a coherent superposition of diffractive eigenstates with
amplitudes $a_i$ and coupling factors $\gamma_i$. A two-channel model is
considered here, including a ground state $\Pp$ and its first excitation
$\mathrm{N}^*$. The coupling of the pomeron to the diffractive eigenstates is
parametrised as in Ref.~\cite{Harland-Lang:2013dia}

\begin{equation}
 F_i(t) = \exp\left[ - (b_i (c_i - t))^{d_i}
                     + (b_i  c_i     )^{d_i} \right],
\end{equation}
where $b_i$, $c_i$, and $d_i$ are parameters. This factor is unity at $t=0$
with $F_i(0) = 1$, and is a more complex version of the simple $F_p(t) =
\exp(B_\P/2 \, t)$ form displayed in Eq.~\eqref{eq:pom_only}.

\begin{figure}[t]

 \centering
 \includegraphics[width=0.5\textwidth]{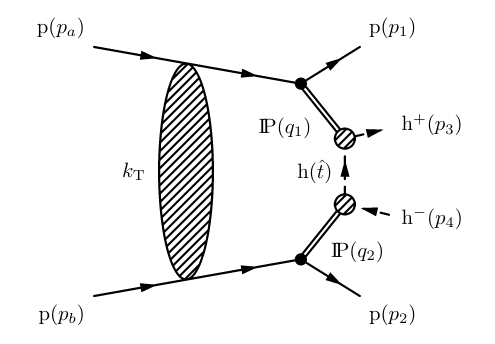}

 \caption{Feynman diagram for the nonresonant continuum of central exclusive
production of hadron pairs via double-pomeron exchange, including the
rescattering correction.}

 \label{fig:feynman_full}

\end{figure}

Additional pomeron exchanges between the incoming (and outgoing) protons cause
interference between the bare $\mathcal{M}$ (Eq.~\eqref{eq:M}) and the rescattered
${\cal M_\text{res}}$ amplitudes (Fig.~\ref{fig:feynman_full}). The resulting
quantity is often referred to as the eikonal survival factor. The interference
is expected to lead to dips in the angular
distributions~\cite{Harland-Lang:2013dia}. The rescattered amplitude is
obtained through a loop integral over $\vec\kt$ as

\ifthenelse{\boolean{cms@external}}{
\begin{multline}
 \mathcal{M}_\text{res} =
   \int \rd^2\vec\kt \,
        \mathcal{M}(\vec{p_1} - \vec\kt, \vec{p_2} + \vec\kt) \\
        \times \sum_{i,j} \gamma_i \abs{a_i}^2 F_i(t_1)
                   \gamma_j \abs{a_j}^2 F_j(t_2) S_{ij}(\kt),
\end{multline}
}{
\begin{equation}
 \mathcal{M}_\text{res} =
   \int \rd^2\vec\kt \,
        \mathcal{M}(\vec{p_1} - \vec\kt, \vec{p_2} + \vec\kt) \,
        \sum_{i,j} \gamma_i \abs{a_i}^2 F_i(t_1)
                   \gamma_j \abs{a_j}^2 F_j(t_2) S_{ij}(\kt),
\end{equation}
}
where $t_1$ and $t_2$ implicitly depend on $\vec\kt$. The screening amplitude
$S$ can come either from a calculation, or from an empirical parametrisation of
a direct measurement of the elastic differential $\Pp\Pp$ cross section. In the
calculation of the cross section, the bare amplitude $\mathcal{M}$ is replaced by
the sum of the bare and rescattered amplitudes, $\mathcal{M} + {\cal
M}_\text{res}$.

\subsection{The CMS and TOTEM detectors}

The central feature of the CMS apparatus is a superconducting solenoid of
6\unit{m} internal diameter, providing a magnetic field of 3.8\unit{T}. Within
the magnetic volume are a silicon pixel and strip tracker, a lead tungstate
crystal electromagnetic calorimeter, and a brass and scintillator hadron
calorimeter, each composed of a barrel and two endcap sections. Forward
calorimeters extend the coverage provided by the barrel and endcap detectors.
Muons are detected in gas-ionisation chambers embedded in the steel flux-return
yoke outside the solenoid. The silicon tracker measures charged particles
within the pseudorapidity range $\abs{\eta} < 2.5$. During the LHC running
period when these measurements were performed, the silicon tracker consisted of
1856 silicon pixel and 15\,148 silicon strip detector modules. For nonisolated
particles with transverse momentum $1 < \pt < 10\GeV$, the track resolutions
are typically 1.5\% in \pt and 20--75\mum in the transverse impact
parameter~\cite{CMS-DP-2020-049}. A more detailed description of the CMS
detector is reported in Ref.~\cite{Chatrchyan:2008zzk}.

The proton spectrometer of the TOTEM experiment consists of two telescope arms,
referred to as ``Arm 1'' and ``Arm 2'' for positive and negative $\eta$,
respectively. The spectrometer is used to detect protons scattered at very
small angles. In each arm, there are two detector ``stations'' located at about
$\pm$213 and $\pm$220\unit{m} relative to the nominal interaction point. Each
station consists of two units. Each unit has two vertical Roman pots (RPs), one
located above, and one below the LHC beam. An RP contains ten layers of silicon
strip detectors, placed alternately in two orthogonal orientations ($u$ and $v$).
They approach the beam to a distance of a few millimetres without affecting the
LHC operation (they are retracted for injection). The RPs are used to detect
protons deflected by only a few microradians relative to the beam, through
reconstruction of their short track segments (``tracklets''). The kinematic
variables of the scattered protons are reconstructed by modelling the transport
of the protons from the interaction point to the RP location, through the
inhomogeneous magnetic fields of several quadrupole and dipole magnets. The
TOTEM detector is described in Refs.~\cite{Anelli:2008zza,Antchev:2013hya}.
 
We use a right-handed coordinate system, with the origin at the nominal
collision point, the $x$ axis pointing to the centre of the LHC ring, the $y$
axis pointing up (perpendicular to the LHC plane), and the $z$ axis along the
anticlockwise beam direction.

\subsection{Analysis outline}
\label{sec:limits}

As discussed above, the RPs measure the direction of the scattered proton. The
transverse momentum is inferred under the assumption that the total momentum of
the proton has not changed in the collision. The longitudinal momentum is
computed from the conservation of energy and momentum, taking into account all
produced particles. The acceptance of the stations is not azimuthally
uniform~\cite{Anelli:2008zza} and the coverage in the $y$ component of the
scattered proton momentum is $0.175 < \abs{p_{1/2,y}} < 0.670\GeV$. The
acceptance maps of the two arms are correlated since signals from both arms are
used for triggering. In addition, their efficiency depends on the individual
silicon strip detector efficiencies, which in some cases change significantly
over time.
 
The coverage of the silicon tracker translates into acceptance for centrally
produced hadrons for rapidities $\abs{y} < y_\text{max}$, where $y_\text{max}
\approx 2.0$ in the case of $\PGpp\PGpm$, and $\approx 1.6$ for
$\PKp\PKm$ and $\Pp\PAp$. Tracking is efficient for $\pt > 0.1\GeV$, but the
particle identification capabilities are substantially reduced at momenta above 1.2\GeV.

The data are analysed as functions of the four-momentum transfers squared
$(t_1,t_2)$ or equivalently $(p_\text{1,T},p_\text{2,T})$, and the azimuthal
angle $\phi$ between the momentum vectors of the two scattered protons in the
transverse plane.
 
The measurement is corrected bin-by-bin and no unfolding is needed. The
corrections are computed without using Monte Carlo simulations, except for
those describing the low-energy phenomena (\GEANTfour~\cite{GEANT4:2002zbu}
v10.4.3) needed for the determination of the tracking efficiency. The results
are corrected for the $p_y$ acceptance and the effect of the RP trigger based
on the collected events. This is done by exploiting the expected azimuthal
symmetry of the scattered protons around the beam axis. The combined correction
is given in bins of $(p_\text{1,T}, p_\text{2,T}, \phi)$. Corrections are also
applied for the proton tracklet reconstruction efficiency in the RPs, based on
the hit structure of each tracklet at the strip level.

The measurement is also corrected for trigger, reconstruction, and particle
identification efficiencies of the charged-hadron pair in the silicon tracker.
The first two are constructed using a realistic detector simulation of
single-track events combined with information on the pixel layer occupancy in
the barrel. This combined tracker correction is applied in bins of
$(p_\text{1,T},p_\text{2,T})$, where in each of them a four-dimensional
correction table $[\phi,m_\mathrm{h^+h^-},(\cos\theta,\varphi)_\text{GJ}]$ is
employed based on a kinematic simulation. Here GJ refers to the
Gottfried--Jackson frame~\cite{Gottfried:1964nx} in the centre-of-mass of the
centrally produced hadron pair, where the $z$ axis is in the direction of the
hadron pair in the laboratory frame, the $y$ axis is perpendicular to both
$\hat{z}$ and the incoming proton direction, $\hat{x} = \hat{y} \times
\hat{z}$, and $\theta$ and $\varphi$ are the polar and azimuthal angles of
the positively charged hadron in that frame.

The corrections are applied for each event separately in the form of products
of independent weight factors. This is possible since there are no efficiency
gaps (zero efficiency) in the
$[\phi,m_\mathrm{h^+h^-},(\cos\theta,\varphi)_\text{GJ}]$ space of the
two-hadron system in the rapidity range.

\section{Data-taking conditions}
\label{sec:datataking}

The data were taken in a special $\beta^* = 90\,\mathrm{m}$ run of the LHC, in
the period 2--7 July, 2018 ($\beta^*$ is the value of the amplitude function at
the collision point~\cite{wiedemann2013particle}). With such a high $\beta^*$
setting, the beam divergence is reduced, and the forward detectors are thus
able to measure elastically scattered protons at low angles and small
transverse momenta.

For this data set, the Level 1 (hardware) trigger requires detected protons in
each RP arm, in various configurations. Only the data from double-pomeron
exchange triggers are used in the analysis, not those from the elastic trigger
(Sections~\ref{sec:rp_trig} and~\ref{sec:rp_eff}). In the following, the term
``parallel'' refers to cases where only RPs above or below the beamline have
signals over predefined thresholds in the two arms (top-top or TT,
bottom-bottom or BB), whereas the ``diagonal'' configuration refers to the
other two cases (top-bottom or TB, bottom-top or BT). The high-level trigger
(HLT) has multiple components~\cite{CMS:2016ngn}. The pixel activity filters
require at least 5 pixel clusters and at least 3 layers with
pixel clusters in the barrel pixel detector. The pixel track filters require at
least one track (having at least 3 pixel clusters) in the entire pixel
detector.

Part of the data set was recorded with a proton bunch spacing of 100\unit{ns},
and part with 50\unit{ns}. To reduce the readout rate, not all bunch crossings
were read out in the case of the diagonal trigger configuration. The reduction
of the selected bunches does not simply translate into the reduction of the
recorded luminosity, since the bunch-by-bunch luminosity can greatly vary. A
detailed recalculation of the integrated luminosity was performed at the level
of bunch crossings and data segments. A data segment is a fraction of a run
that contains $2^{18}$ orbits, \ie, lasting about 23.3 seconds. This yields an
integrated luminosity of $4.7\pbinv$ for the present data.

\begin{figure*}[htb]

 \centering
 \includegraphics[width=\textwidth]{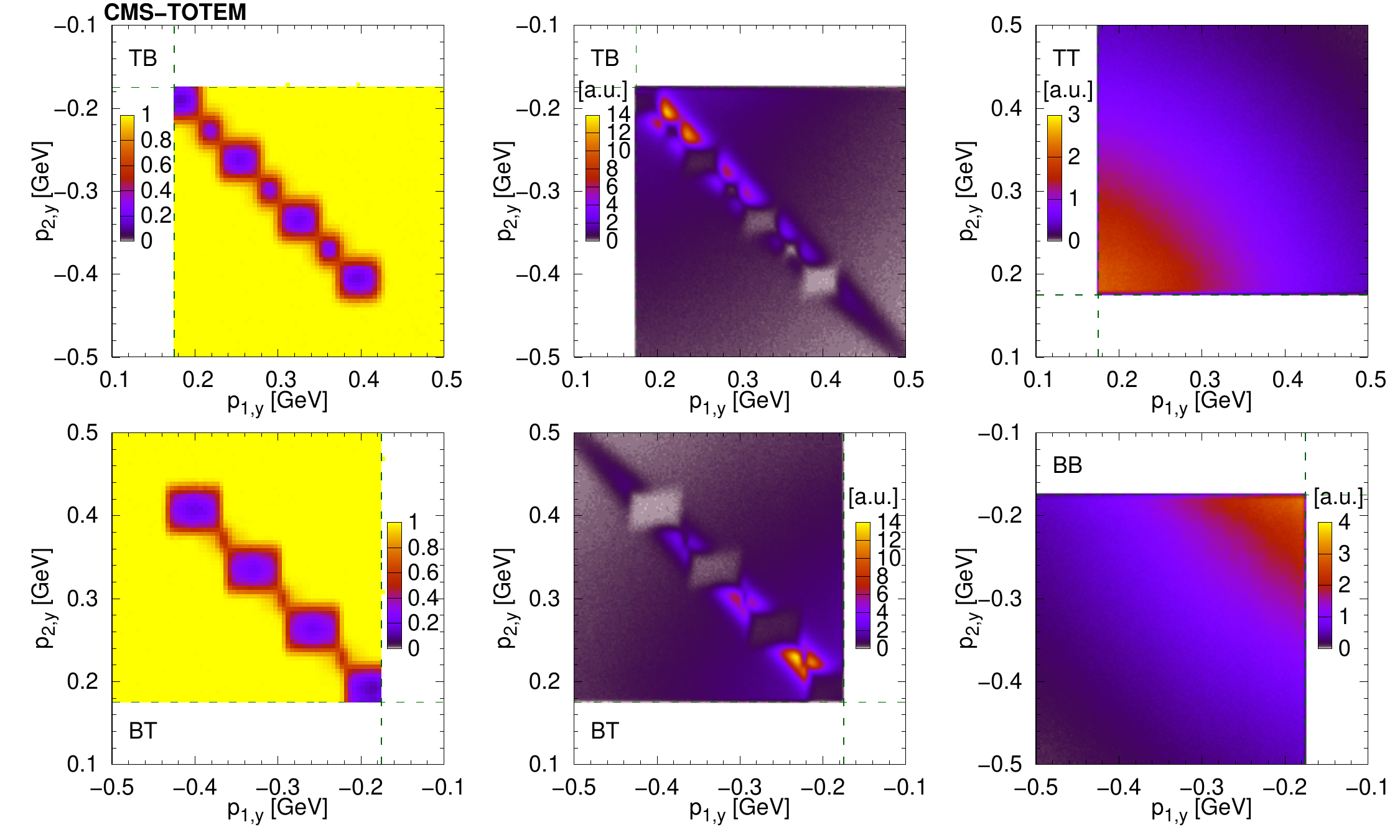}

 \caption{Left: Fraction of elastic-like events left after the veto trigger as
functions of momentum components in the $y$ direction in Arms 1 and 2, shown
here for the two diagonal trigger configurations: bottom pots in Arm 2 with top
pots in Arm 1 (upper left), and top pots in Arm 2 with bottom pots in Arm 1
(lower left). Centre and right: Joint distribution of detected proton momenta
$(p_{1,y},p_{2,y})$ in Arms 1 and 2 for all four trigger configurations
(centre: TB and BT, right: TT and BB). Limits of single-proton acceptance are
indicated with long dashed lines.}

 \label{fig:veto_py12}

\end{figure*}

In practice, sometimes more than one $\Pp\Pp$ collision happens in a bunch
crossing (pileup) with the average in the range $0.1\text{--}0.3$. The
detector-level signatures required for the present study are two slightly
scattered but intact protons, two oppositely charged centrally produced
hadrons, and the sum of the proton and hadron momenta close to zero (within the
combined momentum resolution of 50\MeV in $x$, and 20\MeV in $y$). A \phhp\
event can be correctly recorded if there is exactly one detectable \phhp\ final
state, no other double-pomeron exchange, single-, double-diffractive, or
nondiffractive collision in the same bunch crossing (they would be visible in
the tracker), and no visible elastic collisions (no detectable scattered
protons in RPs). In the analysis, events with more than one $\Pp\Pp$ collision are
rejected.

The average pileup of visible interactions is
$\langle\mu_\text{vis}\rangle
      = L_\text{int} \sigma_\text{vis} / (n_\text{bunch} n_\text{orbit})$,
where $\sigma_\text{vis}$ is the cross section for visible collisions,
$L_\text{int}$ is the average integrated luminosity in a given time period with
$n_\text{orbit}$ orbits, and $n_\text{bunch}$ is the number of selected
colliding bunches. The integrated luminosity loss due to the selection of
events with a single collision per bunch crossing is quantified by the factor
$\exp(-\mu_\text{vis})$. Most of the elastic $\Pp\Pp$ collisions are rejected by the
elastic RP proton pair veto (Section~\ref{sec:rp_trig}), whereas a large
fraction of events with the \phhp\ final state is selected. In this way the
inelastic $\Pp\Pp$ cross section is a good approximation for the visible cross
section, $\sigma_\text{vis} \sim \sigma_\text{inel}.$ There are several
measurements of the inelastic $\Pp\Pp$ cross section at $\sqrt{s} = $ 13\TeV at the
LHC~\cite{Aaboud:2016mmw,Sirunyan:2018nqx,Aaij:2018okq,Antchev:2017dia}, with
an average value of $79 \pm 1\unit{mb}$. For comparison, the recommended cross
section for inelastic events selected with a loose ``minimum bias'' trigger by
CMS is $69.2\unit{mb}$. We use $\sigma_\text{vis} = 79 \pm 5\unit{mb}$.

The values of beam-related quantities (instantaneous luminosity, bunch crossing
selection) and of the detector-related ones (acceptance, trigger efficiency)
are checked by comparing the numbers of observed and expected
events, in all RP trigger configurations (TB, BT, TT, and BB), separately in
each data segment. Overall a good agreement is found.

\section{Scattered protons in the Roman pots}
\label{sec:protons}

Details of proton reconstruction, including specific aspects relevant for the
high-$\beta^*$ run (strip-level efficiencies, local and global alignment,
optics determination), are reported in Ref.~\cite{note:2023}.

\begin{figure*}[htb]

 \centering
  \includegraphics[width=\textwidth]{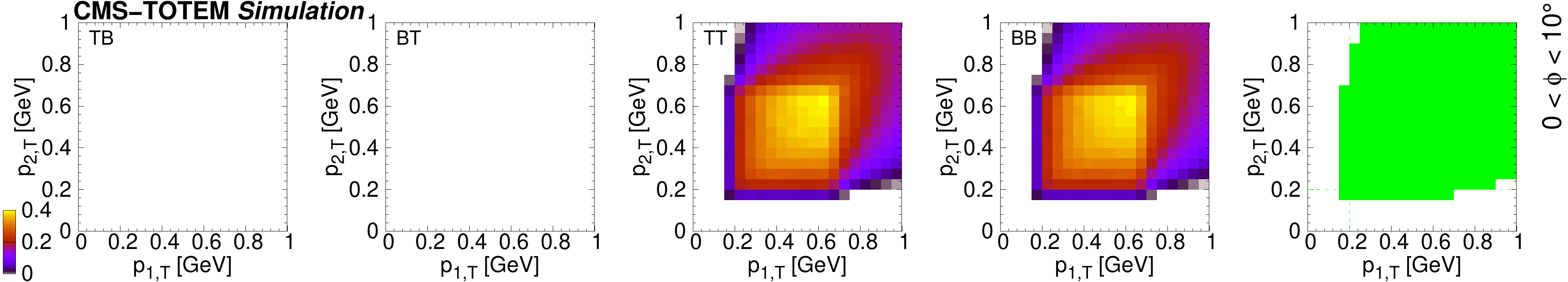}
  \includegraphics[width=\textwidth]{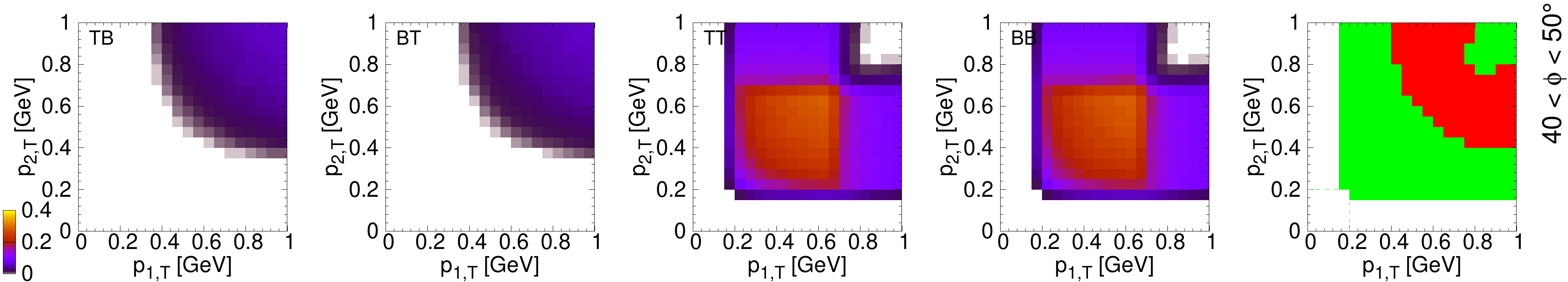}
  \includegraphics[width=\textwidth]{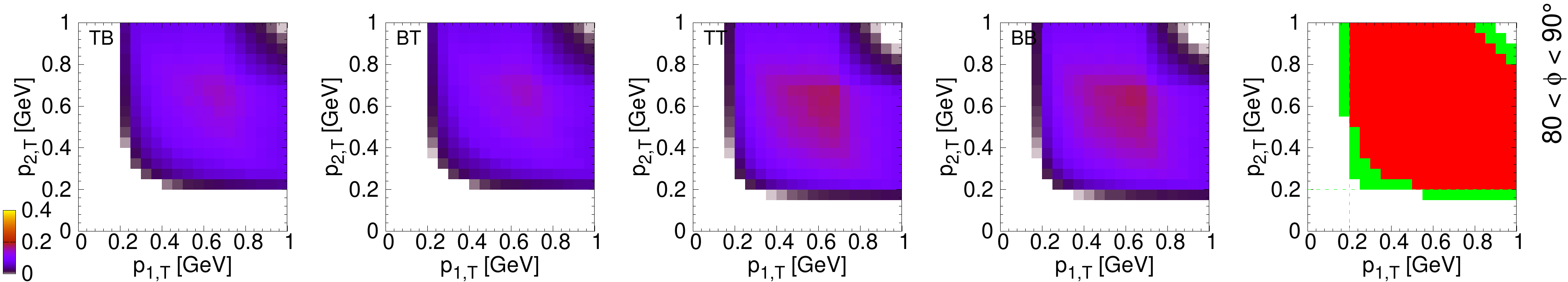}
  \includegraphics[width=\textwidth]{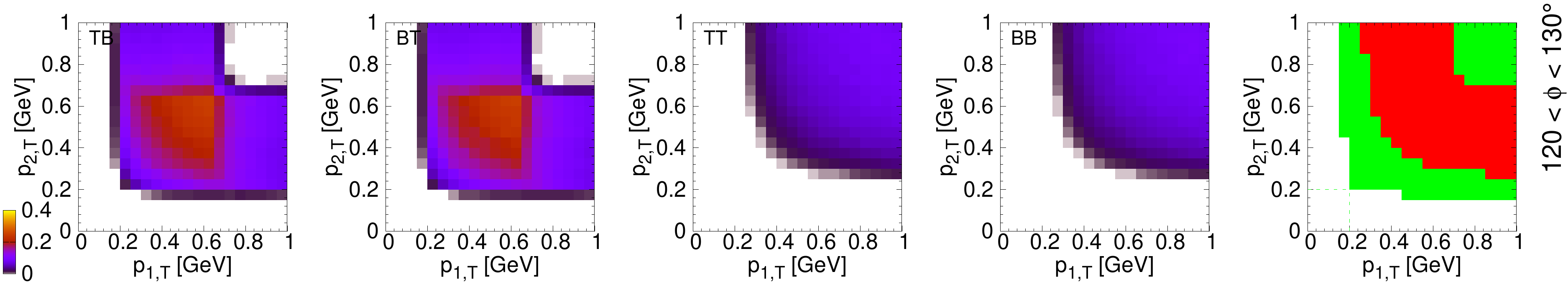}

 \caption{Calculated detection efficiencies for the pair of scattered protons
as functions of their transverse momenta $(p_\text{1,T}, p_\text{2,T})$ in some
selected bins of the $\Pp\Pp$ azimuthal angle $\phi$ (indicated on the right side of
each row). The first four columns show the efficiencies for each trigger
configuration, the four rows show four different angular ranges, and the
rightmost column displays the coverage of the measurement with colour codes
(white: not covered; green: covered by at least one configuration; red: covered
by all configurations). Lines corresponding to 0.2\GeV are drawn in the
rightmost plots.}

 \label{fig:rp_eff_thei}

\end{figure*}

\subsection{Proton-pair trigger acceptance (elastic veto)}
\label{sec:rp_trig}

The elastic $\Pp\Pp$ cross section is much larger than that for central exclusive
production events and, in the absence of an appropriate trigger algorithm,
would saturate the available bandwidth of the data acquisition system.
Therefore, elastic events need to be rejected. Triggering is based on ``trigger
roads'', each consisting of 32 adjacent silicon strips with the aim of reducing
the number of channels to handle. Since there are 512 strips in a plane, there
are 16 trigger roads for each orientation (u and v). The trigger bit is set if
trigger roads at the same location have signals above a predefined threshold in
at least three planes. Suitable combinations of these trigger bits from RPs in
diagonal configuration are used to reject elastic events. This algorithm is
referred to as ``elastic veto''.

To determine the efficiency loss due to the elastic veto algorithm,
proton tracks from events with parallel trigger configurations are used. A
track from a TT event is combined with one from a BB event, and the resulting
diagonal event is used as input to the veto algorithm. These events are used to
build an efficiency table as a function of $(p_{1,y},p_{2,y})$, which is used
later for the calculation of the RP-related corrections
(Section~\ref{sec:rp_eff}). The fraction of elastic-like events left after
veto trigger is shown in Fig.~\ref{fig:veto_py12} (left) as a
function of $p_y$ for Arms 1 and 2. It compares well with the measured
$(p_{1,y},p_{2,y})$ distribution of the events obtained with the diagonal
trigger configurations, as shown in Fig.~\ref{fig:veto_py12} (middle).

\subsection{Coverage and trigger acceptance}
\label{sec:rp_eff}

The combined trigger and detection acceptance, as well as the efficiency of the
RP system are calculated in bins of $(p_\text{1,T},p_\text{2,T},\phi)$ using
400 million simulated two-proton events. These events are generated with
uniform and independent $p_\text{1,T}$ and $p_\text{2,T}$ distributions in the
range $(0, 1)\GeV$ and uniform $\phi$ distribution in the range $(0, \pi)$,
using the single-proton and proton-pair trigger acceptances discussed above.
The detection efficiencies for a pair of scattered protons in selected bins
of the $\Pp\Pp$ azimuthal angle $\phi$ are shown in
Fig.~\ref{fig:rp_eff_thei}, where the rightmost column displays the joint
efficiency using all configurations, \ie, the coverage of the measurement.
In general, large regions are populated by all four configurations. Some
corners of phase space are not covered: they are at high $p_\text{1,T}$ and low
$p_\text{2,T}$ (and vice versa), if $\phi < 20^\circ$ or $\phi > 160^\circ$; at
low $p_\text{1,T}$ and $p_\text{2,T}$, if $70 < \phi < 110^\circ$; and at
high $p_\text{1,T}$ and $p_\text{2,T}$, if $80 < \phi < 100^\circ$.
Regions where the proton detection efficiency is below 2\% are not used in the
analysis.

\begin{figure*}[htb]

 \centering
 \includegraphics[width=0.32\textwidth]{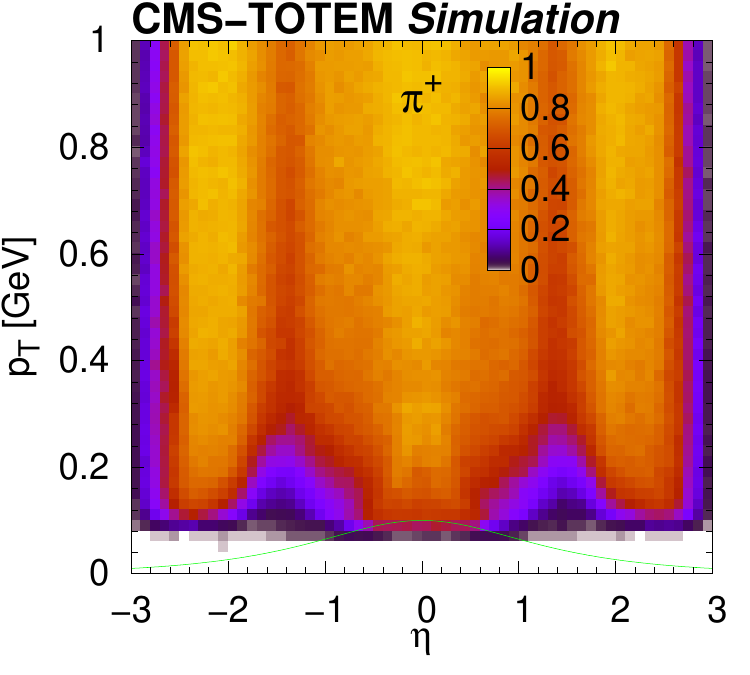}
 \includegraphics[width=0.32\textwidth]{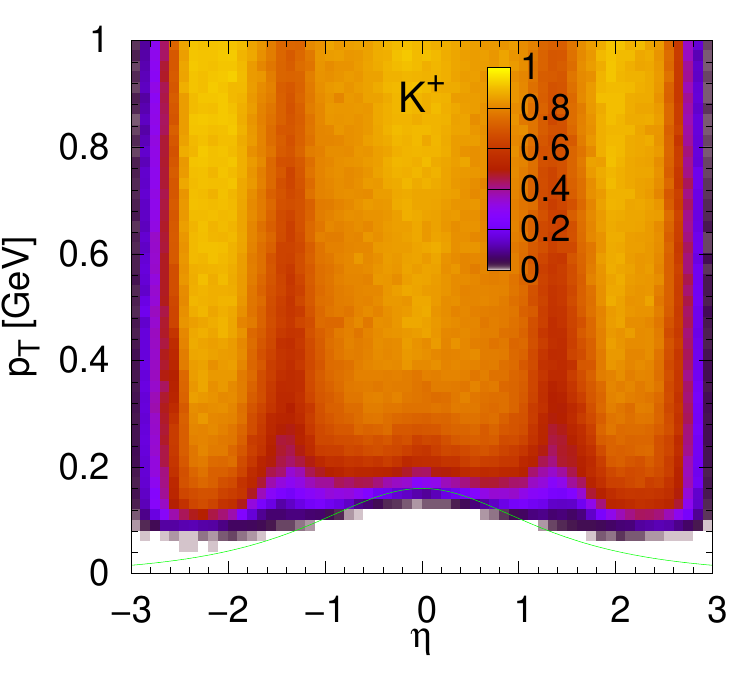}
 \includegraphics[width=0.32\textwidth]{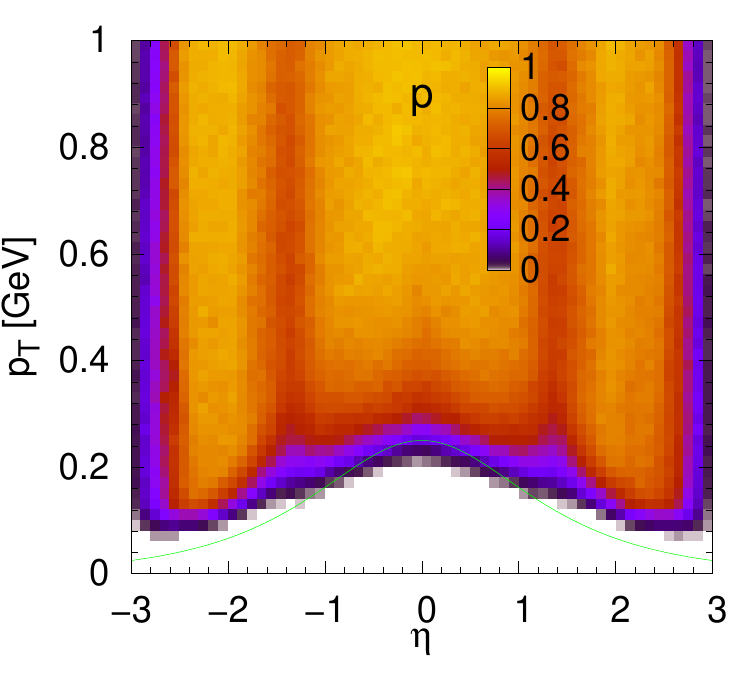}

 \caption{The combined reconstruction and HLT efficiency (reconstructed and
passing the HLT selection) for positively charged pions, kaons, and protons as
functions of $(\eta,\pt)$.
Curves indicate constant total momentum ($p = 0.1\GeV$ for pions, 0.16\GeV for
kaons, 0.25\GeV for protons). Plots for negatively charged particles are
similar.}

 \label{fig:tr_hlt_eff}

\end{figure*}

\section{Particles produced in the central region}
\label{sec:particles}

The finding and fitting of the charged-particle trajectories in the silicon
tracker are performed with well-tested methods~\cite{CMS:2014pgm}, which
include the results of an optimisation for low-momentum particles. In the
following, we discuss some single- and two-track parameters using events with
exactly two oppositely charged reconstructed particles. The distribution of
the $z$ coordinate of the reconstructed charged particles at their closest
approach to the beamline is Gaussian with a fitted mean value of $\langle z
\rangle = -0.37\cm$ and standard deviation $\sigma_z = 4.63\cm$. The
distribution of the transverse impact parameter of the reconstructed charged
particles has a Cauchy distribution with $\Gamma = 0.046\cm$. The distribution
of the difference $z_4 - z_3$ of the reconstructed charged particle pair has
again a Cauchy distribution with $\Gamma = 0.139\cm$. Events are selected if
the $z$ values of both tracks satisfy $\abs{z - \langle z \rangle} < 4
\sigma_z$, and their vertex is closer to the beamline than 1\cm. The latter
requirement eliminates photon conversions in the beam pipe and pixel tracker
layers, while significantly reducing the contribution of long-lived decays.
Low $\pt$ particles looping in the solenoidal magnetic field
might rarely (rate below 0.5\%) be reconstructed as two oppositely charged
particles with opposite momentum vectors. Their contribution is visible in the
distribution of $\abs{\vec{p_3} + \vec{p_4}}/m$ as peaks near zero. In the
analysis, events containing a looper are removed by requiring $\abs{\vec{p_3} +
\vec{p_4}}/m > 0.2$, where $m$ is the mass of the identified hadron. The
corresponding event loss is negligible.

\subsection{High-level trigger and tracking efficiency}
\label{sec:hltrk}

The distributions of charged particles in the $(\eta,\pt)$ plane reveal
``valleys'' corresponding to inefficiencies at low \pt. These are present
because the HLT contains pixel activity filters and various pixel track filters
(Section~\ref{sec:datataking}). Having a simple single-track tracking
efficiency table is not sufficient, and a combined ``track-pair HLT and
tracking'' efficiency is necessary. We have generated, simulated, and
reconstructed single pions, kaons, and protons of both electric charges
uniformly in the kinematic range $-3 < \eta < 3$, $0 < \pt < 2\GeV$ (30 million
events for each species). These events are used to determine the reconstruction
efficiency and to study the distribution of the hit patterns in the pixel
layers. The combination of these is then used to determine the HLT and
reconstruction efficiency of two-track events (Section~\ref{sec:datataking}).
The extracted single-particle reconstruction efficiencies show the acceptance
edge near $\abs{\eta} \approx 2.5$ and the efficiency losses at low total
momenta because of multiple Coulomb scattering and energy loss. Fragments of a
charged-particle trajectory are rarely reconstructed as separate tracks. These
are concentrated at $\eta \approx 0$ for pions (loopers), and around
$\abs{\eta} \sim 2$ for kaons (scatters or decays), but their frequency is at
or below the 1\% level. The combined reconstruction and HLT efficiencies for
single particles are shown in Fig.~\ref{fig:tr_hlt_eff}, projected onto the
$(\eta,\pt)$ plane. These plots show a significant decrease of efficiency in
the region of the barrel-endcap transition because of the small number of
clusters of pixel detector layers with hits.

\begin{figure*}[htb]

 \centering
 \includegraphics[width=0.32\textwidth]{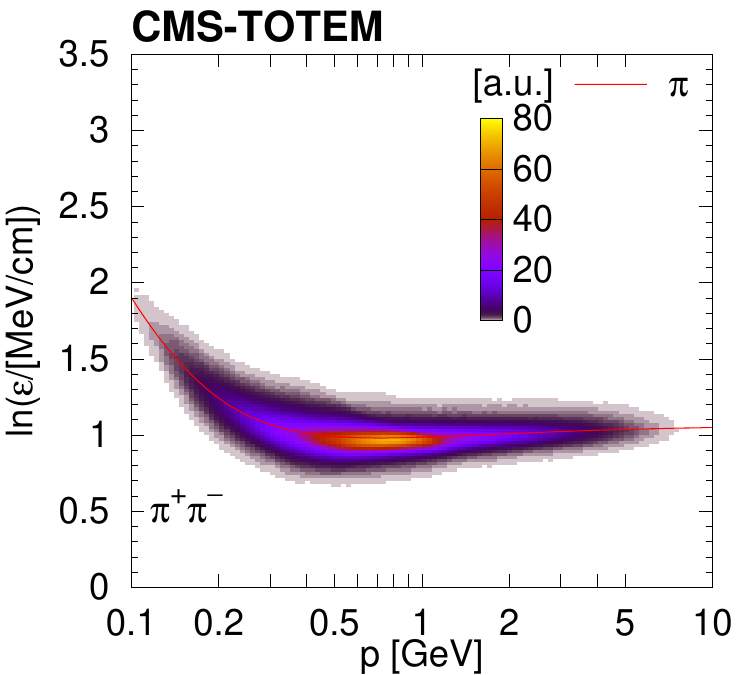}
 \includegraphics[width=0.32\textwidth]{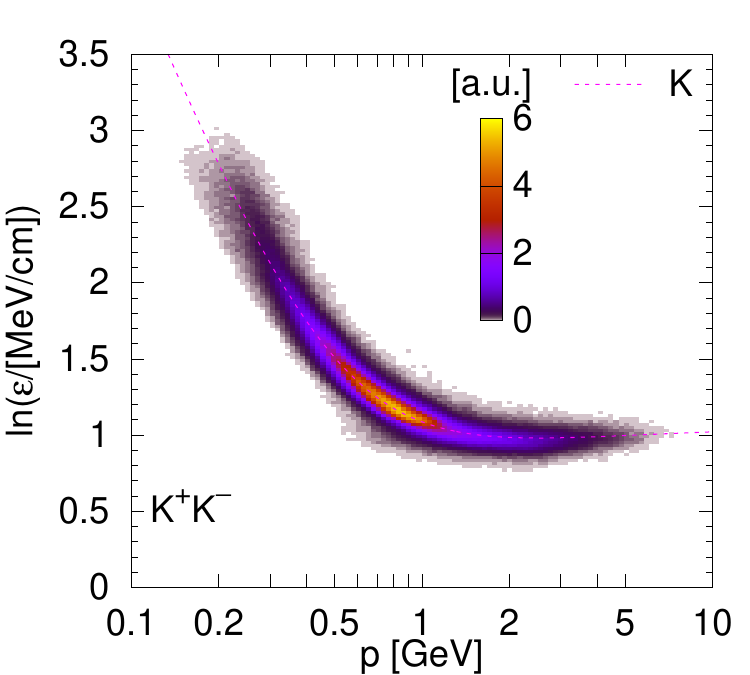}
 \includegraphics[width=0.32\textwidth]{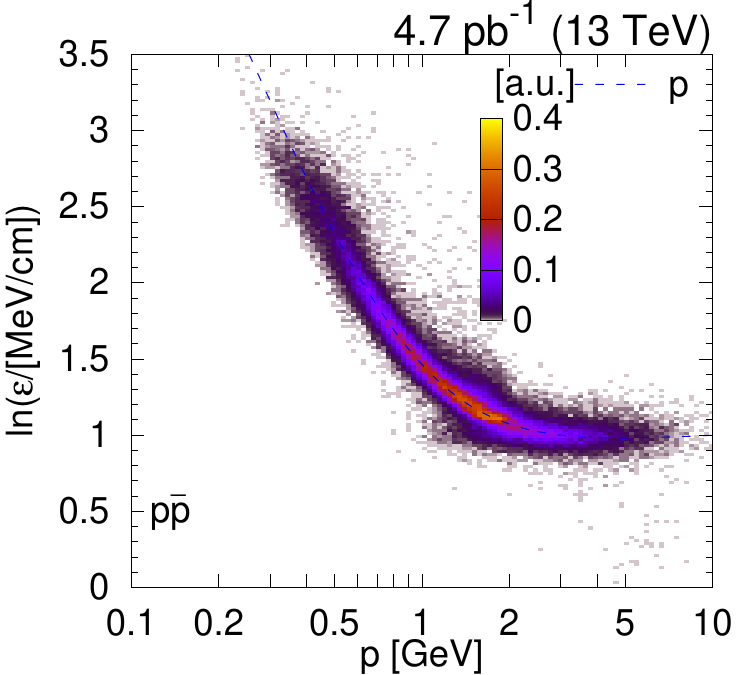}

 \includegraphics[width=0.32\textwidth]{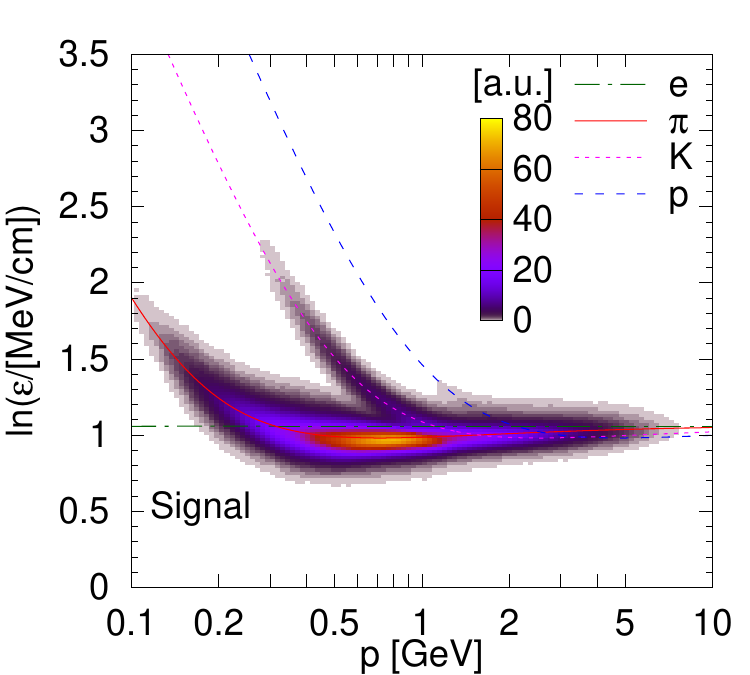}
 \includegraphics[width=0.32\textwidth]{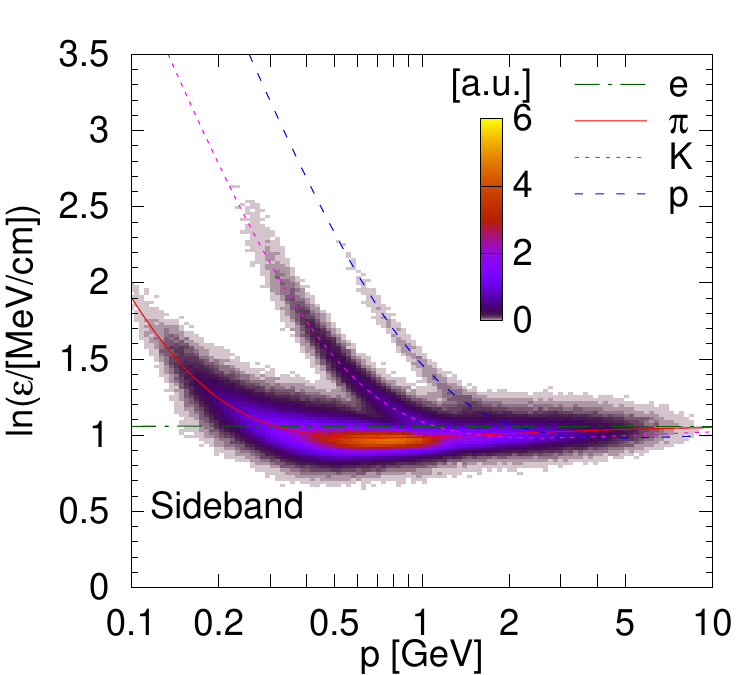}

 \caption{Distribution of $\ln\varepsilon$ as a function of total momentum, for
reconstructed charged particles in selected two-track events (identified
$\PGpp\PGpm$, $\PKp\PKm$, $\Pp\PAp$, signal, and sideband; Section~\ref{sec:classification}). The variable $\varepsilon$ is the most
probable energy loss rate at a reference path length $l_0 = 450\mum$. The
colour scale is shown in arbitrary units and is linear. The curves show the
expected $\ln\varepsilon$ for electrons, pions, kaons, and protons (Eq.~(34.12)
in Ref.~\cite{ParticleDataGroup:2022pth}).}

 \label{fig:eloss}

\end{figure*}

\subsection{Particle identification}
\label{sec:pid}

Details on the determination of the silicon strip properties, detector gain
calibration, model validation, and on the estimation of the most probable
energy loss rate $\varepsilon$ (or its logarithm) and its variance
$\sigma_{\ln\varepsilon}^2$ based on control samples in data are identical to
those already discussed in Ref.~\cite{Sirunyan:2017zmn}. The distributions of
$\ln\varepsilon$ as a function of momentum, for reconstructed charged particles
in selected two-track events are shown in Fig.~\ref{fig:eloss}, including
identified $\PGpp\PGpm$, $\PKp\PKm$, and $\Pp\PAp$ pairs (without track-RP
momentum matching), as well as signal and sideband distributions, which
correspond to events where momentum is not conserved in the transverse plane
(Section~\ref{sec:classification}). The curves show the expected values
according to Eq.~(34.12) in Ref.~\cite{ParticleDataGroup:2022pth}. While the
sideband region displays a reasonable amount of pions, kaons, and protons, the
signal region reveals only few protons in the sample (Fig.~\ref{fig:eloss},
lower left). The exclusive production of $\Pp\PAp$ pairs is suppressed because
of the limited energy and phase space available for pair creation. For
two-track identification, the knowledge of $\sigma_{\ln\varepsilon}$ is
essential, and is extracted from data using identified pions, kaons, and
protons in narrow bins for the ranges $-3 < \eta < 3$ and $\pt < 2\GeV$.

The probability density function $P_k$ for identifying a charged particle with
momentum $p$, and estimated values $\ln\varepsilon$ and
$\sigma_{\ln\varepsilon}^2$ as being of type $k$ (\eg, a pion, kaon, or proton)
is a Gaussian function with mean $\langle\ln\varepsilon\rangle_k(p)$. A
particle pair is identified of type $\mathrm{h^+h^-}$ if that choice is at
least ten times more likely than the others, $P_{1,\mathrm{h}} P_{2,\mathrm{h}}
> 10 P_{1,k} P_{2,k}$, where $k \neq h$. If no type choice fulfils any of
the above conditions, the particle-pair is left unidentified. To verify the
particle identification capabilities, a detailed simulation was set up. Samples
with oppositely charged particles ($\PGpp\PGpm$, $\PKp\PKm$, and $\Pp\PAp$)
were generated uniformly in the range $-3 < \eta < 3$ and in total momentum
$(p_3,p_4)$ up to 2\GeV; each sample has 100~000 events. The most probable
value of $\varepsilon$ was taken from a model~\cite{ParticleDataGroup:2022pth}
using the density correction according to Ref.~\cite{Sternheimer:1983mb}; its
relative standard deviation was sampled from the measured distribution of
$\sigma_{\ln\varepsilon}$. The identification efficiencies are close to 100\%
at low momenta, and slowly decrease above approximately 1\GeV (pions and kaons)
or 2\GeV (protons).

\begin{figure*}[htb]

 \centering
 \includegraphics[width=0.49\textwidth]{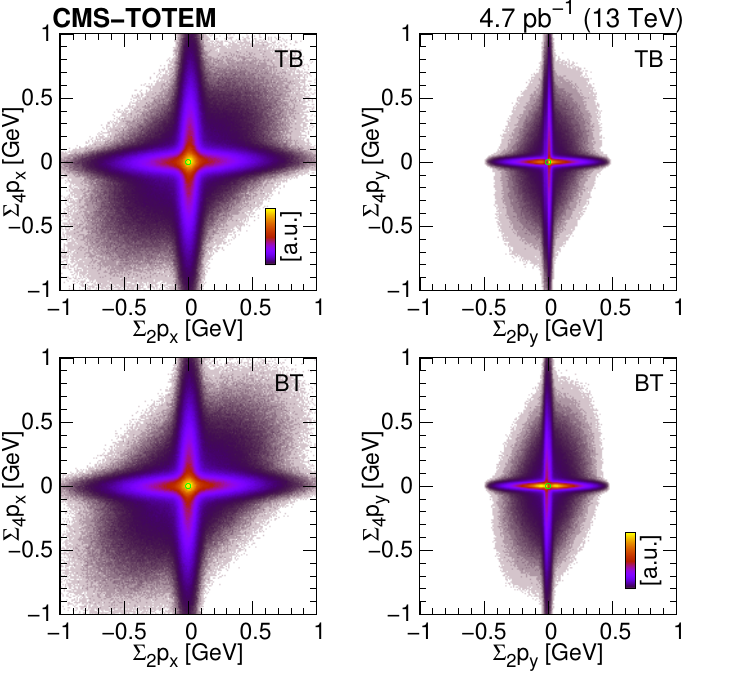}
 \includegraphics[width=0.49\textwidth]{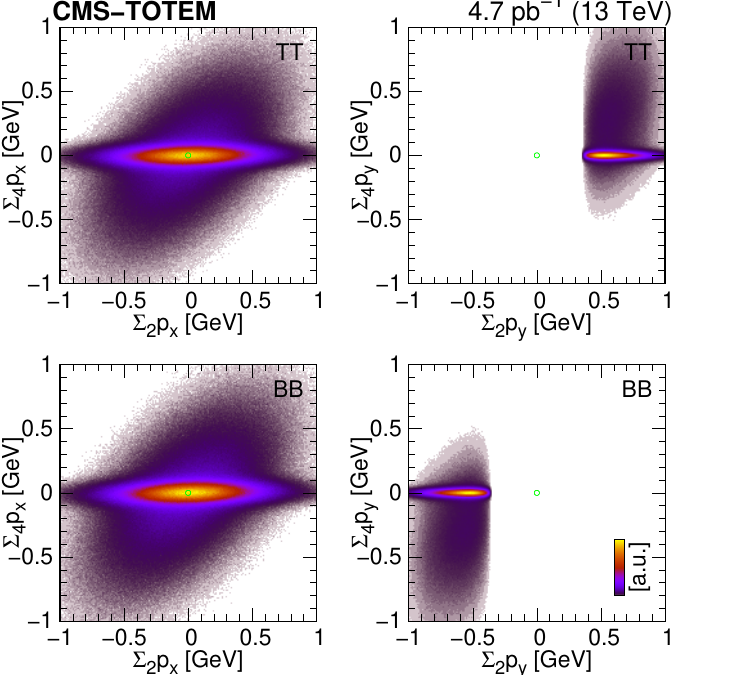}

 \caption{Distribution of the sum of the scattered proton and central hadron
momenta and the sum of the scattered proton momenta ($\sum_4 p_x$ vs. $\sum_2
p_x$, $\sum_4 p_y$ vs. $\sum_2 p_y$) shown for the diagonal trigger
configurations (TB and BT, left) and the parallel ones (TT and BB, right) in
the case of 2-track events.}

 \label{fig:sump}

\end{figure*}
 
\subsection{Combined corrections for the silicon tracker}

The combined HLT, reconstruction, and identification efficiencies are
calculated in the four-dimensional space of
$[\phi,m_\mathrm{h^+h^-},(\cos\theta,\varphi)_\text{GJ}]$
(Section~\ref{sec:limits}), separately for each $(p_\text{1,T},p_\text{2,T})$
bin, using exclusive two-track events generated with uniform distributions.
The combined efficiency is usually above 10\% and can reach 90\%. It is lower
than 1\% only in small regions, specifically in the first 20\MeV wide mass
bin at the threshold $2 m_{\PGp/\PK/\Pp}$ near $\phi \approx \pi$, and in the
case of kaons for $m > 2.6\GeV$ because of the reduced particle identification
capabilities at higher momenta. In summary, we have coverage in
$[\phi,m,(\cos\theta,\varphi)_\text{GJ}]$ through most of the RP trigger
configurations. The corrections described so far are included in the
analysis by weighting events with the product of three factors:

\ifthenelse{\boolean{cms@external}}{
\begin{multline}
 \left(\frac{1}{L_\text{int}}\right)_\text{trigger conf.} \,
 \left(\frac{\prod \text{tracklet weights}}
      {\text{azimuthal accep.}}\right)_\text{RPs} \, \\ 
 \times \left(\frac{1}{\text{combined effic.}}\right)_\text{tracker},
\end{multline}
}{
\begin{equation}
 \left(\frac{1}{L_\text{int}}\right)_\text{trigger conf.} \,
 \left(\frac{\prod \text{tracklet weights}}
      {\text{azimuthal accep.}}\right)_\text{RPs} \,
 \left(\frac{1}{\text{combined effic.}}\right)_\text{tracker},
\end{equation}
}
which refer to the actual RP trigger configuration of the event, the Roman pot
and the tracker related corrections, respectively.

\begin{table}[bh]

 \centering

 \caption{Ranges (in parentheses) for bias and resolution of the reconstructed
transverse momentum and the two-hadron invariant mass, shown for pions, kaons,
and protons, in \MeV units.}

 \label{tab:resol}

 \begin{scotch}{llccc}
  & & \PGp & \PK & \Pp \\
  \hline
  \multirow{2}{*}{Single hadrons} 
  & \pt bias 
  & $(0,+5)$ & $(-10,+5)$ & $(-15,+5)$ \\
  & \pt resolution & $(5,15)$ & $(10,20)$ & $(15,25)$ \\
  \multirow{2}{*}{$\mathrm{h^+h^-}$ pairs}
  & $m$ bias
  & $(0,+4)$ & $(-2,0)$ & $(-2,0)$ \\
  & $m$ resolution
  & \multicolumn{3}{c}{$15$ at $m_{\PGpp\PGpm} \approx 0.6\GeV$}
 \end{scotch}

\end{table}

\subsection{Momentum and mass resolutions}
\label{sec:resol}

The bias and resolution of the reconstructed transverse momentum and the
two-hadron invariant mass are shown in Table~\ref{tab:resol} for pions, kaons,
and protons. The \pt bias for kaons and protons occurs because all particles
are reconstructed with the pion mass assumption but the physical effects of
particle passage through matter are mass and momentum dependent. The \pt
resolutions also show some phase space dependence. The uncertainties seen in
momentum sum distributions below (Section~\ref{sec:classification}) mostly come
from the proton momentum reconstruction of the RP system. The bias of the
two-hadron invariant mass is proportional to the decay momentum of the daughter
in the centre-of-mass frame. In summary, the observed shifts are much smaller
than the bin width of $20\MeV$, but become comparable with it at higher masses.
For uneven mass distributions (narrow resonances), these effects should be
properly unfolded, but for smooth distributions (nonresonant continuum) an
unfolding is not needed.

\section{Event classification}
\label{sec:classification}

The classification of events is largely based on momentum conservation in the
transverse plane. Once corrected for the effects of the beam crossing angle
($120\,\mu\mathrm{rad}$), the system of the colliding protons has zero momentum.
Therefore, the sum of the momenta of the scattered protons and the other
particles created in the collision should also be zero. Momentum conservation
in the $z$ direction is already utilised for the calculation of longitudinal
momenta of the scattered protons, since the resolution from a direct
measurement would be poor.

The sum of scattered proton transverse momenta ($\sum_2 \vec{p_\text{T}}$ for
these two particles) and the sum of scattered proton and central hadron
transverse momenta ($\sum_4 \vec{p_\text{T}}$ for these four particles) play an
important role. The joint distributions of their components ($\sum_4 p_x$ vs.
$\sum_2 p_x$, $\sum_4 p_y$ vs. $\sum_2 p_y$) are shown for each trigger
configuration for 2-track events in Fig.~\ref{fig:sump}. The contributions of
elastic pileup (two scattered protons and two unrelated central charged
hadrons, vertical band) and central exclusive events (two scattered protons and
two central charged hadrons, horizontal band) are visible. In addition, a
slanted area of nonexclusive or inelastic background is present. In the case
of TT and BB trigger configurations, the vertical band is absent, since such
configurations ($\sum_2 p_y > 2 \times 0.2\GeV$) prevent the recording of
elastic events.

\begin{figure*}[htb]

 \centering
 \includegraphics[width=0.49\textwidth]{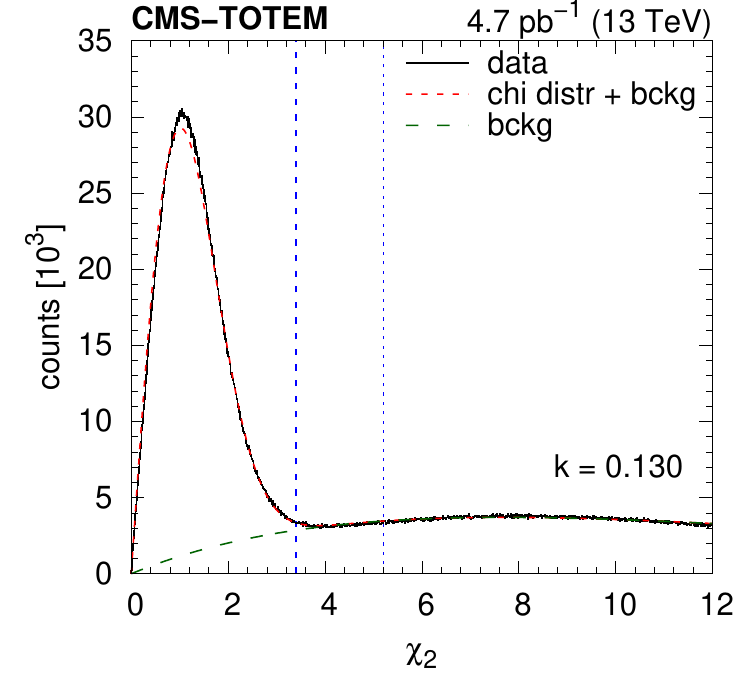}
 \includegraphics[width=0.49\textwidth]{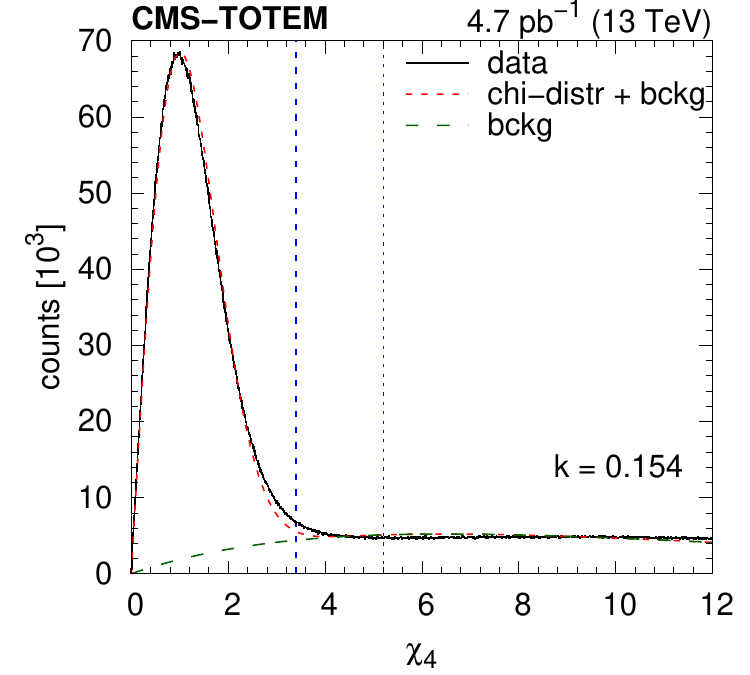}

 \includegraphics[width=0.49\textwidth]{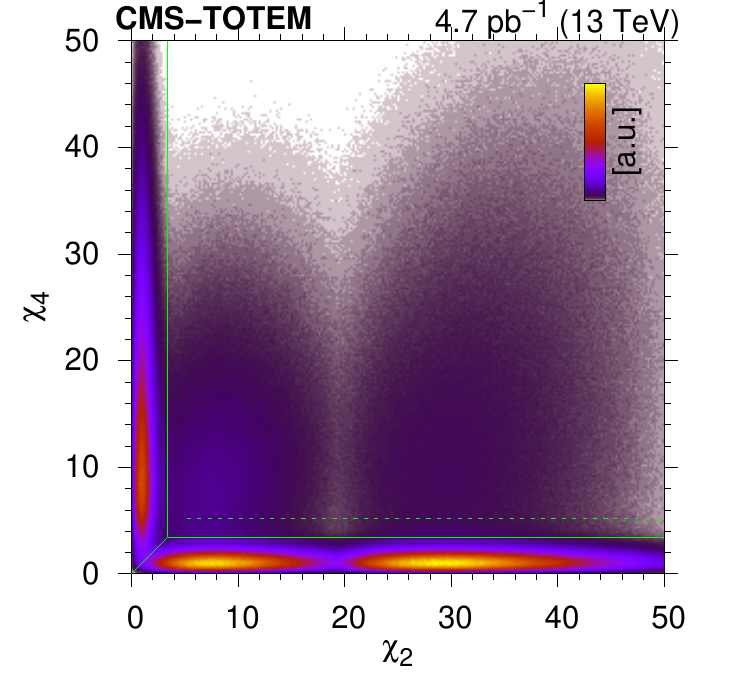}
 \includegraphics[width=0.49\textwidth]{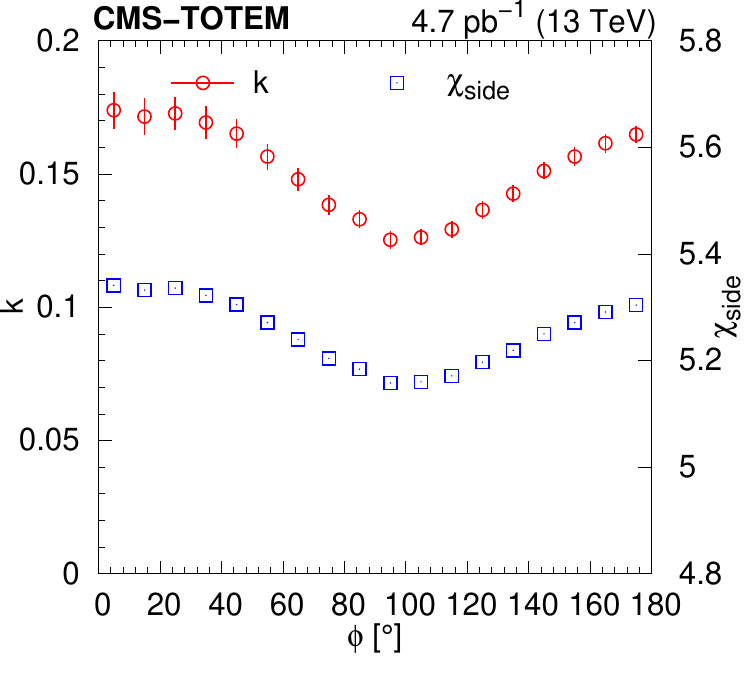}

 \caption{Distributions of the classification variables $\chi_2$ (upper left)
and $\chi_4$ (upper right). Fits using a two-component model are indicated. The
distributions are integrated over the angle between the scattered protons
$\phi$. Selection lines at $\chi_\text{sign} = 3.4$ (green solid) and
$\chi_\text{side} \approx 5.2$ (green dotted) are also plotted. Lower left:
Joint distributions of classification variables $\chi_2$ and $\chi_4$. Central
exclusive signal events are at the bottom whereas elastic events are at the left
margin. Lower right: Coefficient $k$ and the position of the upper cutoff
$\chi_\text{side}$ for $\chi_4$ to describe the background component as a
function of the angle between the scattered protons $\phi$, in the plane
transverse to the beam direction.}

 \label{fig:chi}

\end{figure*}

The classification variable is the Mahalanobis distance $\chi$, which is a
multidimensional generalisation of the standard deviation taking into account
internal correlations~\cite{chandra1936generalised}. It is based on the value
and covariance of momentum sums, defined in the multivariate normal case as
$\chi(\vec{s}) = (\vec{s}^T V^{-1} \vec{s})^{1/2}$ where $V(\vec{s})$ is the
covariance matrix. For two-dimensional vectors $(s_x,s_y)$ it can be written as

\begin{equation}
 \chi(\vec{s}) = \left(\frac{V_{yy} s_x^2 - 2 V_{xy} s_x s_y + V_{xx} s_y^2}
                            {V_{xx} V_{yy} - V_{xy}^2}\right)^{1/2}.
\end{equation}

For each event, the classification variable $\chi_2$ is based on $\vec{s} =
\sum_2 \vec\pt$, whereas the value of $\chi_4$ is computed from $\vec{s} =
\sum_4 \vec\pt$. They both follow the chi distribution of two degrees of
freedom (dof) with a parameterless probability density function $\chi
\exp(-\chi^2/2)$. Distributions and joint distributions of $\chi_2$ and
$\chi_4$ are shown in Fig.~\ref{fig:chi} without any preselection (integrated
over the azimuthal angle between the scattered protons $\phi$). The
distributions are fitted with a two-component model: a sum of a chi
distribution (signal) and a phase space motivated term (background) as $A \chi
\exp(-\chi^2/2) + B \chi \exp(- k \chi)$. The functional form fits the measured
distribution well, with two relevant parameters only ($B/A$ and $k$).

An event is used in the analysis if it is more likely a central exclusive
signal than an elastic+pileup background, $\chi_4 < \chi_2$. Signal events are
defined by $\chi_4 \le \chi_\text{sign} = 3.4$, where the numerical choice for
$\chi_\text{sign}$ translates into a signal loss below 0.5\%.
Sideband events in the region $\chi_\text{sign} \le \chi_4 < \chi_\text{side}$
are subtracted with weight of $-1$ to compensate for nonexclusive events in the
signal region. For this subtraction scheme to work, we must have equal
numbers of nonexclusive events in the signal and sideband regions. We need a
$\chi_\text{side}$ value that satisfies $\int_0^{\chi_\text{sign}} \chi \exp(-k
\chi) \rd\chi = \int_{\chi_\text{sign}}^{\chi_\text{side}} \chi \exp(-k \chi)
\rd\chi$. Although the choice of signal cutoff is fixed, the value of the
sideband cutoff depends on the actual distribution of $\chi_4$. The fitted
coefficient $k$ and the value of $\chi_\text{side}$ for $\chi_4$ as a function
$\phi$ are shown in Fig.~\ref{fig:chi} (lower right).

\section{Results}
\label{sec:results}

The measured distributions are the following: distribution of the
azimuthal angle $\phi$ between the scattered proton momenta,
       $\rd^3\sigma/\rd p_\text{1,T}\,\rd p_\text{2,T}\,\rd\phi$;
invariant mass of two hadrons $m$,
       $\rd^3\sigma/\rd p_\text{1,T}\,\rd p_\text{2,T}\,\rd m$;
squared four-momentum $\max(\hat{t},\hat{u})$ of the virtual meson,
       $\rd^3\sigma/\rd p_\text{1,T}\,\rd p_\text{2,T}\,\rd \max(\hat{t},\hat{u})$,
in the range $0.2 < (p_\text{1,T},p_\text{2,T}) < 0.8\GeV$. In the plots the
scattered protons are indexed such that $p_\text{1,T} \ge p_\text{2,T}$.
Tabulated results are provided in HEPData~\cite{HEPData}.

\subsection{Systematic uncertainties}

The relevant systematic uncertainties are listed in Table~\ref{tab:syst}, where
values propagated to the final differential cross sections are given. The
sources are the pileup correction (because of the uncertainty in the visible
cross section), the reduced RPs availability, the integrated luminosity, the
efficiency of the RPs, the removal of the nonexclusive background, the fraction
of signal events mistaken as background or lost because of the selection
against looping particles, and the efficiency of single-particle tracking.

The systematic uncertainty of the visible cross section $\sigma_\text{vis}$ is
estimated as half the difference between the value used and the minimum bias
value. This uncertainty is propagated to the final differential cross sections
through the pileup correction factor, and its contribution to the systematic
uncertainty is 1.0\%. A few short periods with reduced data taking efficiency
are due to the loss of one out of the four data streams. About 2--3\% of the
data segments are affected, leading to about 0.5\% systematic uncertainty. The
systematic uncertainty in the integrated luminosity (per data segment) is
2.5\%~\cite{CMS:2019jhq}. The efficiency of tracklet reconstruction in the RPs
depends on the knowledge of the individual strip efficiencies. These were
determined by a tag-and-probe method~\cite{CMS:2010svw} using 0-track and
2-track data sets. The effect on proton reconstruction efficiency is about 3\%.
To estimate the systematics related to the removal of nonexclusive background
($\chi_4$ distribution), the functional form of the background model was varied
(uncertainty below $0.5\%$). Signal events are lost because of the sideband
removal procedure. Their fraction is estimated with the help of the signal and
the background components of the $\chi_4$ distributions, resulting in an
uncertainty of $0.16\%$, which is neglected. The fraction of events lost
because of the looper removal procedure is estimated by fitting the
$\abs{\vec{p_3} + \vec{p_4}}/m$ distribution with a signal and a background
component. The estimated value is below $0.5\%$, which is again neglected. The
systematic uncertainty in the single-particle tracking efficiency of centrally
produced hadrons in the relevant low-momentum regions is 1.4\%, according to a
study based on control samples in data~\cite{CMS:2010mua}. To demonstrate the
particle identification capabilities for particle- antiparticle hadron pairs, a
detailed simulation was set up. The probability of misidentification is low,
and in the most populated low-momentum regions it is below 1\%.

The uncertainty of the RP and single-particle tracking efficiencies are
included for both arms, and for both centrally reconstructed charged hadrons.
The estimated total systematic uncertainty of the differential cross section
measurements is 5.4\%, with the various contributions treated as independent
and added in quadrature.

\begin{table}[htb]

 \caption{Systematic uncertainties of the differential cross sections.}

 \label{tab:syst}

 \centering
 \begin{scotch}{lrl}
  Source & Value & Remark \\
  \hline
  Pileup correction                      &  $1.0\%$ &
                Through $\sigma_\text{vis}$ \\
  Periods with reduced RP availability   &  $0.5\%$ \\
  Integrated luminosity ($L_\text{int}$) &  $2.5\%$ \\
  HLT efficiency                         &  $<0.1\%$ & Neglected \\
  Total normalisation-type               &  $2.7\%$ \\
  [\cmsTabSkip]
  Roman pot efficiency                   &  $3.0\%$
                                                     & Twice \\
  Background removal                     &  $<0.5\%$ & Neglected \\
  Lost events during background removal  &  $0.16\%$ & Neglected \\
  Lost events due to looper removal      &  $<0.5\%$ & Neglected \\
  Single-particle tracking efficiency    &  $1.4\%$  & Twice \\
  Particle identification efficiency     &  $<1.0\%$ & Neglected \\
  Total efficiency-type                  &  $4.7\%$ \\
  [\cmsTabSkip]
  Total systematics                      &  $5.4\%$
 \end{scotch}
\end{table}

 \begin{figure*}[htb]

 \centering
 \includegraphics[width=0.65\textwidth]{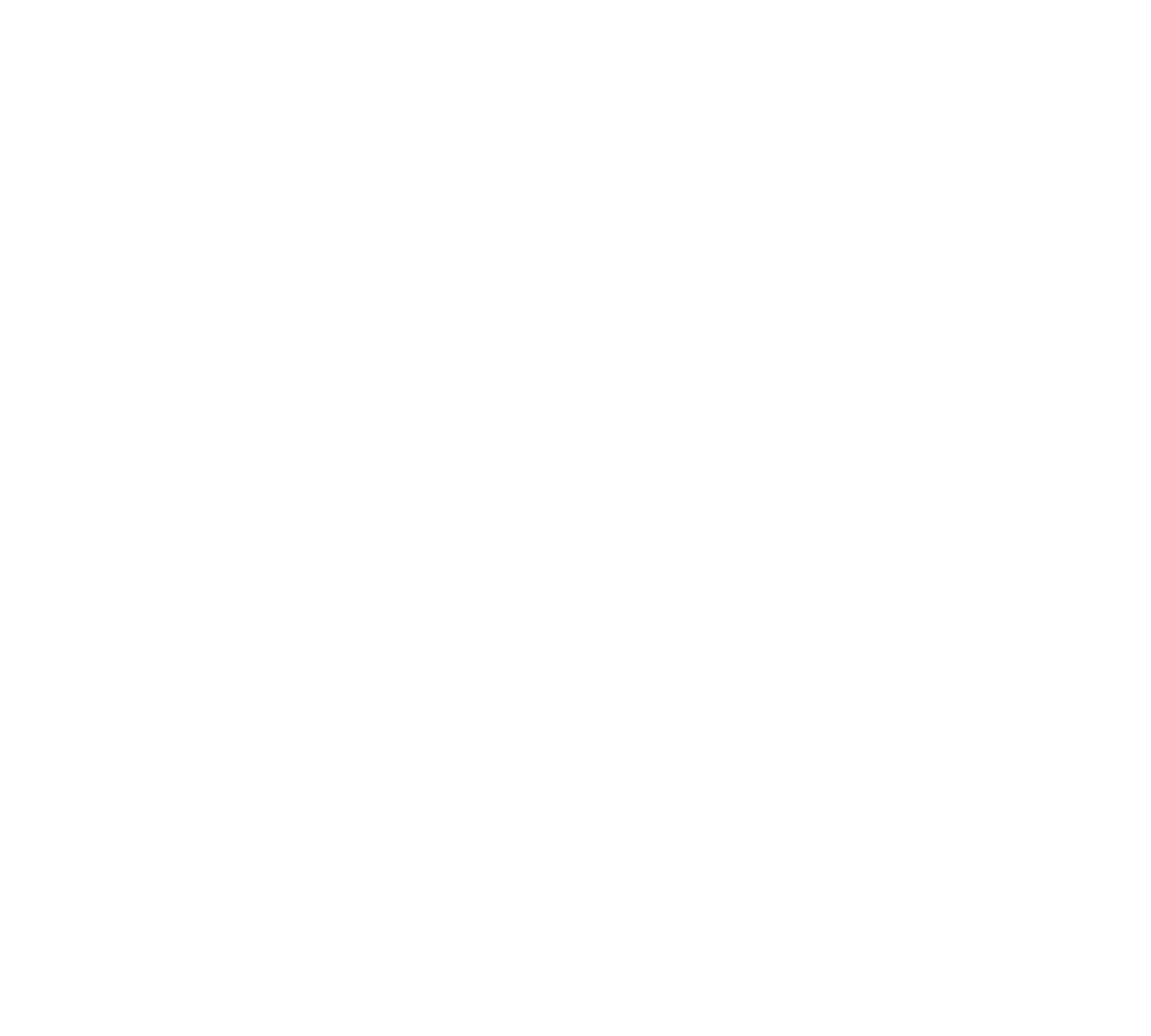}
 \includegraphics[width=0.65\textwidth]{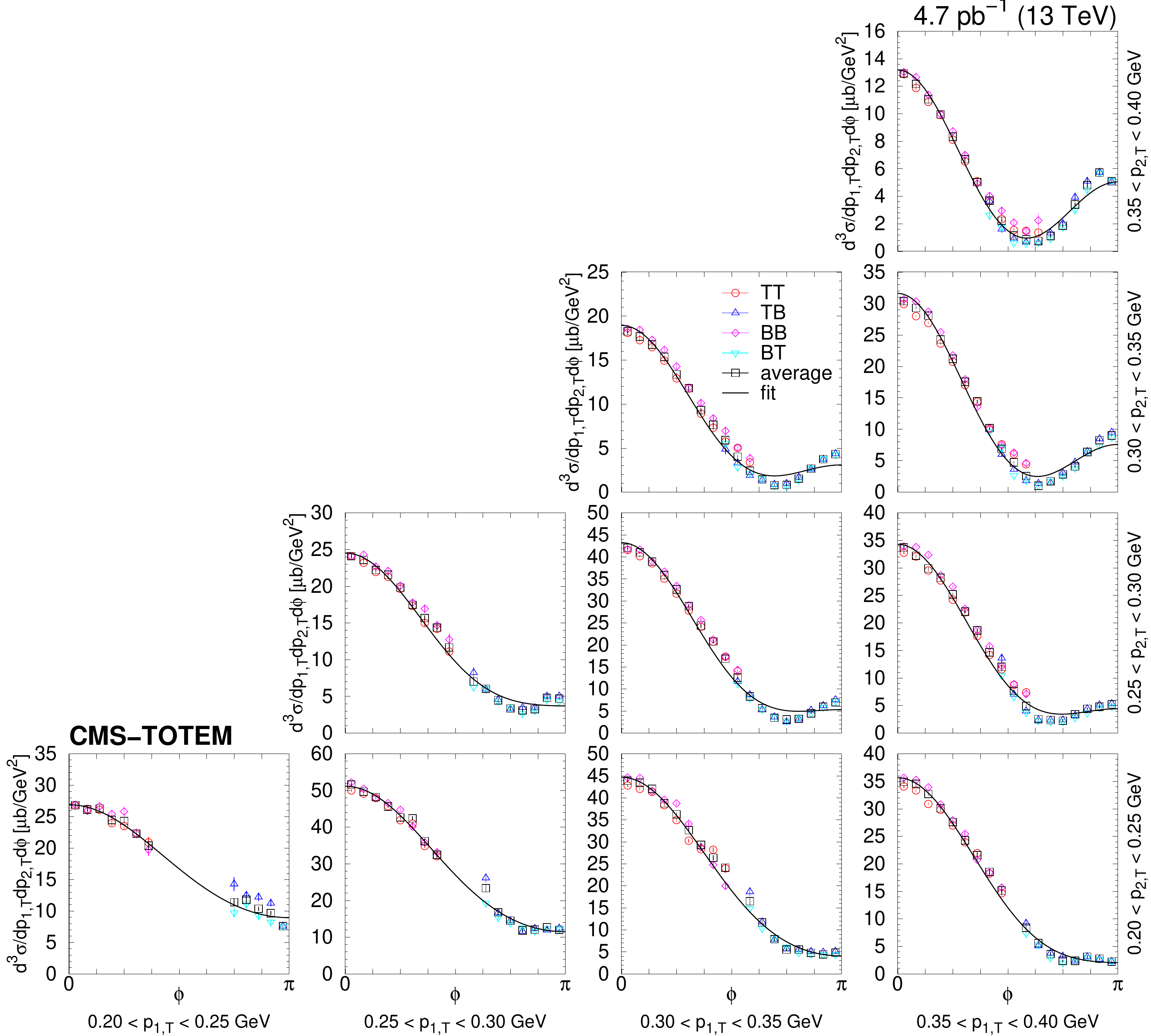}

 \caption{Distributions of $\rd^3\sigma/\rd p_\text{1,T} \rd p_\text{2,T} \rd\phi$
as functions of $\phi$ in the $\PGpp\PGpm$ nonresonant region ($0.35 <
m_{\PGpp\PGpm} < 0.65\GeV$) in several $(p_\text{1,T},p_\text{2,T})$ bins in
the range $0.20<p_\text{1,T}<0.40\GeV$ and $0.20<p_\text{2,T}<0.40\GeV$, in units of
$\mu\mathrm{b}/\GeVns^2$. Values based on data from each RP trigger configuration
(TB, BT, TT, and TT) are shown separately with coloured symbols, whereas the
weighted average is indicated with black symbols. Results of individual fits
with the form $[A(R - \cos\phi)]^2 + c^2$ (Eq.~\eqref{eq:ARc}) are plotted with
the curves. The error bars indicate the statistical uncertainties.}

 \label{fig:dndphi_detail_1_1}

 \end{figure*}

 \begin{figure*}[htb]

 \centering
 \includegraphics[width=0.65\textwidth]{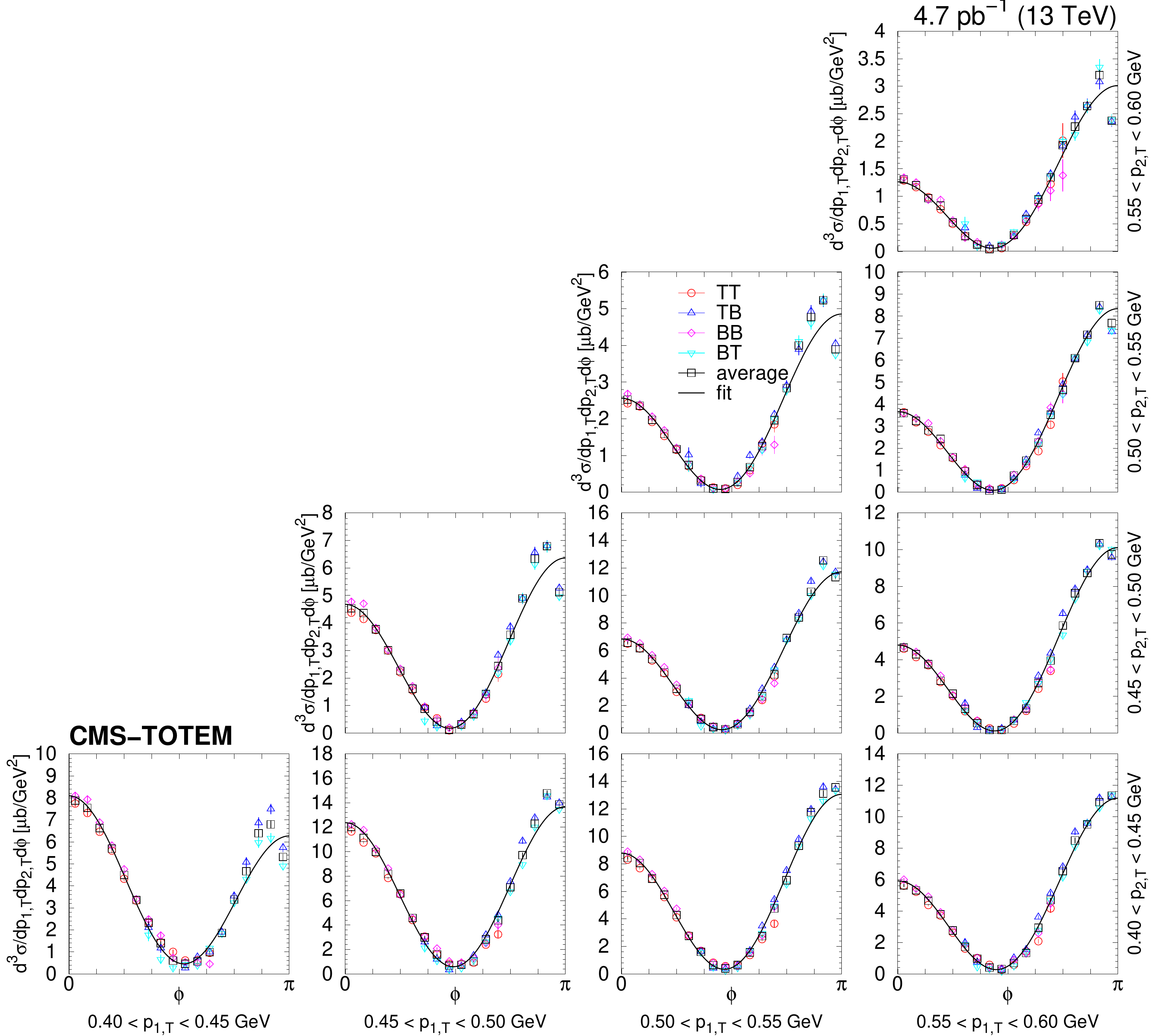}
 \includegraphics[width=0.65\textwidth]{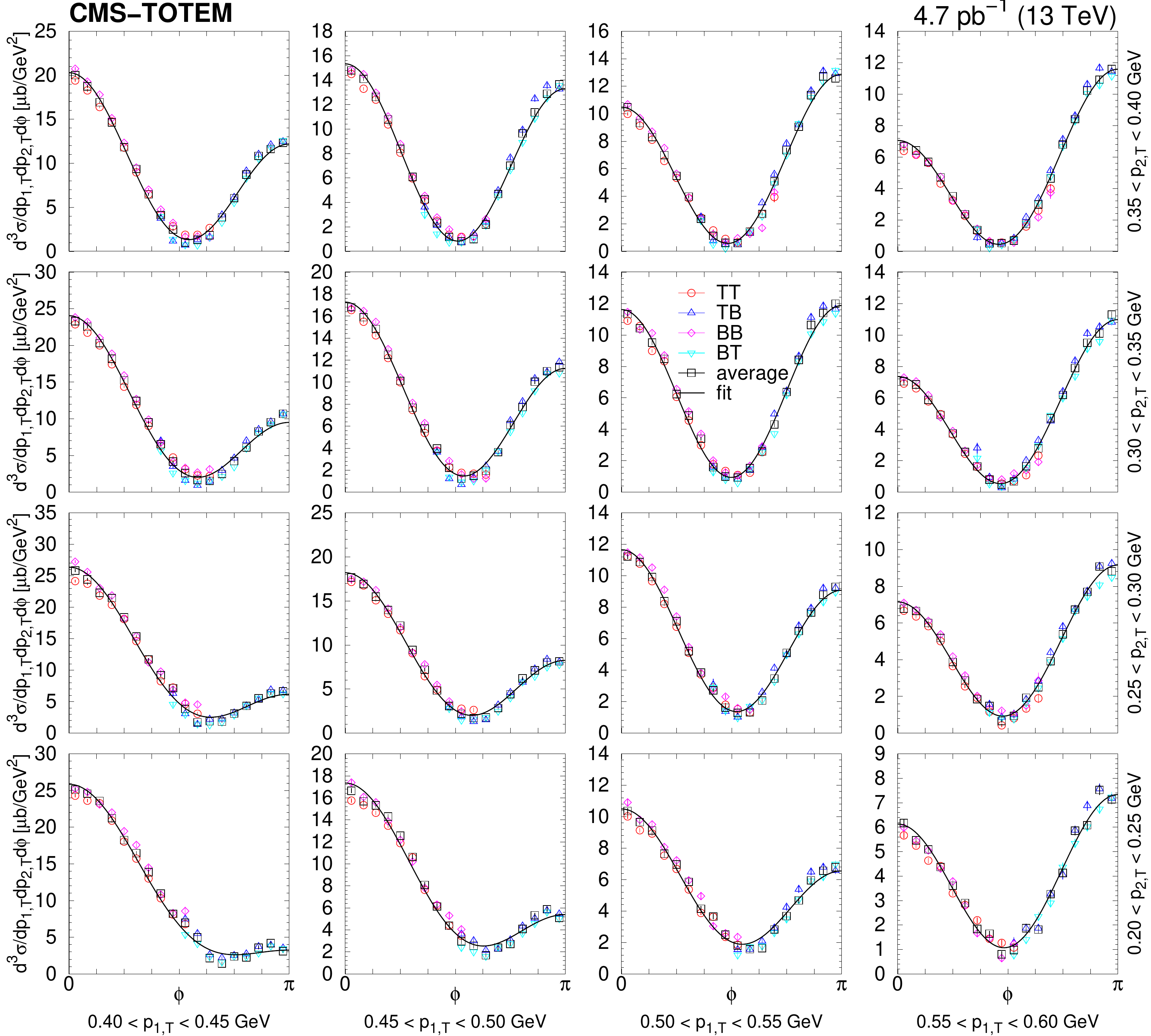}

 \caption{Distributions of $\rd^3\sigma/\rd p_\text{1,T} \rd p_\text{2,T} \rd\phi$
as functions of $\phi$ in the $\PGpp\PGpm$ nonresonant region ($0.35 <
m_{\PGpp\PGpm} < 0.65\GeV$) in several $(p_\text{1,T},p_\text{2,T})$ bins in
the range $0.40<p_\text{1,T}<0.60\GeV$ and $0.20<p_\text{2,T}<0.60\GeV$, in units of
$\mu\mathrm{b}/\GeVns^2$. Values based on data from each RP trigger configuration
(TB, BT, TT, and TT) are shown separately with coloured symbols, whereas the
weighted average is indicated with black symbols. Results of individual fits
with the form $[A(R - \cos\phi)]^2 + c^2$ (Eq.~\eqref{eq:ARc}) are plotted with
the curves. The error bars indicate the statistical uncertainties.}

 \label{fig:dndphi_detail_1_2}

 \end{figure*}

 \begin{figure*}[htb]

 \centering
 \includegraphics[width=0.65\textwidth]{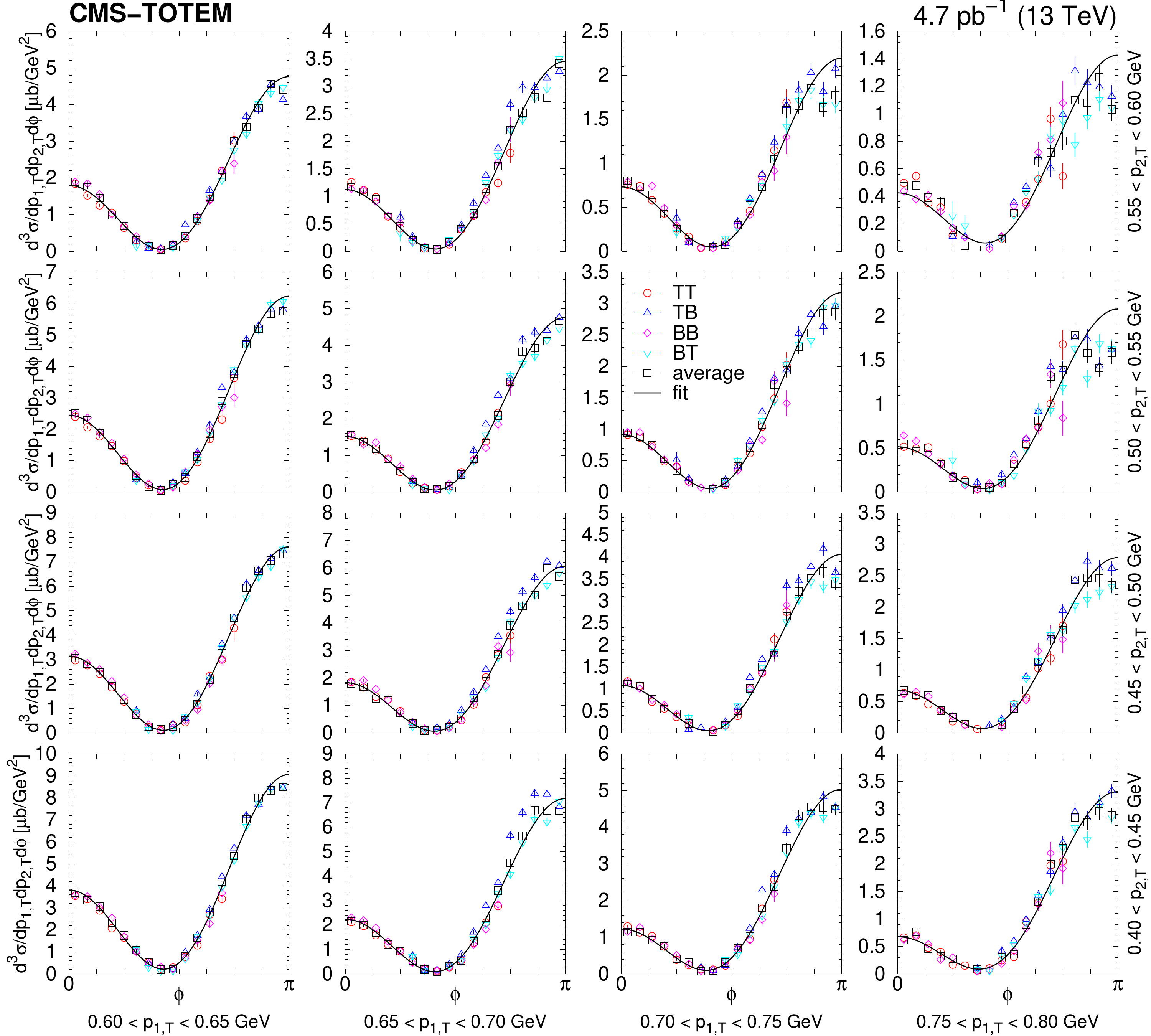}
 \includegraphics[width=0.65\textwidth]{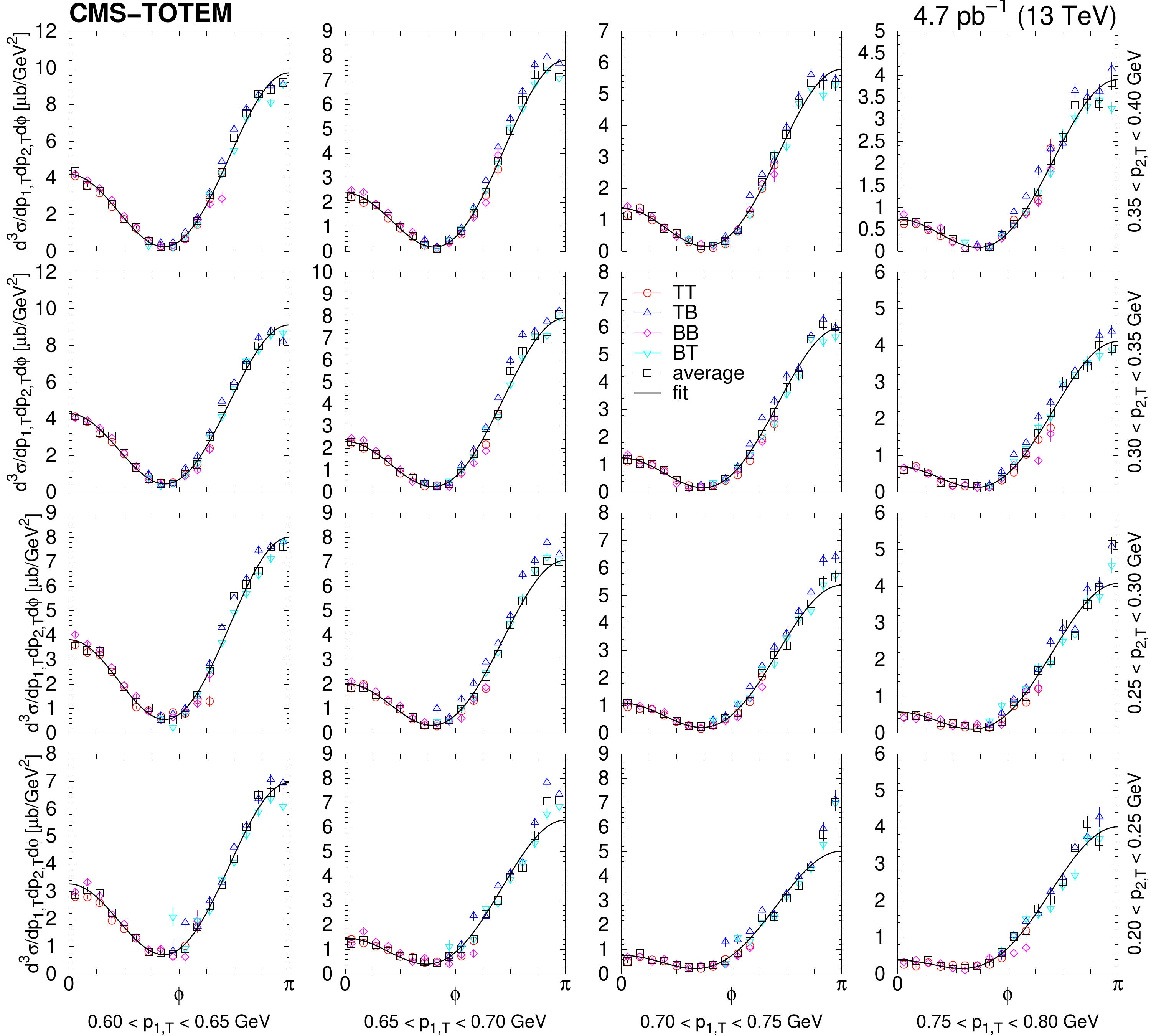}

 \caption{Distributions of $\rd^3\sigma/\rd p_\text{1,T} \rd p_\text{2,T} \rd\phi$
as functions of $\phi$ in the $\PGpp\PGpm$ nonresonant region ($0.35 <
m_{\PGpp\PGpm} < 0.65\GeV$) in several $(p_\text{1,T},p_\text{2,T})$ bins in
the range $0.60<p_\text{1,T}<0.80\GeV$ and $0.20<p_\text{2,T}<0.60\GeV$, in units of
$\mu\mathrm{b}/\GeVns^2$. Values based on data from each RP trigger configuration
(TB, BT, TT, and TT) are shown separately with coloured symbols, whereas the
weighted average is indicated with black symbols. Results of individual fits
with the form $[A(R - \cos\phi)]^2 + c^2$ (Eq.~\eqref{eq:ARc}) are plotted with
the curves. The error bars indicate the statistical uncertainties.}

 \label{fig:dndphi_detail_1_3}

 \end{figure*}

\clearpage

\subsection{The \texorpdfstring{$\phi$}{phi} distributions}

We focus on $\PGpp\PGpm$ production because this system has a wide invariant
mass window ($0.35 < m_{\PGpp\PGpm} < 0.65\GeV$) without resonance
contributions, contrary to the $\PKp\PKm$ case. The distributions of
$\rd^3\sigma/\rd p_\text{1,T} \rd p_\text{2,T} \rd\phi$ as functions of $\phi$ in
several $(p_\text{1,T},p_\text{2,T})$ bins are shown in
Figs.~\ref{fig:dndphi_detail_1_1}--\ref{fig:dndphi_detail_1_3}. The
differential cross sections are given in units of $\mu\mathrm{b}/\GeVns^2$.
Values based on data from each RP trigger configuration (TB, BT, TT, and BB)
are shown separately with coloured symbols, whereas the weighted average is
indicated with black symbols. The error bars show the statistical
uncertainties. Gaps (missing data points) in some $(p_\text{1,T})$ bins of
Fig.~\ref{fig:dndphi_detail_1_1} are due to the vanishing acceptance in those
bins.

\begin{figure*}[hbt]

 \centering
 \includegraphics[width=0.49\textwidth]{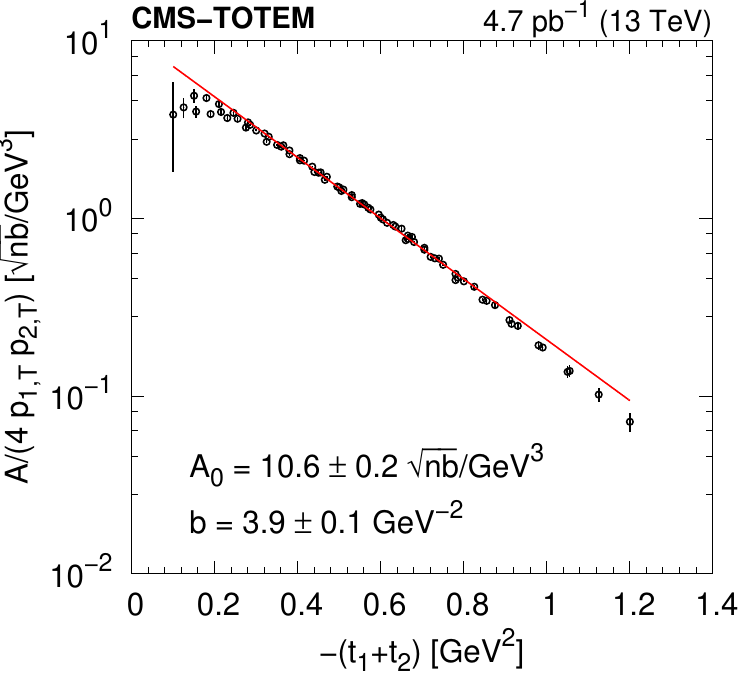}
 \includegraphics[width=0.49\textwidth]{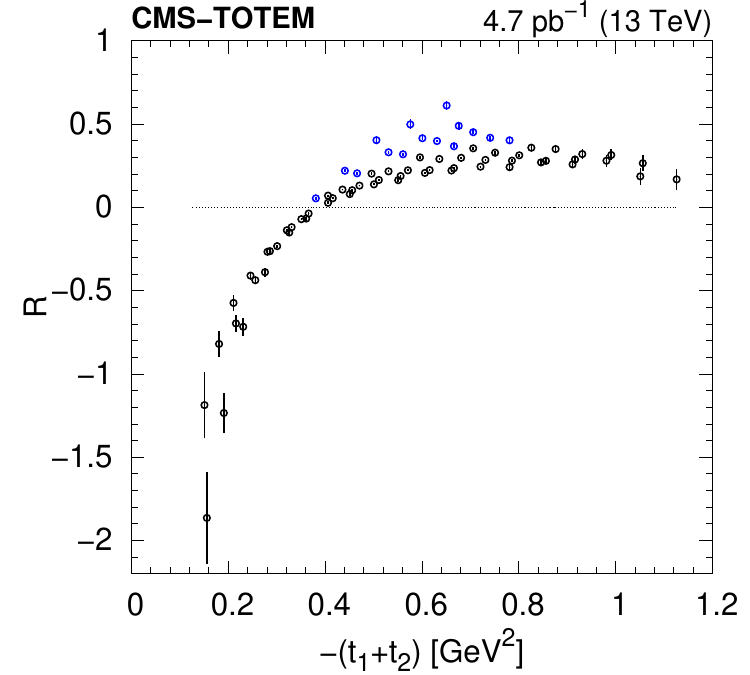}

 \includegraphics[width=0.49\textwidth]{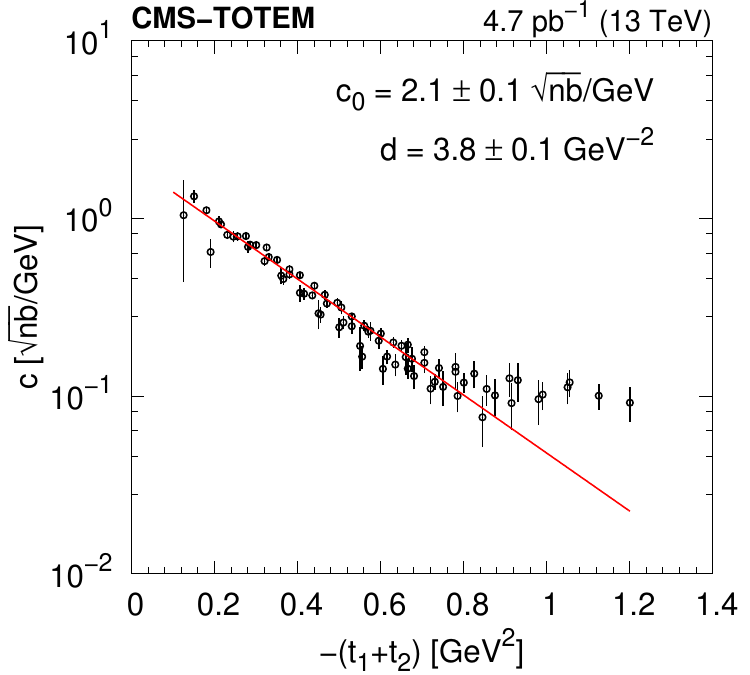}
 \includegraphics[width=0.49\textwidth]{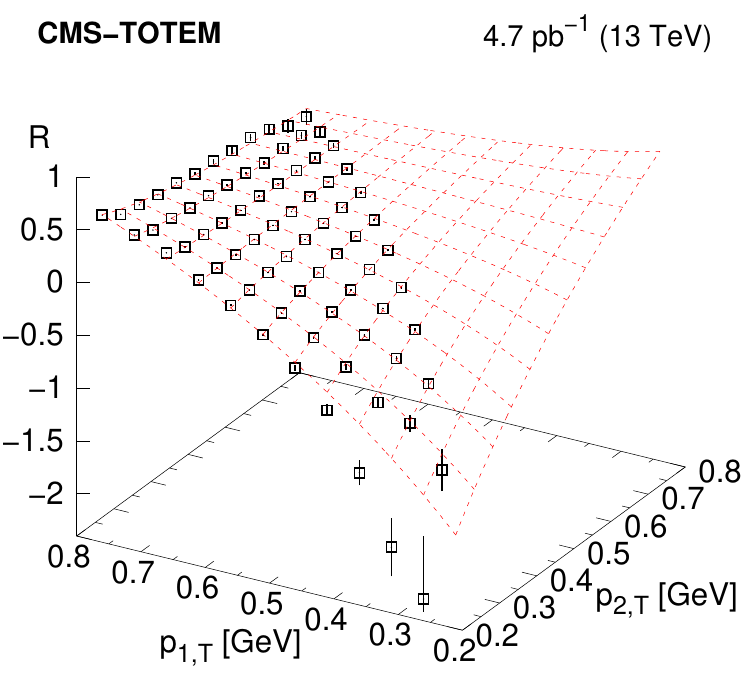}

 \caption{Dependence of the parameters $A$, $R$, and $c$ (Eq.~\eqref{eq:ARc})
on $(t_1,t_2)$. The fits correspond to the functional forms displayed in
Eqs.~\eqref{eq:A}. In the upper right plot, points with
significantly different proton transverse momenta ($\abs{p_\text{1,T} -
p_\text{2,T}} > 0.35\GeV$) are coloured blue. The lower right plot shows the
dependence of $R$ on the two scattered proton transverse momenta.}

 \label{fig:ARc-dep}

\end{figure*}

The distributions from the different trigger configurations are consistent.
Except for the lowest $\pt$ bins, they feature a minimum where
the differential cross section gets close to zero, while local maxima at $\phi
= 0$ and $\pi$ are also present. The distributions are asymmetric at low and
high transverse momenta, but there exists a symmetric region around
$p_\text{1,T} + p_\text{2,T} \approx 0.8\text{--}0.9\GeV$.

The data can be fitted with a simple functional form

\begin{equation}
 \frac{d^3\sigma}{\rd p_\text{1,T} \rd p_\text{2,T} \rd\phi} = [A(R - \cos\phi)]^2 + c^2,
 \label{eq:ARc}
\end{equation}
where $A$, $R$, and $c$ are functions of $(p_\text{1,T},p_\text{2,T})$. The
$\chi^2/\text{dof}$ values corresponding to individual fits are mostly in the
range $1\text{--}10$. The formula features a sum of squared amplitudes
(Section~\ref{sec:theory}) and is inspired by previous theoretical and
experimental studies~\cite{Close:1999bi,Close:2000dx}, where $A(R - \cos\phi)$
is connected to the quantum mechanical amplitude of the process. If the total
amplitude crosses zero at a given $\phi$, its squared value will have a
parabolic minimum. Such a dip at $\phi = \arccos R$ can be understood as an
effect of additional pomeron exchanges between the incoming protons, resulting
from the interference between the bare and the rescattered
amplitudes~\cite{Harland-Lang:2013dia}. The term containing $c$ is added
incoherently; it is small and is present to improve the quality of the fit.

The dependences of the parameters $A$, $R$, and $c$ on $(t_1,t_2)$ are shown in
Fig.~\ref{fig:ARc-dep}. Although $A$ and $c$ can be approximated with a simple
$t_1+t_2$ dependence, the results for $R$ do not all have the same dependence.
The measured points are fitted with
model-motivated~\cite{Close:1999bi,Close:2000dx} functional forms

\begin{equation}
 \begin{aligned}
 A(t_1,t_2) &= 4 \sqrt{t_1 t_2} A_0 \re^{b(t_1 + t_2)},
 \label{eq:A}
 \\
 R(t_1,t_2) &\approx
  \frac{1.2(\sqrt{-t_1} + \sqrt{-t_2}) - 1.6 \sqrt{t_1 t_2} - 0.8}
       {\sqrt{t_1 t_2} + 0.1},
 \\
 c(t_1,t_2) &= c_0 \re^{d(t_1 + t_2)}.
\end{aligned}
\end{equation}

Although the parametrisation overall gives a good description of the data, there
are some deviations at low and high $-(t_1+t_2)$ values for $A$ and $c$,
respectively. Both $A$ and $c$ have very similar dependence on $(t_1+t_2)$
since $b \approx d$.

The parametrisation of the invariant triple-differential cross section is:

\ifthenelse{\boolean{cms@external}}{
\begin{multline}
 \frac{\rd^3\sigma}{\rd t_1 \rd t_2 \rd\phi} =
   4 \sqrt{t_1 t_2} A_0^2 \re^{2b(t_1 + t_2)}
                         [R(t_1,t_2) - \cos\phi]^2 + \\
 + \frac{1}{4 \sqrt{t_1 t_2}}
                    c_0^2 \re^{2d(t_1 + t_2)},
\end{multline}
}{
\begin{equation}
 \frac{\rd^3\sigma}{\rd t_1 \rd t_2 \rd\phi} =
   4 \sqrt{t_1 t_2} A_0^2 \re^{2b(t_1 + t_2)}
                         [R(t_1,t_2) - \cos\phi]^2 
 + \frac{1}{4 \sqrt{t_1 t_2}}
                    c_0^2 \re^{2d(t_1 + t_2)},
\end{equation}
}
where $A_0 = 10.6 \pm 0.2 \sqrt{\mathrm{nb}}/\GeVns^3$, $b = 3.9 \pm 0.1
\GeVns^{-2}$, $c_0 = 2.1 \pm 0.1 \sqrt{\mathrm{nb}}/\GeVns$, and $d = 3.8 \pm 0.1
\GeVns^{-2}$.

\subsection{Available MC event generators}

The {\dime}~\cite{Harland-Lang:2013dia} (v1.07) MC event generator for
exclusive meson pair production via double-pomeron exchange was used in
previous STAR~\cite{Adam:2020sap} and CMS~\cite{Sirunyan:2020cmr} analyses. It
generates central exclusive nonresonant $\PGpp\PGpm$ and $\PKp\PKm$ production
events via the double-pomeron exchange mechanism. This generator is based on
previous work on proton opacity~\cite{Ryskin:2012az}, and the two-channel
model~\cite{Khoze:2013dha} (Good--Walker approach). It includes a fully
differential treatment of the eikonal survival factor. The {\supc}
generator~\cite{Harland-Lang:2015cta} can be regarded as an evolution of
{\dime}, but only provides events with hadron pair invariant masses above
$2\GeV$. The {\genex} code~\cite{Kycia:2014hea} follows an approach quite
similar to that of {\dime}, but does not include the absorption
corrections~\cite{Adam:2020sap}. The model developers estimated the
corresponding suppression factor to be large, of the order of $2\text{--}5$,
and cross sections have to be scaled down by such suppression factor at masses
$m < 1.2\GeV$. The {\gran} (v1.051) MC event
generator~\cite{Mieskolainen:2019jpv} for high energy diffraction includes
differential screening, an extendable set of scattering amplitudes with
adaptive MC sampling and spin effects.

In the following we consider three models (empirical, one-channel, and
two-channel), and three parametrisations (exponential, Orear-type, and
power-law, Eq.~\eqref{eq:meson-pomeron}) of the proton-pomeron form factor. In
the case of the empirical model the rescattering amplitude comes from a simple
parametrisation of the elastic differential pp
amplitude~\cite{Grau:2012wy,Fagundes:2013aja}, fitted to the measured cross
sections~\cite{TOTEM:2017sdy,TOTEM:2018hki}. The one-channel model assumes
ground state protons (one eigenstate), whereas the two-channel model works with
two diffractive proton eigenstates (Section~\ref{sec:theory}). The one-channel
model is disfavoured in describing the elastic differential cross section
data~\cite{Khoze:2014nia}.

\begin{table*}[htb]

 \caption{Values and statistical uncertainties of the parameters tuned using the
\prof\ tool, given for the empirical, one-channel, and two-channel models along
with the \dime\ soft models 1 and 2 with the exponential, power-law, and the
Orear-type parametrisations of the proton-pomeron form factor. Goodness-of-fit
($\chi^2/\mathrm{dof}$) values are also listed.}

 \label{tab:tunes_result}

 \centering
 \begin{scotch}{lcccc}
  Parameter & Exponential & Orear-type & Power-law & {\dime} 1 / 2 \\
 \hline
 Empirical model & & & &  \\
 $a_\text{ore} [\GeVns{}]$ & --- & $0.735 \pm 0.015$ & --- &  \\
 $b_\text{exp/ore/pow} [\GeVns^{-2~\text{or}~-1}]$ & $1.084 \pm 0.004$ & $1.782 \pm 0.014$ & $1.356 \pm 0.001$ &  \\
 $B_\P~[\GeVns^{-2}]$ & $3.757 \pm 0.033$ & $3.934 \pm 0.027$ & $4.159 \pm 0.019$ &  \\
 $\chi^2/\mathrm{dof}$ & $9470/5796$ & $10059/5795$ & $11409/5796$ &  \\
[\cmsTabSkip]
 One-channel model & & & &  \\
 $\sigma_0 [\unit{mb}]$ & $34.99 \pm 0.79$ & $27.98 \pm 0.40$ & $26.87 \pm 0.30$ &  \\
 $\alpha_P - 1$ & $0.129 \pm 0.002$ & $0.127 \pm 0.001$ & $0.134 \pm 0.001$ &  \\
 $\alpha_P'~[\GeVns^{-2}]$ & $0.084 \pm 0.005$ & $0.034 \pm 0.002$ & $0.037 \pm 0.002$ &  \\
 $a_\text{ore} [\GeVns{}]$ & --- & $0.578 \pm 0.022$ & --- &  \\
 $b_\text{exp/ore/pow} [\GeVns^{-2~\text{or}~-1}]$ & $0.820 \pm 0.011$ & $1.385 \pm 0.015$ & $1.222 \pm 0.004$ &  \\
 $B_\P~[\GeVns^{-2}]$ & $2.745 \pm 0.046$ & $4.271 \pm 0.021$ & $4.072 \pm 0.017$ &  \\
 $\chi^2/\mathrm{dof}$ & $7356/5793$ & $7448/5792$ & $8339/5793$ &  \\
[\cmsTabSkip]
 Two-channel model & & & &  \\
 $\sigma_0 [\unit{mb}]$ & $20.97 \pm 0.48$ & $22.89 \pm 0.17$ & $23.02 \pm 0.23$ & 23 / 33 \\
 $\alpha_P - 1$ & $0.136 \pm 0.001$ & $0.129 \pm 0.001$ & $0.131 \pm 0.001$ & 0.13 / 0.115 \\
 $\alpha_P'~[\GeVns^{-2}]$ & $0.078 \pm 0.001$ & $0.075 \pm 0.001$ & $0.071 \pm 0.001$ & 0.08 / 0.11 \\
 $a_\text{ore} [\GeVns{}]$ & --- & $0.718 \pm 0.012$ & --- &  \\
 $b_\text{exp/ore/pow} [\GeVns^{-2~\text{or}~-1}]$ & $0.917 \pm 0.007$ & $1.517 \pm 0.008$ & $0.931 \pm 0.002$ & 0.45 \\
 $\Delta \abs{a}^2$ & $0.070 \pm 0.026$ & $-0.058 \pm 0.009$ & $0.042 \pm 0.011$ & $-0.04$ / $-0.25$ \\
 $\Delta \gamma$ & $0.052 \pm 0.042$ & $0.131 \pm 0.018$ & $0.273 \pm 0.023$ & 0.55 / 0.4 \\
 $b_1~[\GeVns^2]$ & $8.438 \pm 0.108$ & $8.951 \pm 0.041$ & $8.877 \pm 0.040$ & 8.5 / 8.0 \\
 $c_1~[\GeVns^2]$ & $0.298 \pm 0.012$ & $0.278 \pm 0.004$ & $0.266 \pm 0.006$ & 0.18 / 0.18 \\
 $d_1$ & $0.472 \pm 0.007$ & $0.465 \pm 0.002$ & $0.465 \pm 0.003$ & 0.45 / 0.63 \\
 $b_2~[\GeVns^2]$ & $4.982 \pm 0.133$ & $4.222 \pm 0.052$ & $4.780 \pm 0.060$ & 4.5 / 6.0 \\
 $c_2~[\GeVns^2]$ & $0.542 \pm 0.015$ & $0.522 \pm 0.006$ & $0.615 \pm 0.006$ & 0.58 / 0.58 \\
 $d_2$ & $0.453 \pm 0.009$ & $0.452 \pm 0.003$ & $0.431 \pm 0.004$ & 0.45 / 0.47 \\
 $\chi^2/\mathrm{dof}$ & $5741/5786$ & $6415/5785$ & $7879/5786$ &  
 \end{scotch}

\end{table*}

\subsection{Model tuning}

There exist loose estimates of the numerous parameters of the theoretical model
illustrated in Section~\ref{sec:theory}, including cross sections, slopes,
proton-pomeron and meson-pomeron form factors. Our aim here is to determine the
best values and corresponding uncertainties in these quantities with the
present data. For the tuning of physics parameters the tool
{\prof}~\cite{Buckley:2009bj} (v2.3.3) is employed. It quantifies the per-bin
generator response to parameter variations and numerically optimises the
behaviour of the generator.

The measured distributions included in model tuning are from events with
$\PGpp\PGpm$ central final state in the resonance-free region:

\begin{itemize}
\item
that of azimuthal angle between the scattered proton momenta,
       $\rd^3\sigma/\rd p_\text{1,T}\,\rd p_\text{2,T}\,\rd\phi$, if $0.35 < m < 0.65\GeV$;
\item
that of two-hadron invariant mass at low masses, 
       $\rd^3\sigma/\rd p_\text{1,T}\,\rd p_\text{2,T}\,\rd m$, if $m < 0.7\GeV$;
\item
the distribution of the two-hadron invariant mass at high masses, 
       $\rd^3\sigma/\rd p_\text{1,T}\,\rd p_\text{2,T}\,\rd m$, if $1.8 < m < 2.2\GeV$;
\item
that of squared four-momentum of the virtual meson,
       $\rd^3\sigma/\rd p_\text{1,T}\,\rd p_\text{2,T}\,\rd \max(\hat{t},\hat{u})$,
                                                  if $1.8 < m < 2.2\GeV$;
\end{itemize}
all of them in the range $0.2 < (p_\text{1,T},p_\text{2,T}) < 0.8\GeV$. The
$\phi$ distributions are sensitive to any resonant contribution; for this
reason their mass window is a bit more restricted ($0.35 < m < 0.65\GeV$) than
that of the invariant mass distribution. Information about the exchanged
virtual meson is present in the distribution of $\max(\hat{t},\hat{u})$ and can
be exploited if one of $\hat{t}$ or $\hat{u}$ is close to zero, when the hadron
is nearly on-shell. Such cases occur if the difference of $q_1$ and $q_2$ is
large. In other words, the invariant mass of the two-hadron system should be
large, hence the choice of a resonance-free window at higher mass values.

The parameter space has many dimensions, and there are up to 13 parameters to
tune. The ranges of the parameters are chosen such that the variation of the
model output, \ie, its envelope, covers the measured distributions. The
sampling is random and uniform, and 512 pseudo-experiments are performed for
each form factor parametrisation. Each pseudo-experiment contains one million
generated events, whose average cross section per event is recorded. The
distributions from these MC pseudo-experiments are interpolated with a
first-order polynomial in the parameter space. The tuning, \ie, the
minimisation of the global goodness-of-fit, converges to unique minima for each
model and all three form factor options. The $\chi^2/\text{dof}$ values are in
the range 1.6--2.0 (empirical), 1.2--1.5 (one-channel), and 1.0--1.3
(two-channel). Good fits are achieved with the one-channel or the two-channel
models using exponential or Orear-type form factors, whereas the numerically
best one is the two-channel model with the exponential parametrisation of the
proton-pomeron form factor. It is clear that the power-law parametrisation of
the proton-pomeron form factor is least favoured by our results. The empirical
model also has a poor goodness-of-fit value, although with fairly few
parameters.

\begin{figure*}[htb]

 \centering
 \includegraphics[height=0.40\textheight]{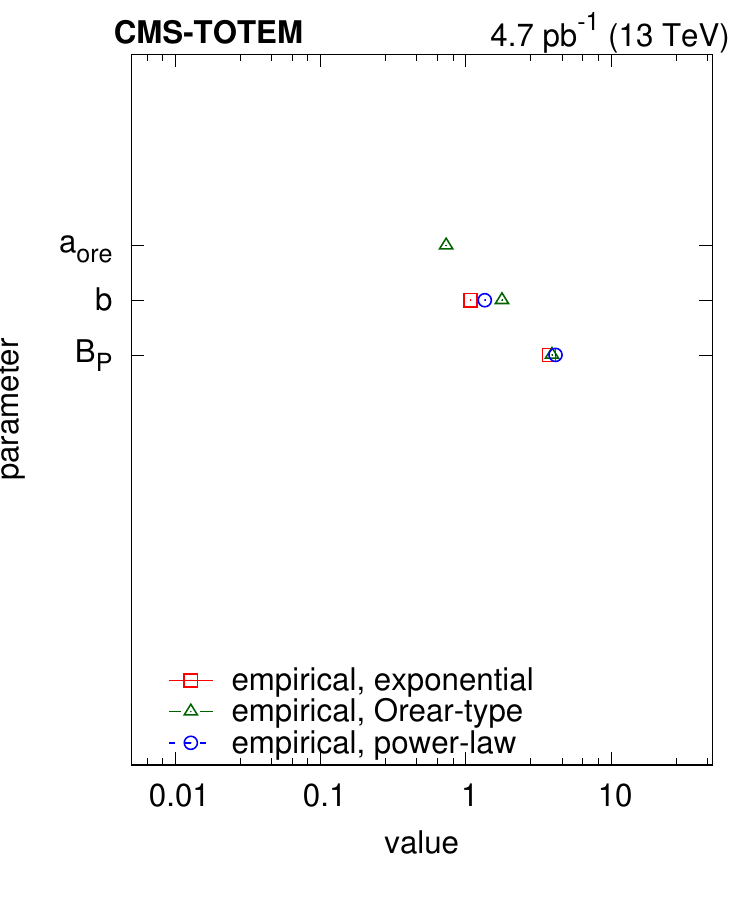}
 \includegraphics[height=0.40\textheight]{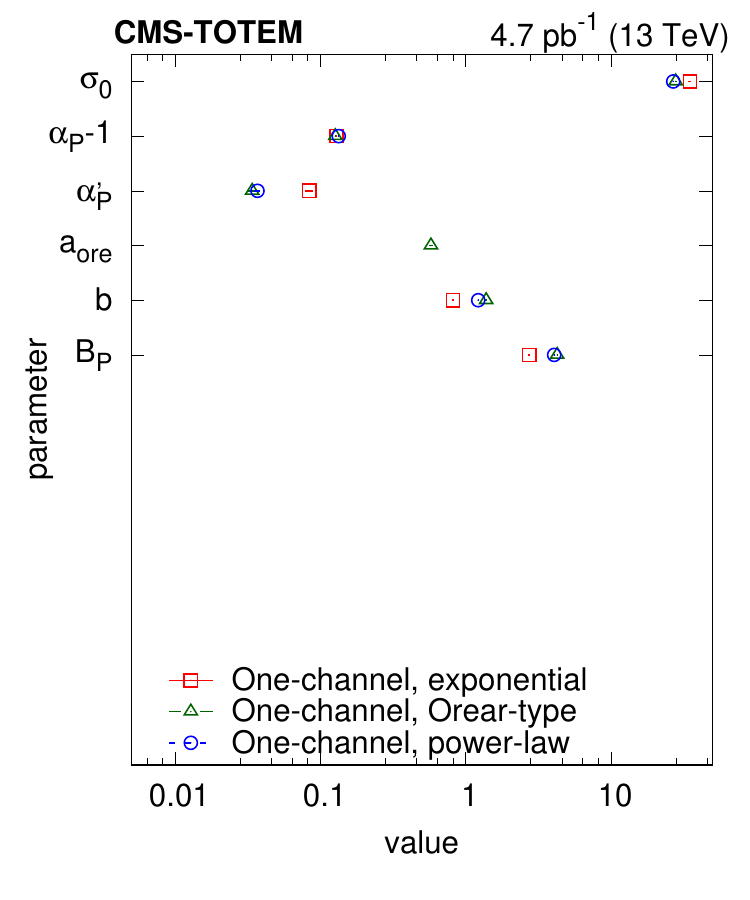}

 \includegraphics[height=0.40\textheight]{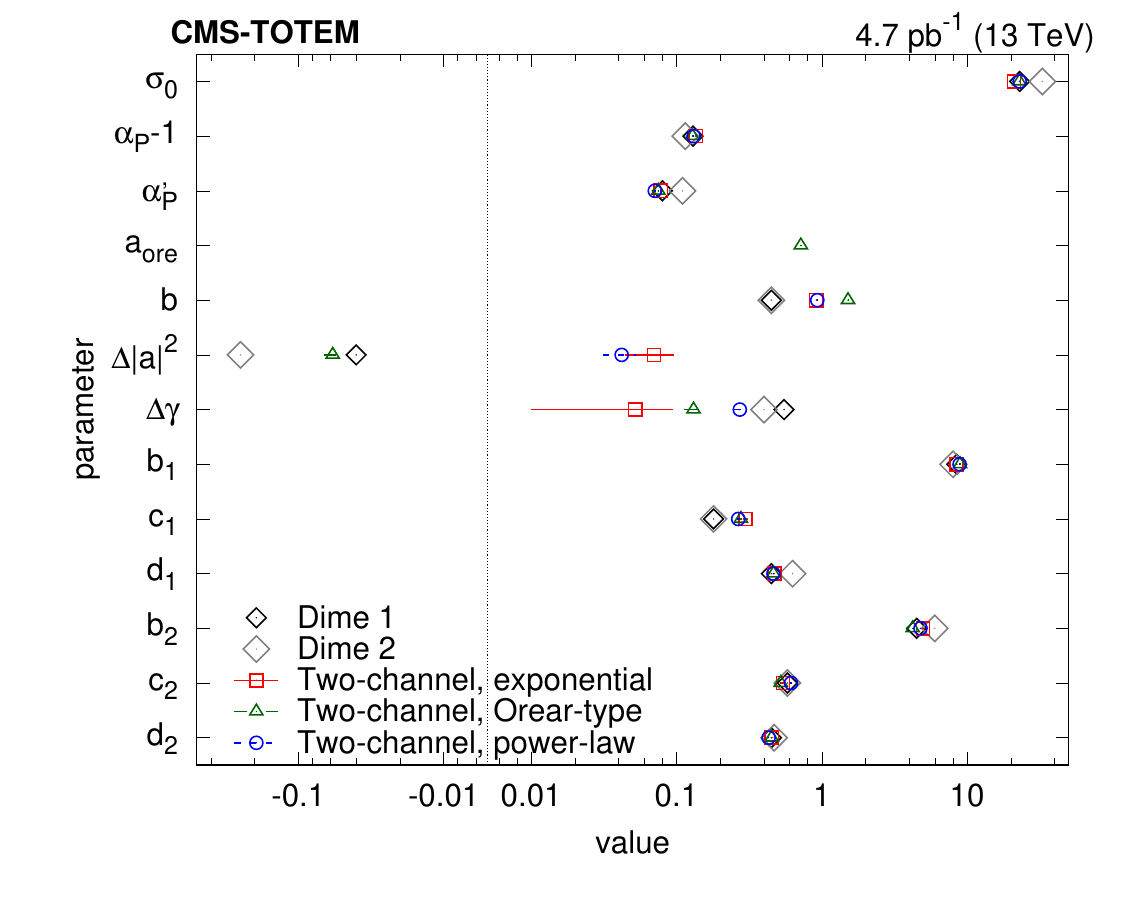}

 \caption{Values of best parameters for the empirical (upper left), one-channel
(upper right), and two-channel (lower) models with several choices of the
proton-pomeron form factor (exponential, Orear-type, power-law). In the case of
the two-channel model, parameter values of models describing the elastic
differential $\Pp\Pp$ cross section from Ref.~\cite{Khoze:2013dha} are also indicated
({\dime} 1 and 2).}
 \label{fig:tune_results}
\end{figure*}

\begin{figure*}[htb!]

 \centering
 \includegraphics[width=0.32\textwidth]{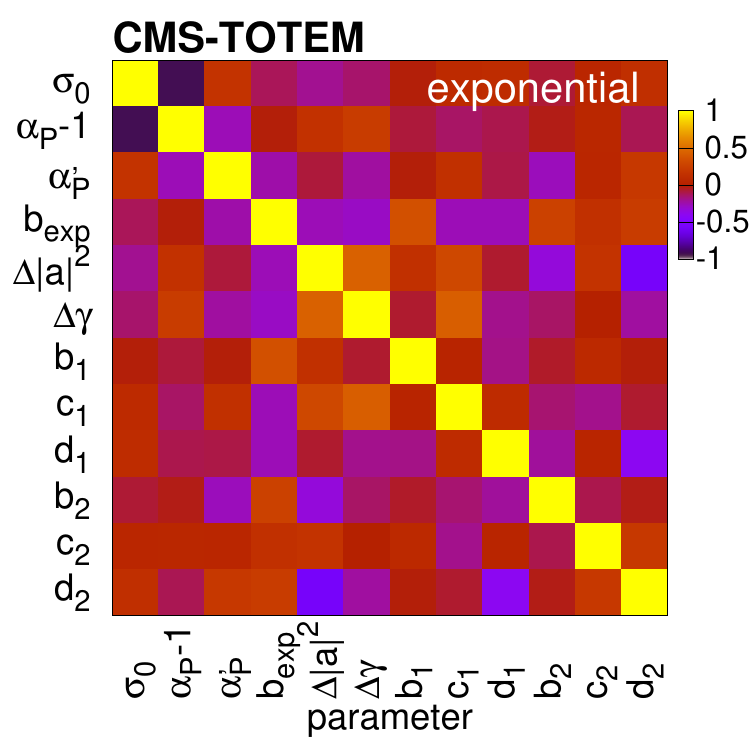}
 \includegraphics[width=0.32\textwidth]{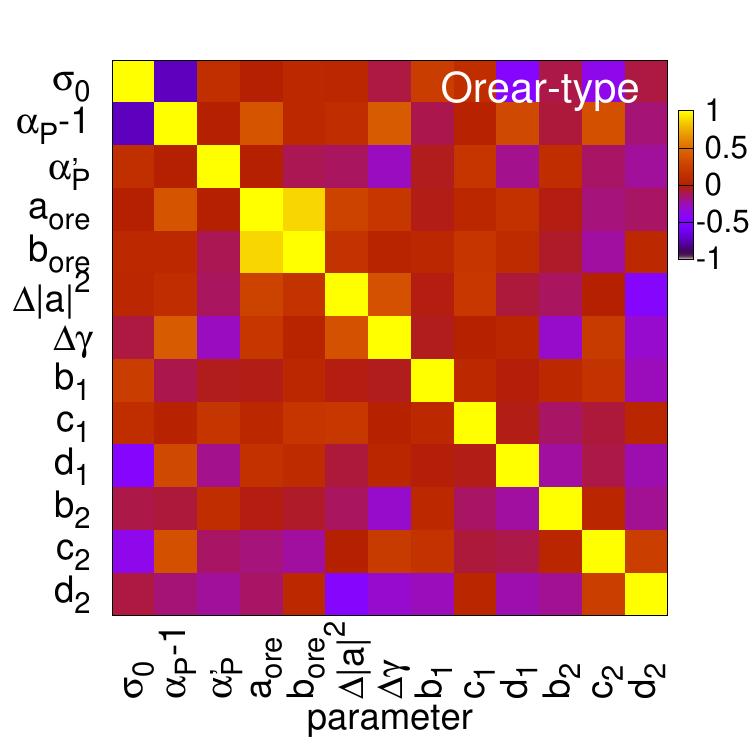}
 \includegraphics[width=0.32\textwidth]{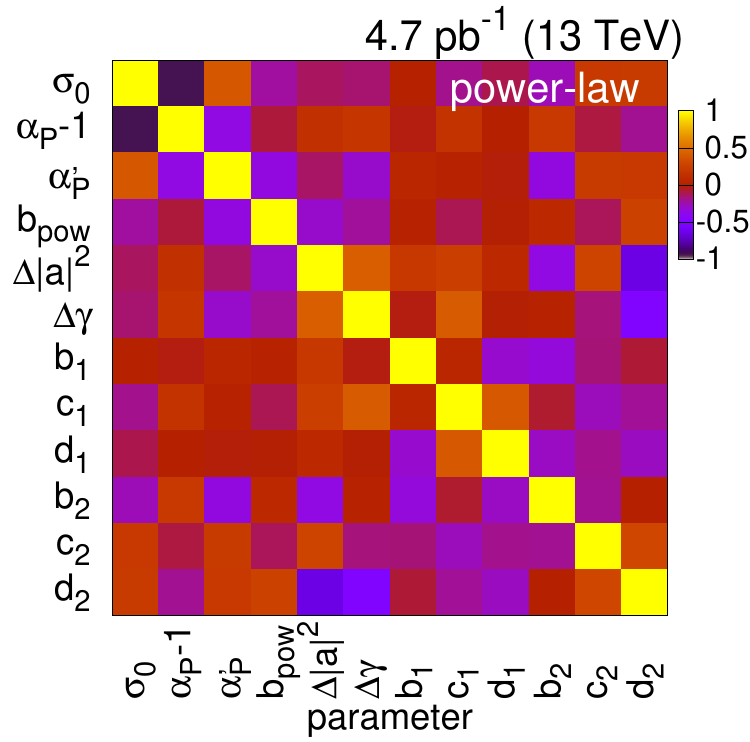}
 \caption{Correlation coefficients among values of best parameters for the
two-channel model, in the case of the exponential (left), Orear-type (centre),
and power-law (right) parametrisations of the proton-pomeron form factor.}

 \label{fig:tune_results_corr}

\end{figure*}

The values of the fitted parameters are listed in Table~\ref{tab:tunes_result}
and shown in Fig.~\ref{fig:tune_results}, with notations $\Delta\abs{a}^2 =
(\abs{a_1}^2 - \abs{a_2}^2)/2$ and $\Delta\gamma = (\gamma_1 - \gamma_2)/2$. In
the case of the two-channel model, parameter values of two models describing
the elastic differential $\Pp\Pp$ cross section from Ref.~\cite{Khoze:2013dha} are
also indicated ({\dime}~1 and 2). Settings of {\dime}~1 agree well with the
best tuned values of the parameters, with the exception of the
$\Delta\abs{a^2}$, $\Delta\gamma$, and $b$ factors. The extracted values of
$\sigma_0$ are stable, except for the one-channel exponential case, and
disagree with earlier estimates (Section~\ref{sec:theory}). The matrix of
correlation coefficients for the two-channel model is displayed in
Fig.~\ref{fig:tune_results_corr} for several choices of the proton-pomeron form
factor.

\subsection{Data-MC comparisons}

\begin{figure*}[htb]
 \centering
 \includegraphics[width=0.65\textwidth]{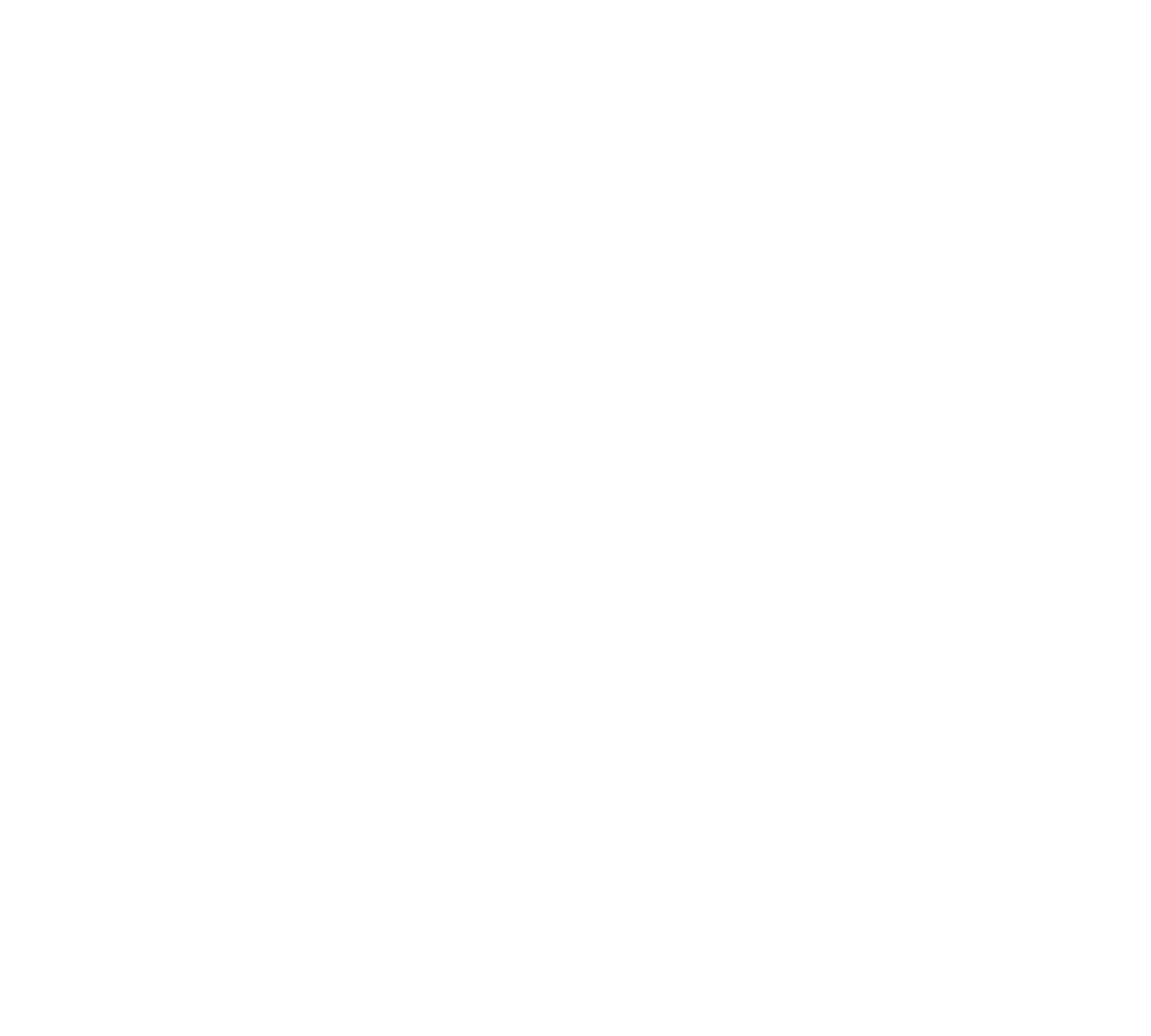}
 \includegraphics[width=0.65\textwidth]{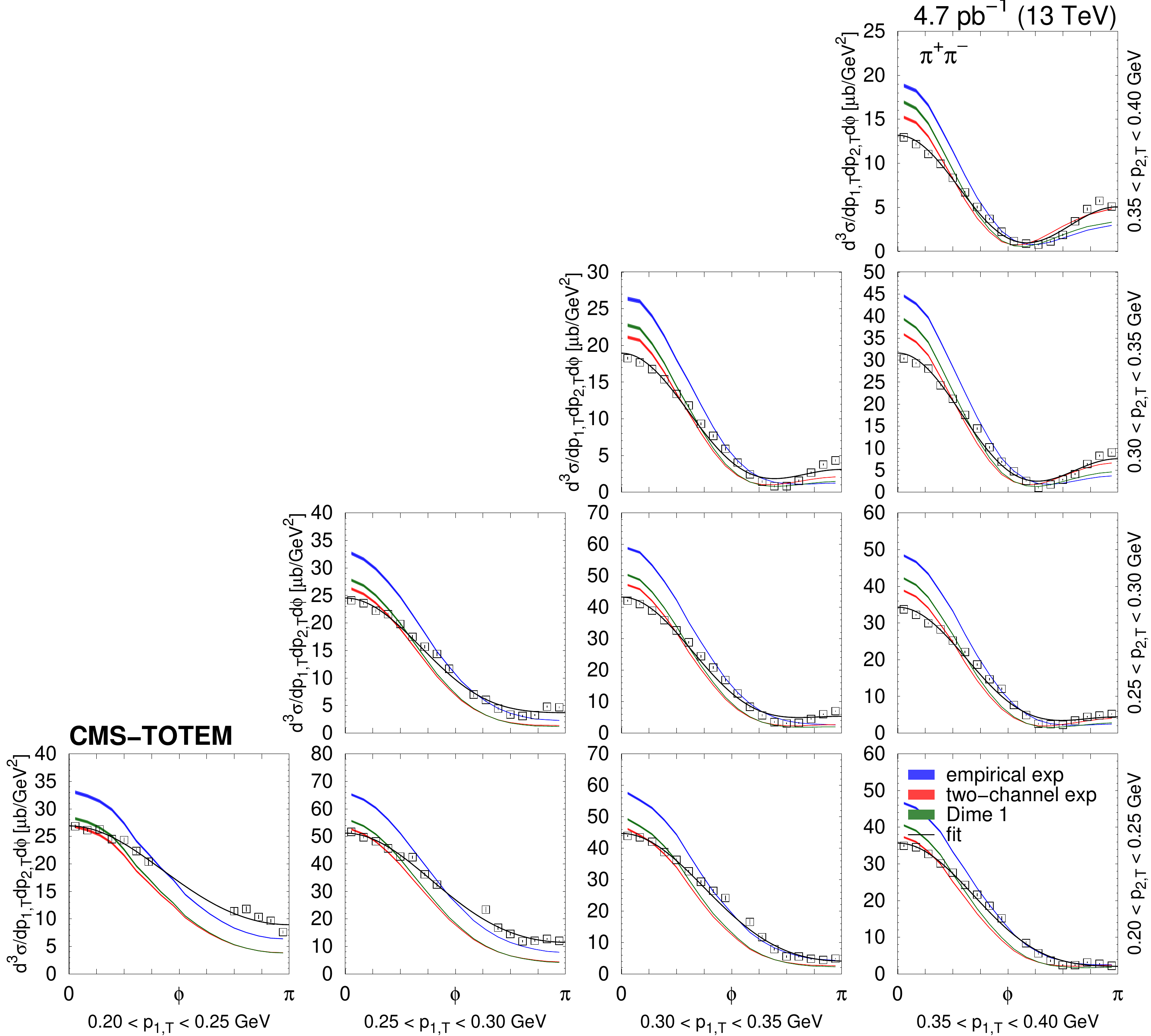}
 \caption{Distribution of $\rd^3\sigma/\rd p_\text{1,T} \rd p_\text{2,T} \rd\phi$
as a function of $\phi$ in several $(p_\text{1,T},p_\text{2,T})$ bins in the
range $0.20<p_\text{1,T}<0.40\GeV$ and $0.20<p_\text{2,T}<0.60\GeV$, in units of
$\mu\mathrm{b}/\GeVns^2$, for the mass range $0.35 < m_{\PGpp\PGpm} < 0.65\GeV$.
Measured values (black symbols) are shown together with the predictions of the
empirical and the two-channel models (coloured curves) using the tuned
parameters for the exponential proton-pomeron form factors (see text for
details). Curves corresponding to {\dime} (model 1) are also plotted. Results
of individual fits with the form $[A(R - \cos\phi)]^2 + c^2$
(Eq.~\eqref{eq:ARc}) are plotted with the curves. The error bars indicate the
statistical uncertainties.}
\label{fig:dndphi_withmc_1_1}
\end{figure*}

\begin{figure*}[htb]
 \centering
 \includegraphics[width=0.65\textwidth]{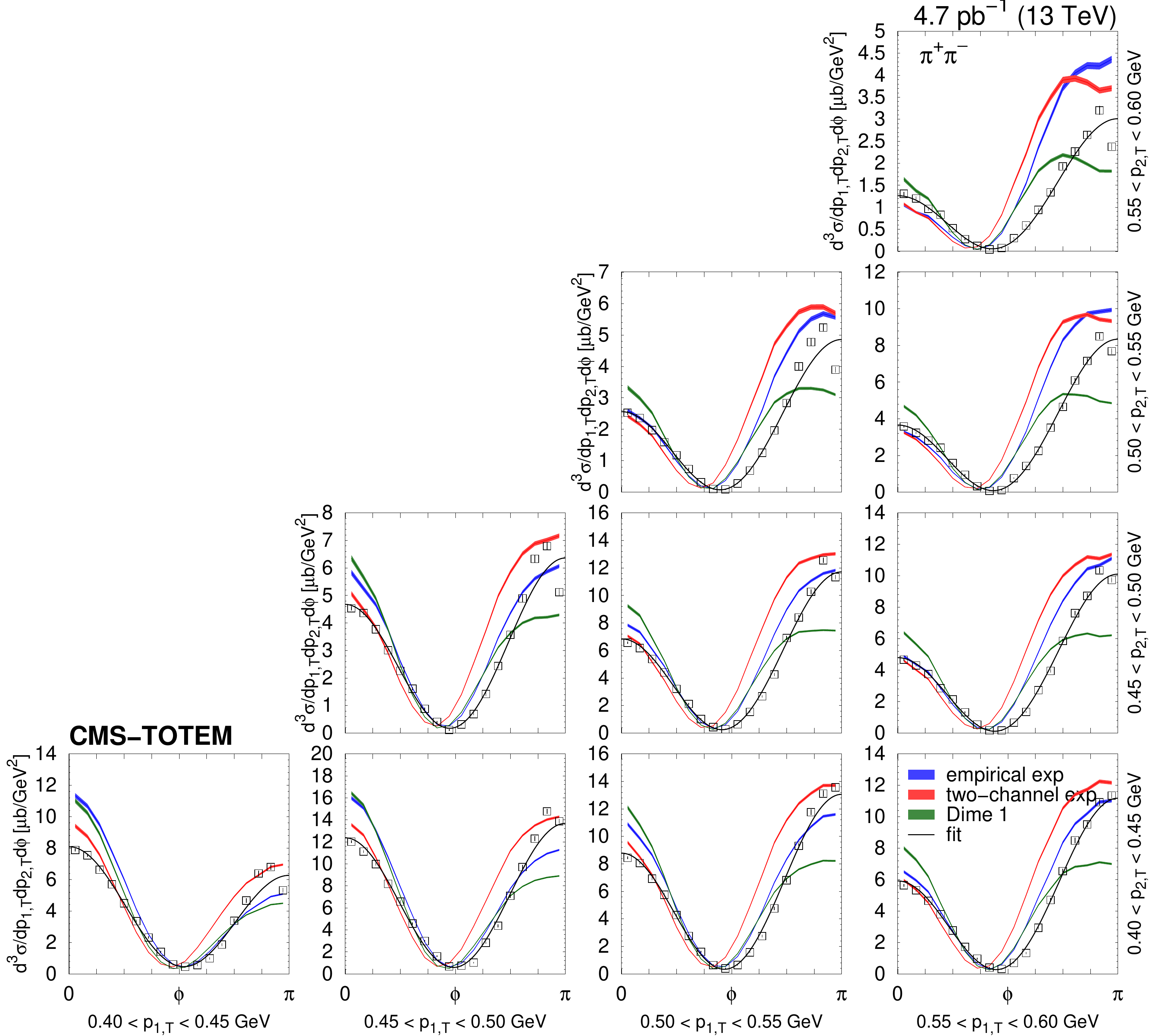}
 \includegraphics[width=0.65\textwidth]{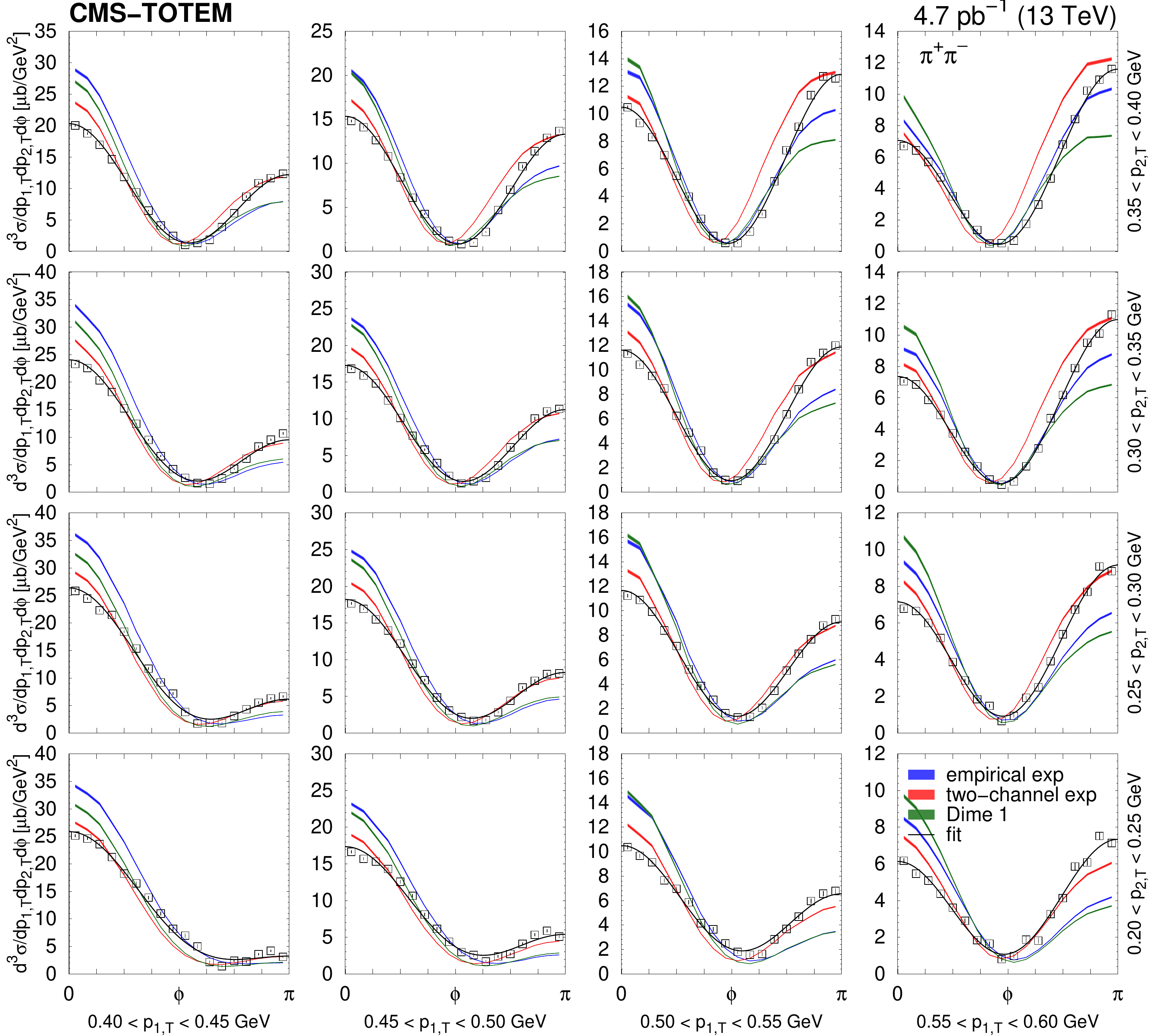}
 \caption{Distribution of $\rd^3\sigma/\rd p_\text{1,T} \rd p_\text{2,T} \rd\phi$
as a function of $\phi$ in several $(p_\text{1,T},p_\text{2,T})$ bins in the
range $0.40<p_\text{1,T}<0.60\GeV$ and $0.20<p_\text{2,T}<0.60\GeV$, in units of
$\mu\mathrm{b}/\GeVns^2$, for the mass range $0.35 < m_{\PGpp\PGpm} < 0.65\GeV$.
Measured values (black symbols) are shown together with the predictions of the
empirical and the two-channel models (coloured curves) using the tuned
parameters for the exponential proton-pomeron form factors (see text for
details). Curves corresponding to {\dime} (model 1) are also plotted. Results
of individual fits with the form $[A(R - \cos\phi)]^2 + c^2$
(Eq.~\eqref{eq:ARc}) are plotted with the curves. The error bars indicate the
statistical uncertainties.}
\label{fig:dndphi_withmc_1_2}
\end{figure*}

\begin{figure*}[htb]
 \centering
 \includegraphics[width=0.65\textwidth]{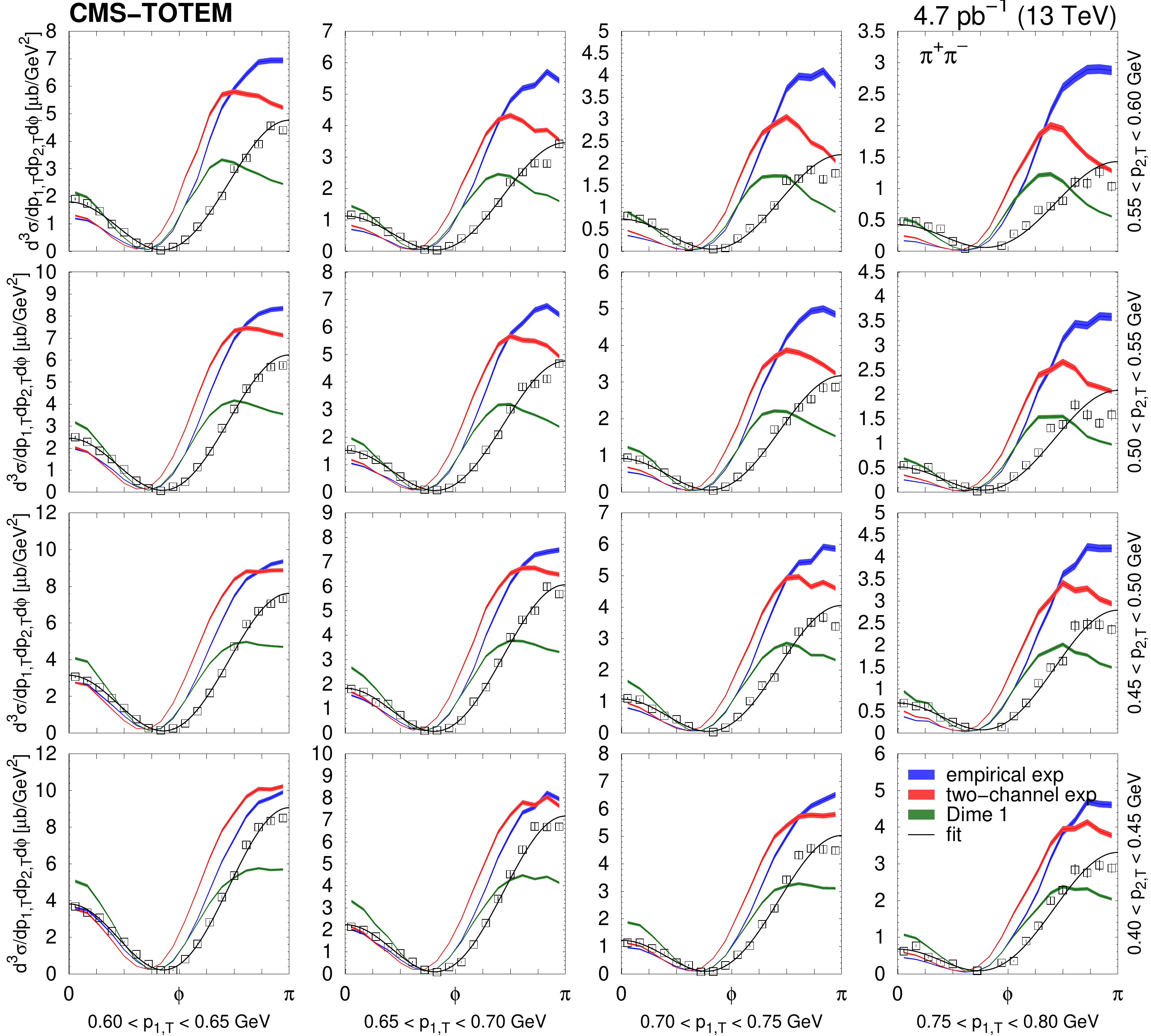}
 \includegraphics[width=0.65\textwidth]{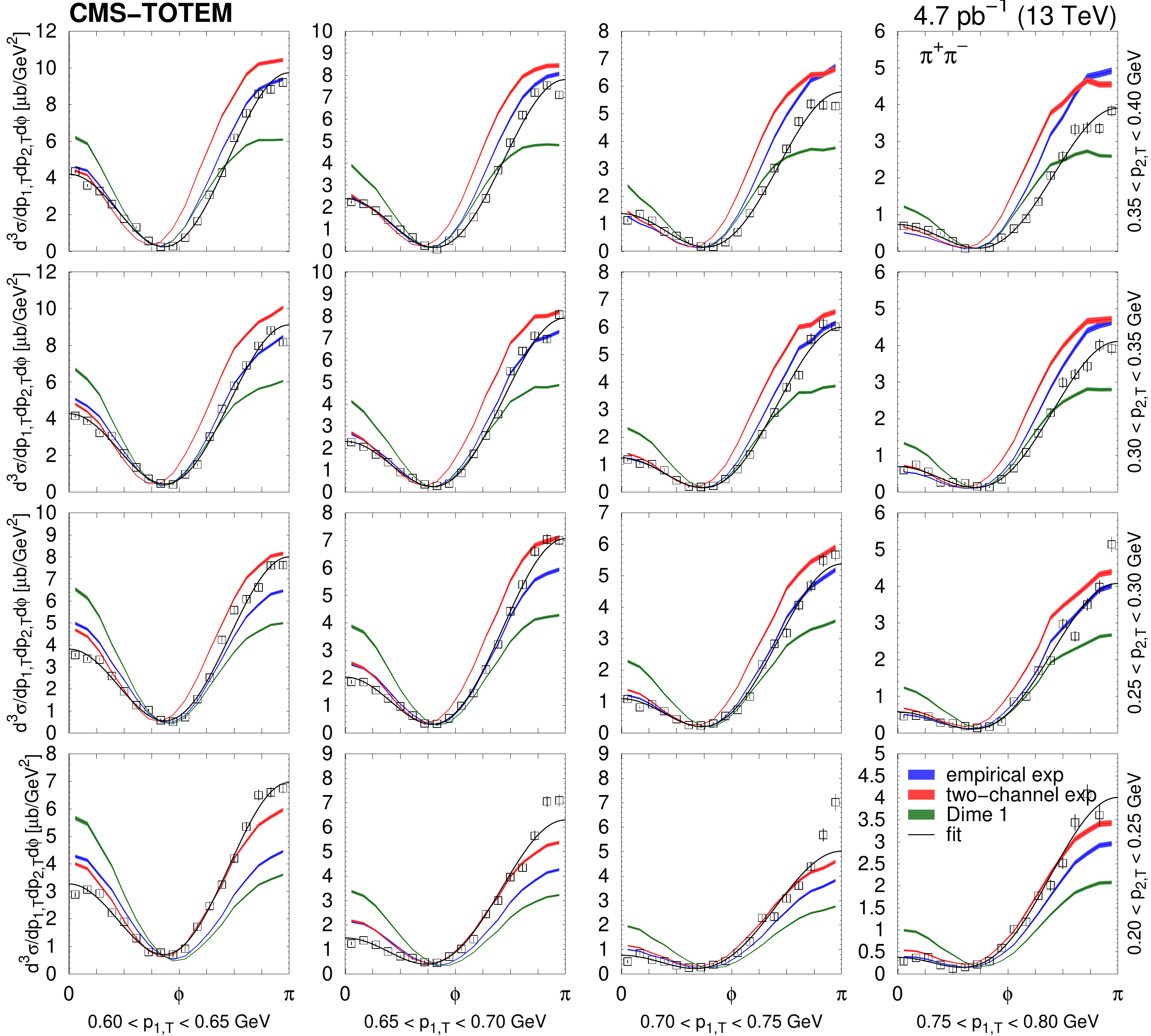}
 \caption{Distribution of $\rd^3\sigma/\rd p_\text{1,T} \rd p_\text{2,T} \rd\phi$
as a function of $\phi$ in several $(p_\text{1,T},p_\text{2,T})$ bins in the
range $0.60<p_\text{1,T}<0.80\GeV$ and $0.20<p_\text{2,T}<0.60\GeV$, in units of
$\mu\mathrm{b}/\GeVns^2$, for the mass range $0.35 < m_{\PGpp\PGpm} < 0.65\GeV$.
Measured values (black symbols) are shown together with the predictions of the
empirical and the two-channel models (coloured curves) using the tuned
parameters for the exponential proton-pomeron form factors (see text for
details). Curves corresponding to {\dime} (model 1) are also plotted. Results
of individual fits with the form $[A(R - \cos\phi)]^2 + c^2$
(Eq.~\eqref{eq:ARc}) are plotted with the curves. The error bars indicate the
statistical uncertainties.}
\label{fig:dndphi_withmc_1_3}
\end{figure*}

The distributions of $\rd^3\sigma/\rd p_\text{1,T} \rd p_\text{2,T} \rd\phi$ in the
nonresonant region ($0.35 < m_{\PGpp\PGpm} < 0.65\GeV$) as a function of $\phi$
in several $(p_\text{1,T},p_\text{2,T})$ bins are shown in
Figs.~\ref{fig:dndphi_withmc_1_1}--\ref{fig:dndphi_withmc_1_3}. The measured
values are shown together with the predictions of the empirical and the
two-channel models using the tuned parameters for the exponential
proton-pomeron form factors. Curves corresponding to {\dime} (model 1) are also
plotted: this generator gives a poor description of the data for $\phi >
\pi/2$. Although the various tuned models give a better description, there are
some regions with sizeable disagreements pointing to the need for further
theoretical developments, specially for low $p_\text{1,T}$ and low
$p_\text{2,T}$, as well as high $p_\text{1,T}$ and high $p_\text{2,T}$
combinations.

\begin{figure*}[htb]

 \centering
 \includegraphics[width=0.49\textwidth]{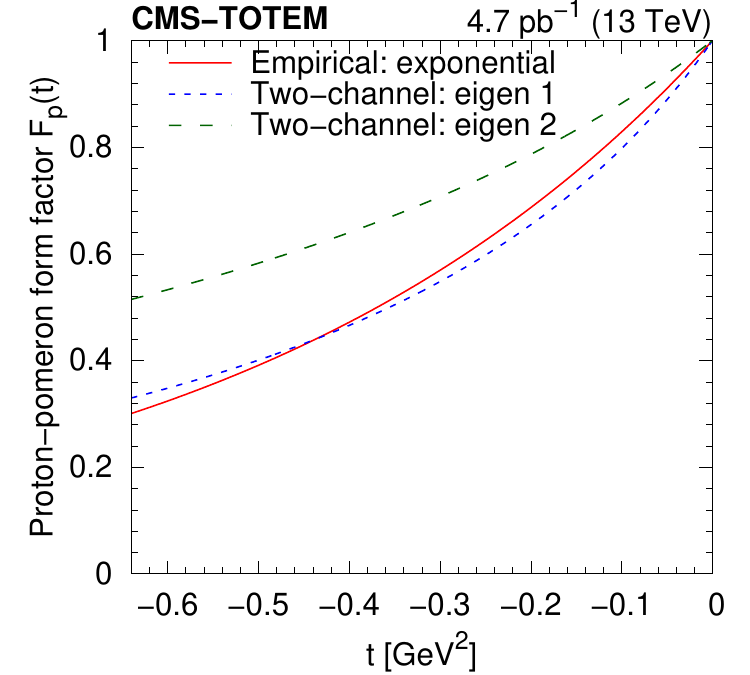}
 \includegraphics[width=0.49\textwidth]{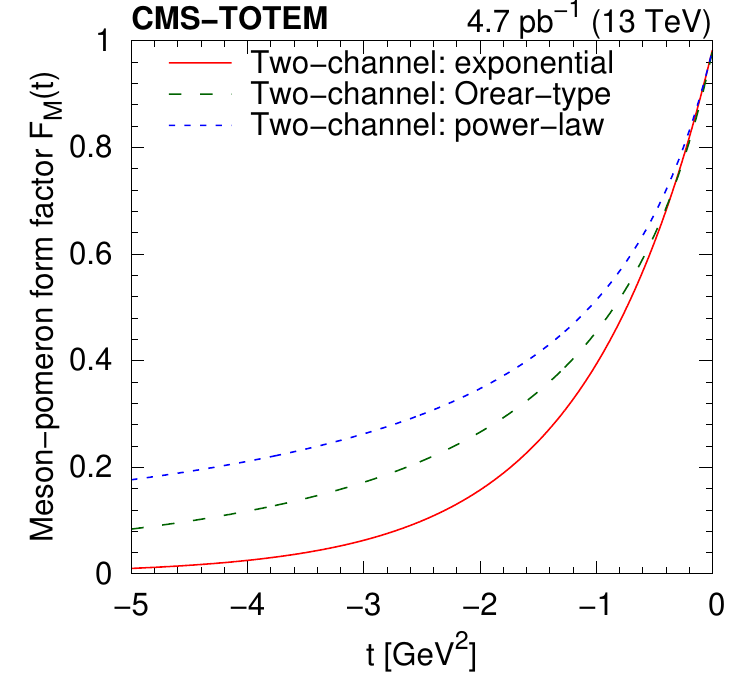}

 \caption{Results of model tuning. Left: The plain exponential proton-pomeron
form factor compared with those of the two diffractive proton eigenstates.
Right: Various options of the meson-pomeron form factor, shown for the
exponential, Orear-type, and power-law parametrisations.}

 \label{fig:fac}

\end{figure*}

The plain exponential proton-pomeron form factor (from the fit with the
empirical model) and those of the two diffractive proton eigenstates are shown
in Fig.~\ref{fig:fac}~(left). One of the eigenstates is quite close to the
exponential form. Various options of the meson-pomeron form factor in the
two-channel model, for the exponential, power-law, and the Orear-type
parametrisations are shown in Fig.~\ref{fig:fac}~(right).

\begin{figure*}[htb]

 \centering
 \includegraphics[width=0.65\textwidth]{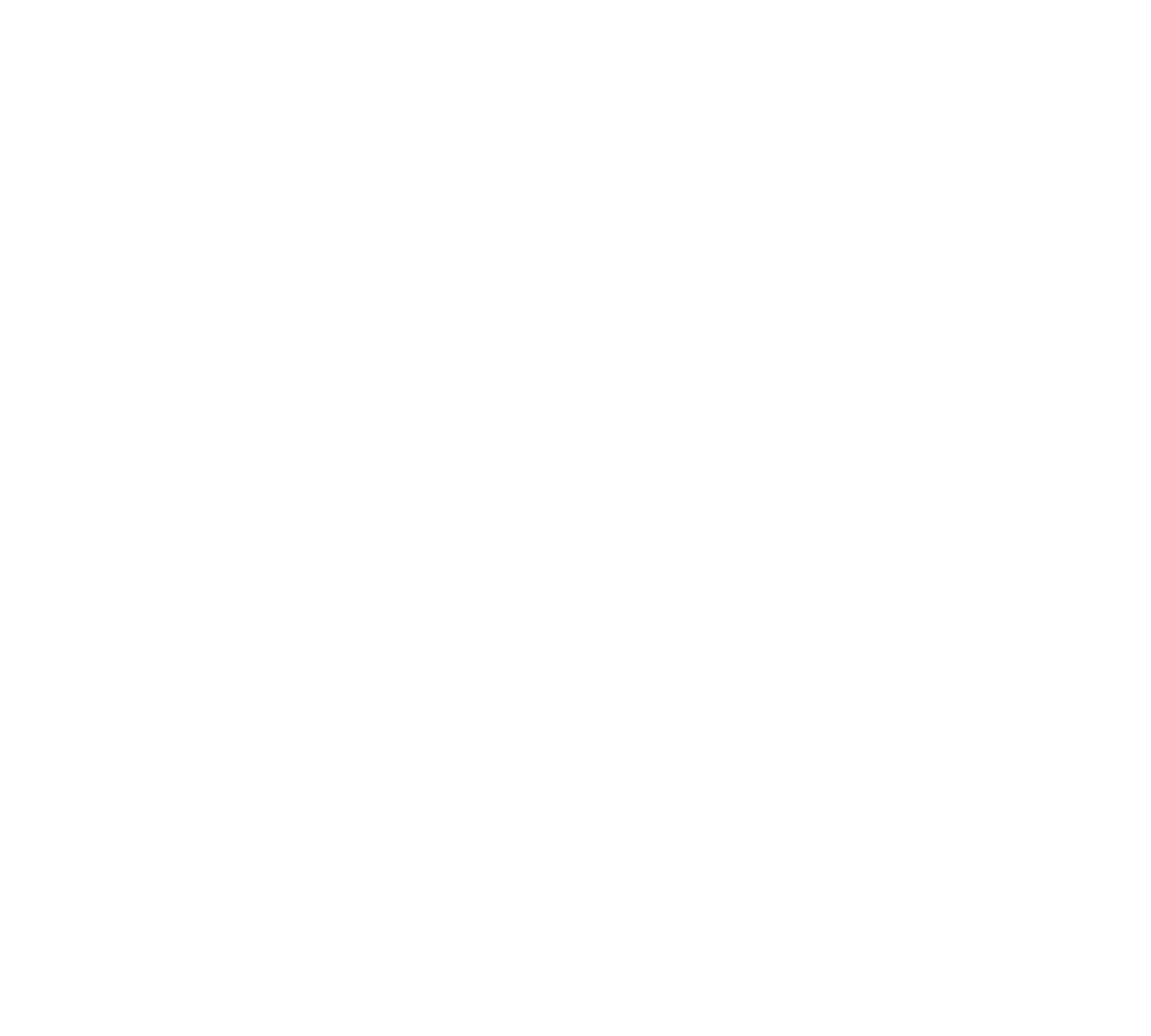}
 \includegraphics[width=0.65\textwidth]{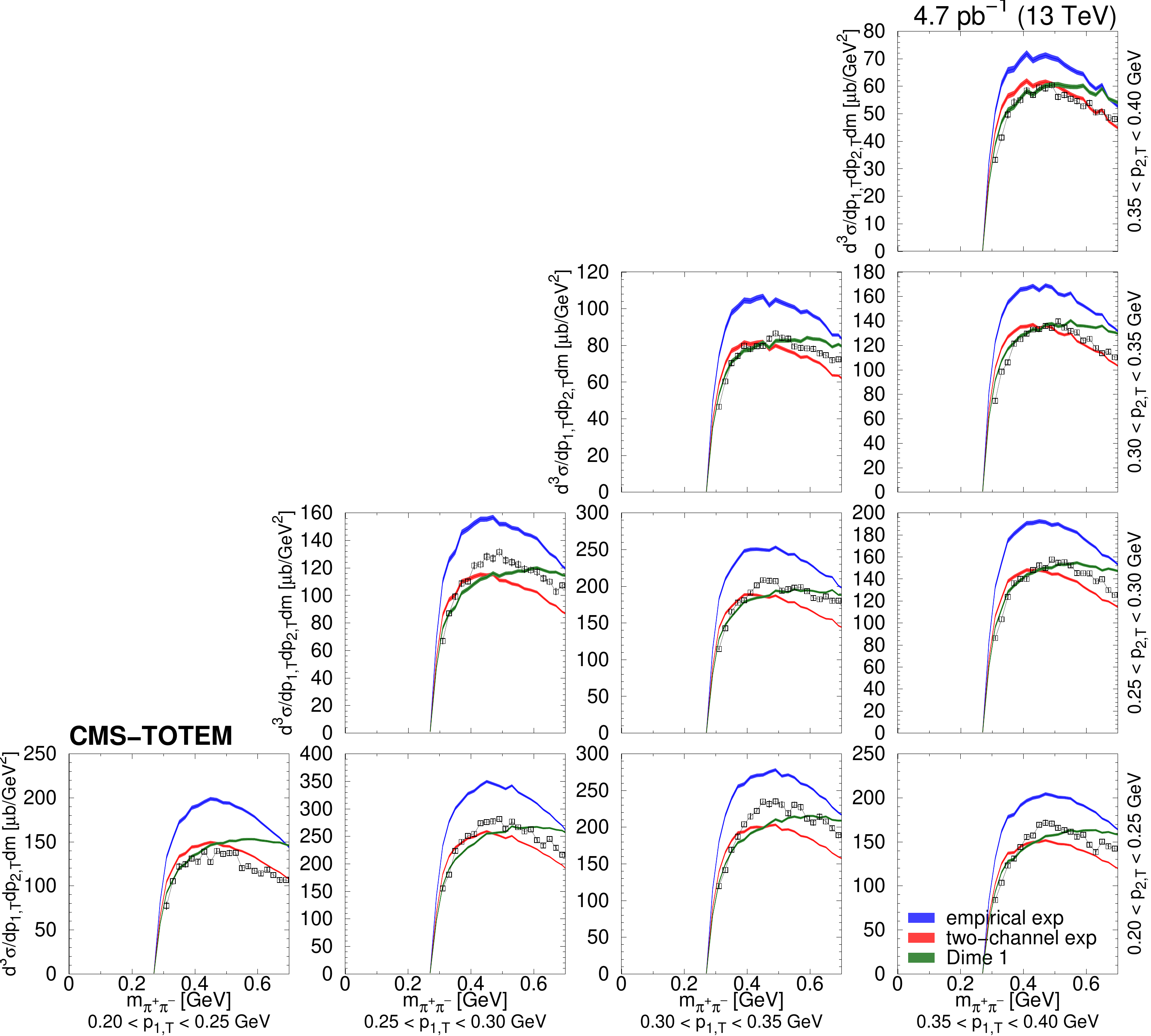}
 \caption{Distribution of $\rd^3\sigma/\rd p_\text{1,T} \rd p_\text{2,T} \rd m$ as
a function of $m$ for $\PGpp\PGpm$ pairs in several $(p_\text{1,T},p_\text{2,T})$ bins,
in units of $\mu\mathrm{b}/\GeVns^3$, for the mass range $0.35 < m_{\PGpp\PGpm} <
0.65\GeV$. Measured values (black symbols) are shown together with the
predictions of the empirical and the two-channel models (coloured curves) using
the tuned parameters for the exponential proton-pomeron form factors (see text
for details). Curves corresponding to {\dime} (model 1) are also plotted. The
error bars indicate the statistical uncertainties. Lines connecting the data
points are drawn to guide the eye.}
\label{fig:dndm_019_1_1}
\end{figure*}

\begin{figure*}[htb]

 \centering
 \includegraphics[width=0.65\textwidth]{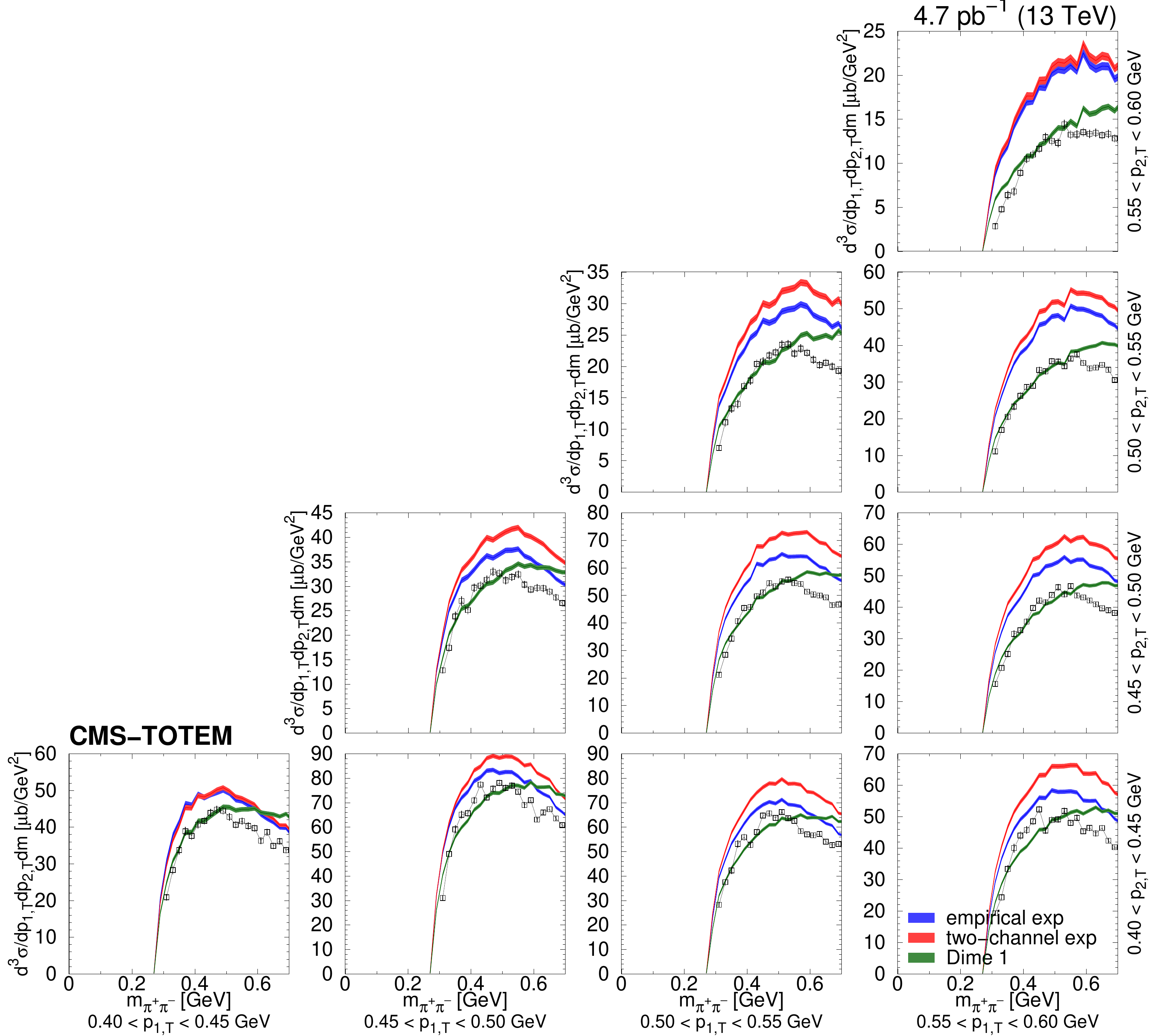}
 \includegraphics[width=0.65\textwidth]{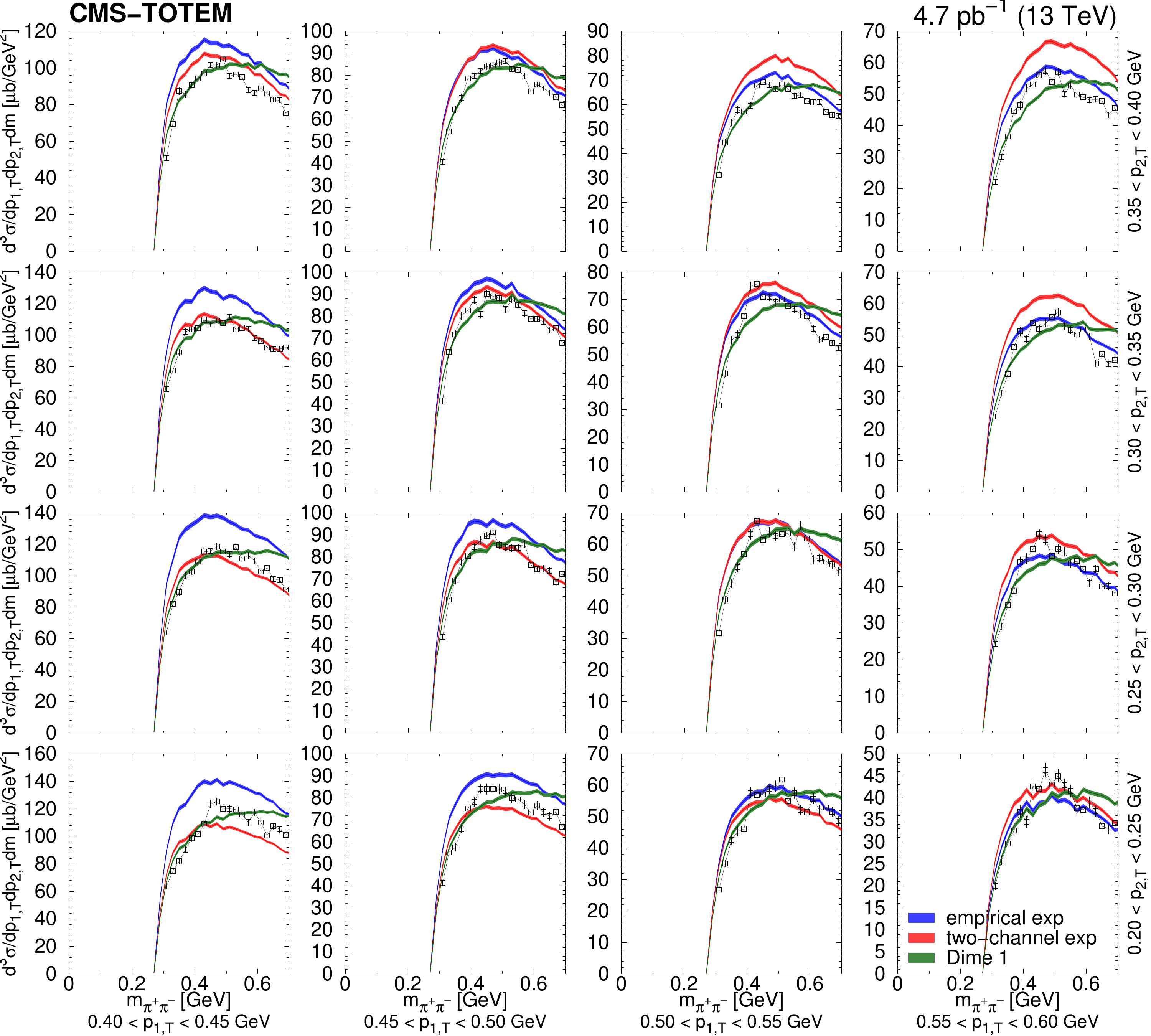}
 \caption{Distribution of $\rd^3\sigma/\rd p_\text{1,T} \rd p_\text{2,T} \rd m$ as
a function of $m$ for $\PGpp\PGpm$ pairs in several $(p_\text{1,T},p_\text{2,T})$ bins,
in units of $\mu\mathrm{b}/\GeVns^3$, for the mass range $0.35 < m_{\PGpp\PGpm} <
0.65\GeV$. Measured values (black symbols) are shown together with the
predictions of the empirical and the two-channel models (coloured curves) using
the tuned parameters for the exponential proton-pomeron form factors (see text
for details). Curves corresponding to {\dime} (model 1) are also plotted. The
error bars indicate the statistical uncertainties. Lines connecting the data
points are drawn to guide the eye.}
\label{fig:dndm_020_1_2}
\end{figure*}

\begin{figure*}[htb]

 \centering
 \includegraphics[width=0.65\textwidth]{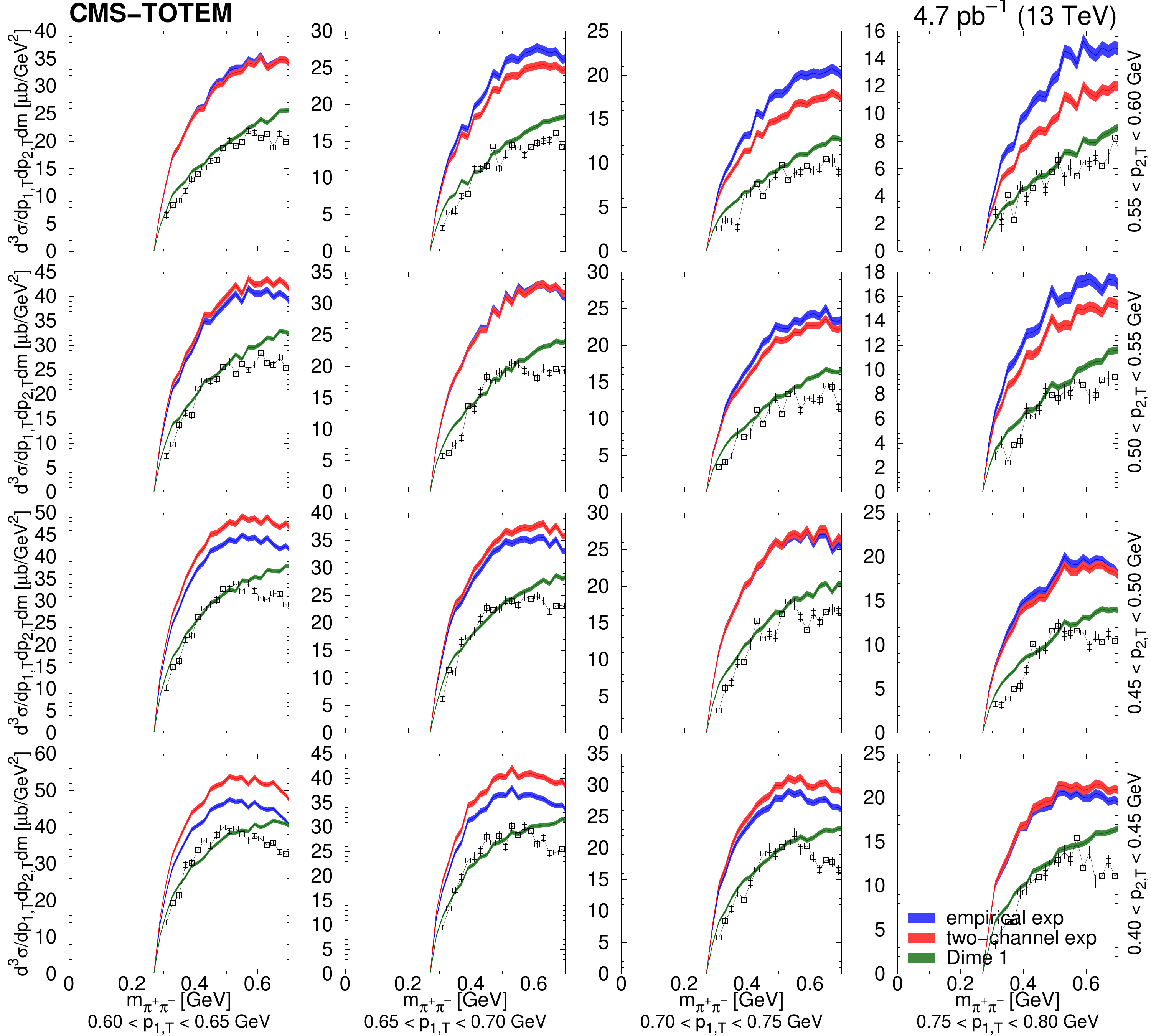}
 \includegraphics[width=0.65\textwidth]{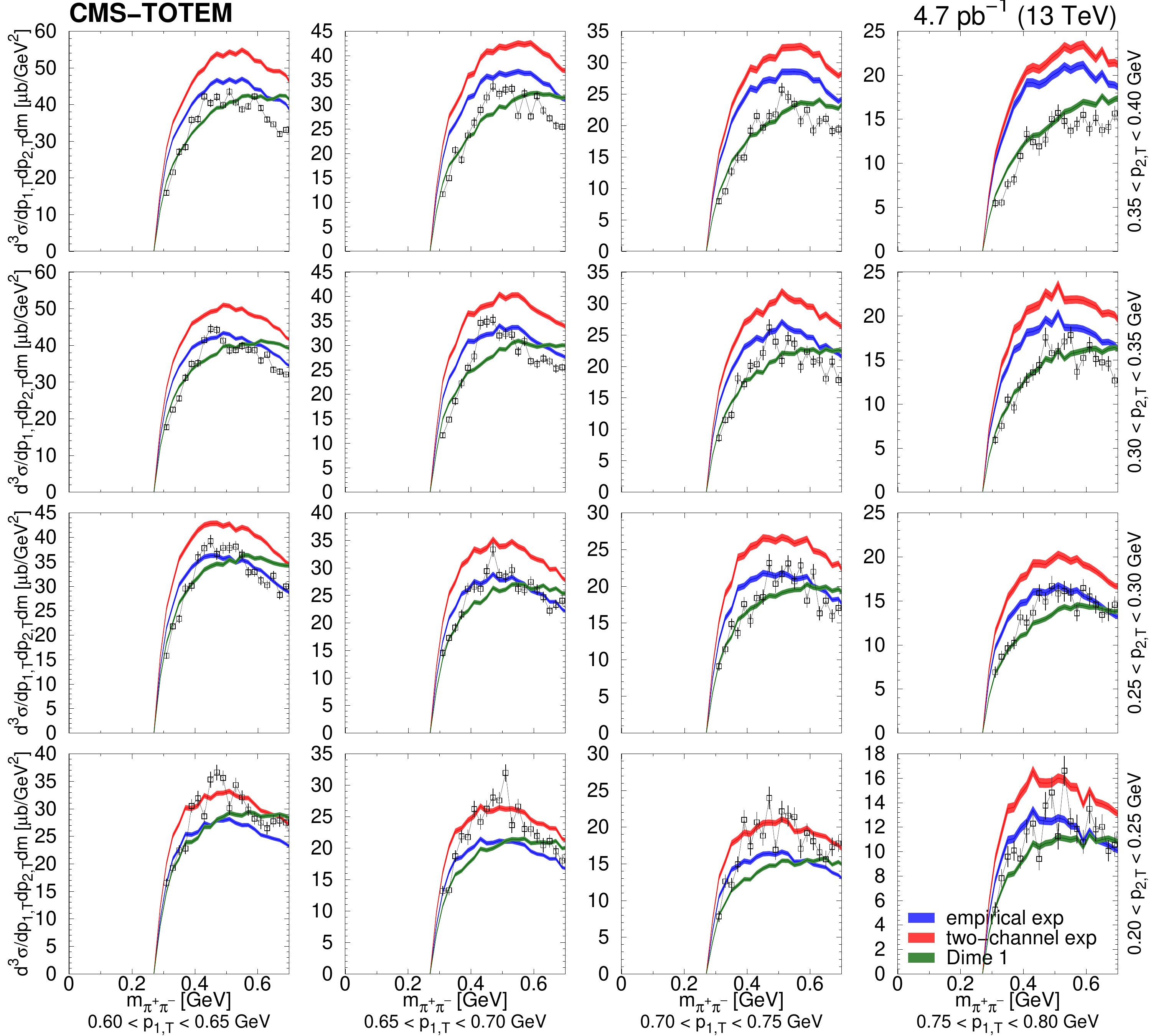}
 \caption{Distribution of $\rd^3\sigma/\rd p_\text{1,T} \rd p_\text{2,T} \rd m$ as
a function of $m$ for $\PGpp\PGpm$ pairs in several $(p_\text{1,T},p_\text{2,T})$ bins,
in units of $\mu\mathrm{b}/\GeVns^3$, for the mass range $0.35 < m_{\PGpp\PGpm} <
0.65\GeV$. Measured values (black symbols) are shown together with the
predictions of the empirical and the two-channel models (coloured curves) using
the tuned parameters for the exponential proton-pomeron form factors (see text
for details). Curves corresponding to {\dime} (model 1) are also plotted. The
error bars indicate the statistical uncertainties. Lines connecting the data
points are drawn to guide the eye.}
\label{fig:dndm_021_1_3}
\end{figure*}

The distributions of $\rd^3\sigma/\rd p_\text{1,T} \rd p_\text{2,T} \rd m$ as a
function of $m$ in several $(p_\text{1,T},p_\text{2,T})$ bins are shown for
$\PGpp\PGpm$ in Figs.~\ref{fig:dndm_019_1_1}--\ref{fig:dndm_021_1_3}. Measured
values are shown together with the predictions of models using the tuned
parameters for the exponential proton-pomeron form factors. The tuned models do
not satisfactorily describe the regions with high $p_\text{1,T}$ and high
$p_\text{2,T}$ combinations, whereas the empirical model is not able to
describe data well at low transverse momenta.

\begin{figure*}[htb]

 \centering
 \includegraphics[width=0.65\textwidth]{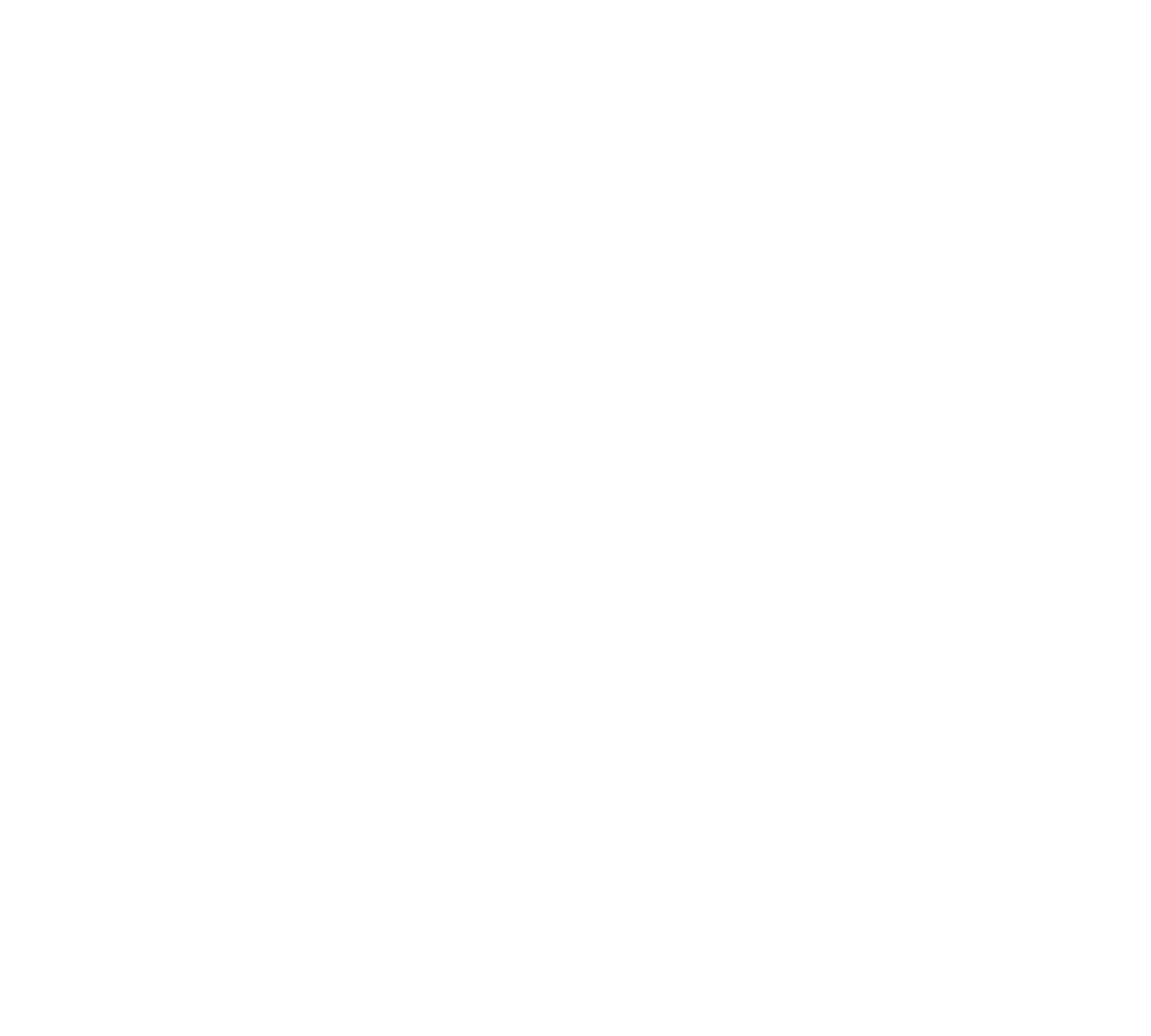}
 \includegraphics[width=0.65\textwidth]{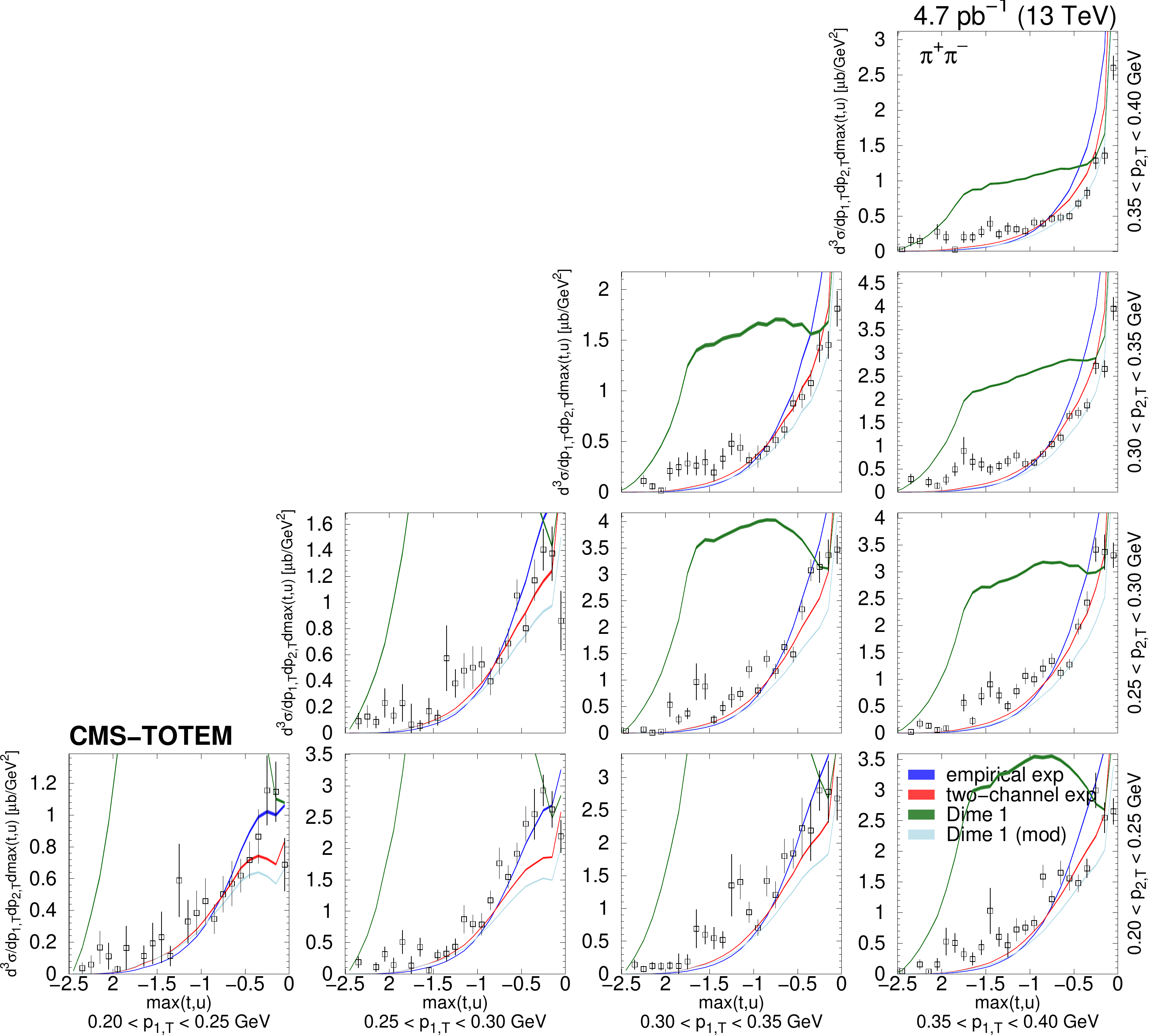}

 \caption{Distribution of the squared momentum transfer of the virtual pion in
several $(p_\text{1,T},p_\text{2,T})$ bins, in units of $\mu\mathrm{b}/\GeVns^3$,
for the mass range $1.8 < m_{\PGpp\PGpm} < 2.2\GeV$. Measured values (black
symbols) are shown together with the predictions of the empirical and the
two-channel models (coloured curves) using the tuned parameters for the
exponential proton-pomeron form factors (see text for details). Curves
corresponding to {\dime} (model 1, ``Dime 1'') and its modification (labelled
``Dime 1 (mod)'') with $b_\text{exp} = 0.9\GeV^{-2}$ are also plotted. The
error bars indicate the statistical uncertainties.}

 \label{fig:dndsha_022_1_1}

 \end{figure*}

\begin{figure*}[htb]

 \centering
 \includegraphics[width=0.65\textwidth]{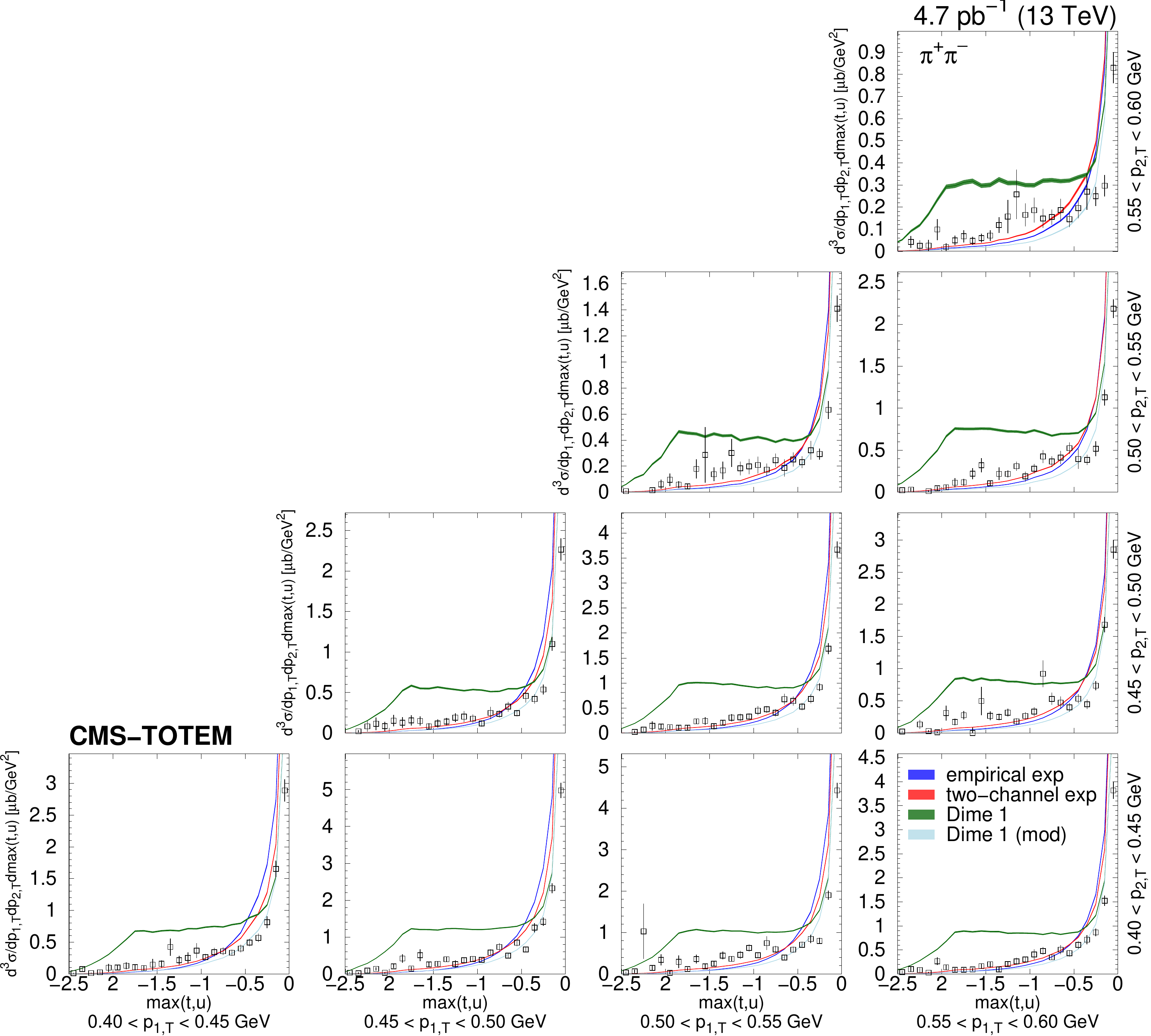}
 \includegraphics[width=0.65\textwidth]{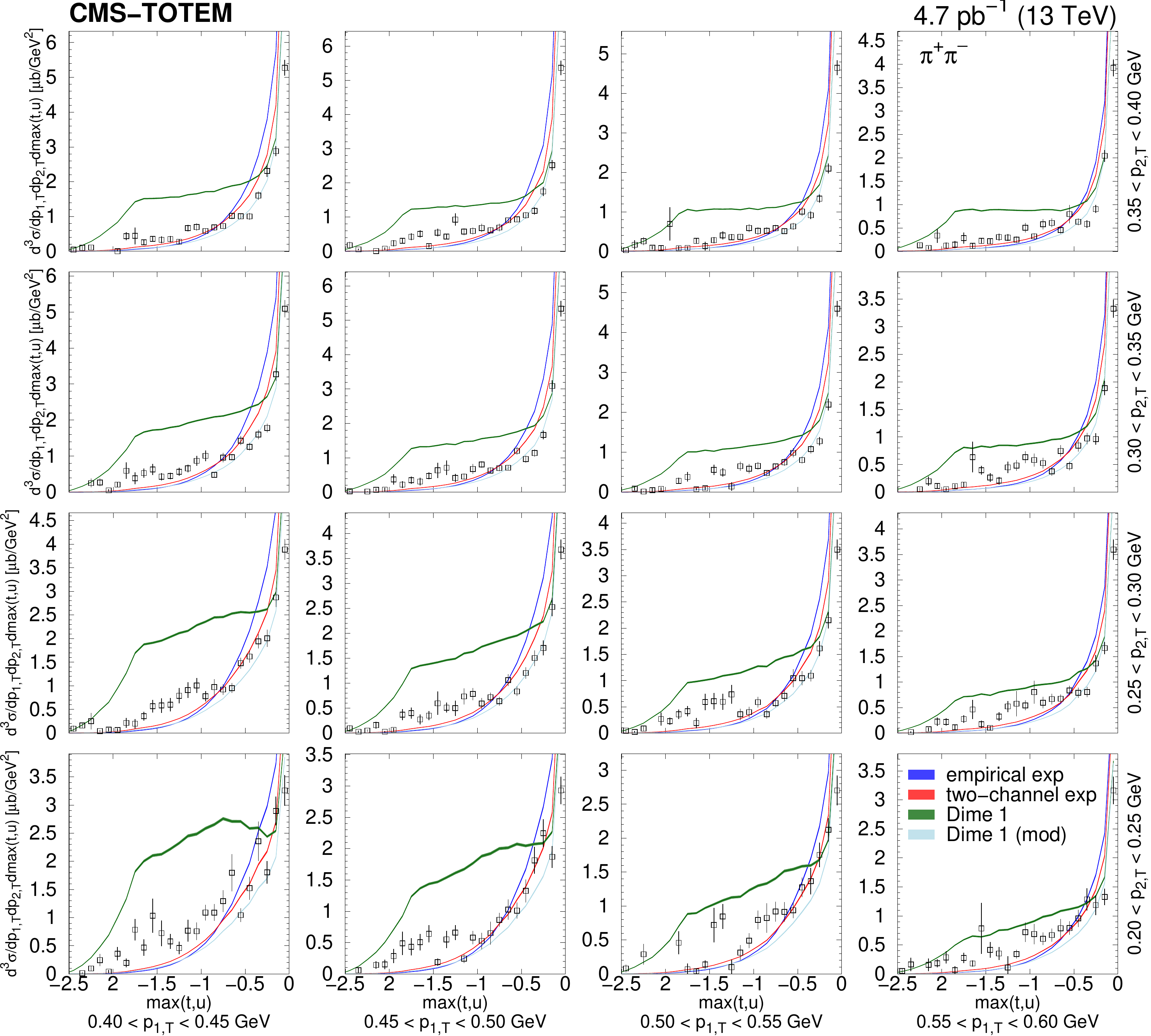}

 \caption{Distribution of the squared momentum transfer of the virtual pion in
several $(p_\text{1,T},p_\text{2,T})$ bins, in units of $\mu\mathrm{b}/\GeVns^3$,
for the mass range $1.8 < m_{\PGpp\PGpm} < 2.2\GeV$. Measured values (black
symbols) are shown together with the predictions of the empirical and the
two-channel models (coloured curves) using the tuned parameters for the
exponential proton-pomeron form factors (see text for details). Curves
corresponding to {\dime} (model 1, ``Dime 1'') and its modification (labelled
``Dime 1 (mod)'') with $b_\text{exp} = 0.9\GeV^{-2}$ are also plotted. The
error bars indicate the statistical uncertainties.}

 \label{fig:dndsha_023_1_2}

 \end{figure*}

The distributions of the squared momentum transfer of the virtual pion at
invariant masses $1.8 < m_{\PGpp\PGpm} < 2.2\GeV$ in several
$(p_\text{1,T},p_\text{2,T})$ bins are shown for $\PGpp\PGpm$ in
Figs.~\ref{fig:dndsha_022_1_1}--\ref{fig:dndsha_023_1_2}. Measured values are
shown together with the predictions of the empirical and the two-channel models
using the tuned parameters for the exponential proton-pomeron form factors.
Curves corresponding to {\dime} (model 1, ``Dime 1'') and its modification
(labelled ``Dime 1 (mod)'') with $b_\text{exp} = 0.9\GeV^{-2}$ are also
plotted.

\section{Summary}
\label{sec:summary}

We examined the central exclusive production of charged hadron pairs in
proton-proton collisions at a centre-of-mass energy of 13\TeV. Events are
selected by requiring both scattered protons to be detected in the TOTEM Roman
pots and exactly two oppositely charged identified pions in the CMS silicon
tracker. The process is studied in the resonance-free region, for invariant
masses of the centrally produced two-pion system $m_{\PGpp\PGpm} < 0.7\GeV$ or
$m_{\PGpp\PGpm} > 1.8\GeV$. Differential cross sections are measured as a
function of the azimuthal angle between the surviving protons in a wide region
of scattered proton transverse momenta, between $0.2$ and $0.8\GeV$, and for
pion rapidities $\abs{y} < 2$. A rich structure of nonperturbative interactions
related to double-pomeron exchange emerges and is measured with good precision.
The parabolic minimum in the distribution of the two-proton azimuthal angle is
observed for the first time. It can be understood as an effect of additional
pomeron exchanges between the incoming protons, resulting from the interference
of the bare and the rescattered amplitudes. With model tuning, various physical
quantities related to the pomeron cross section, proton-pomeron, and
meson-pomeron form factors, pomeron trajectory and intercept, as well as
coefficients of diffractive eigenstates of the proton are determined.

\clearpage

\begin{acknowledgments}
We congratulate our colleagues in the CERN accelerator departments for the excellent performance of the LHC and thank the technical and administrative staffs at CERN and at other CMS sand TOTEM institutes for their contributions to the success of the common CMS-TOTEM effort. In addition, we gratefully acknowledge the computing centres and personnel of the Worldwide LHC Computing Grid and other centres for delivering so effectively the computing infrastructure essential to our analyses. Finally, we acknowledge the enduring support for the construction and operation of the LHC, the CMS and TOTEM detectors, and the supporting computing infrastructure provided by the following funding agencies: SC (Armenia), BMBWF and FWF (Austria); FNRS and FWO (Belgium); CNPq, CAPES, FAPERJ, FAPERGS, and FAPESP (Brazil); MES and BNSF (Bulgaria); CERN; CAS, MoST, and NSFC (China); MINCIENCIAS (Colombia); MSES and CSF (Croatia); RIF (Cyprus); SENESCYT (Ecuador); MoER, ERC PUT and ERDF (Estonia); Academy of Finland, Magnus Ehrnrooth Foundation, MEC, HIP, and and Waldemar
von Frenckell Foundation (Finland); CEA and CNRS/IN2P3 (France); SRNSF (Georgia); BMBF, DFG, and HGF (Germany); GSRI (Greece); NKFIH (Hungary); DAE and DST (India); IPM (Iran); SFI (Ireland); INFN (Italy); MSIP and NRF (Republic of Korea); MES (Latvia); LAS (Lithuania); MOE and UM (Malaysia); BUAP, CINVESTAV, CONACYT, LNS, SEP, and UASLP-FAI (Mexico); MOS (Montenegro); MBIE (New Zealand); PAEC (Pakistan); MES and NSC (Poland); FCT (Portugal); MESTD (Serbia); MCIN/AEI and PCTI (Spain); MOSTR (Sri Lanka); Swiss Funding Agencies (Switzerland); MST (Taipei); MHESI and NSTDA (Thailand); TUBITAK and TENMAK (Turkey); NASU (Ukraine); STFC (United Kingdom); DOE and NSF (USA).

\hyphenation{Rachada-pisek} Individuals have received support from the Marie-Curie programme and the European Research Council and Horizon 2020 Grant, contract Nos.\ 675440, 724704, 752730, 758316, 765710, 824093, and COST Action CA16108 (European Union); the Leventis Foundation; the Alfred P.\ Sloan Foundation; the Alexander von Humboldt Foundation; the Science Committee, project no. 22rl-037 (Armenia); the Belgian Federal Science Policy Office; the Fonds pour la Formation \`a la Recherche dans l'Industrie et dans l'Agriculture (FRIA-Belgium); the Agentschap voor Innovatie door Wetenschap en Technologie (IWT-Belgium); the F.R.S.-FNRS and FWO (Belgium) under the ``Excellence of Science -- EOS" -- be.h project n.\ 30820817; the Beijing Municipal Science \& Technology Commission, No. Z191100007219010 and Fundamental Research Funds for the Central Universities (China); the Ministry of Education, Youth and Sports (MEYS) of the Czech Republic; the Shota Rustaveli National Science Foundation, grant FR-22-985 (Georgia); the Deutsche Forschungsgemeinschaft (DFG), under Germany's Excellence Strategy -- EXC 2121 ``Quantum Universe" -- 390833306, and under project number 400140256 - GRK2497; the Hellenic Foundation for Research and Innovation (HFRI), Project Number 2288 (Greece); the Hungarian Academy of Sciences, the New National Excellence Program - \'UNKP, the NKFIH research grants K 124845, K 124850, K 128713, K 128786, K 129058, K 131991, K 133046, K 138136, K 143460, K 143477, K 146913, K 146914, K 147048, 2020-2.2.1-ED-2021-00181, and TKP2021-NKTA-64 (Hungary); the Council of Science and Industrial Research, India; ICSC -- National Research Centre for High Performance Computing, Big Data and Quantum Computing, funded by the NextGenerationEU program (Italy); the Latvian Council of Science; the Ministry of Education and Science, project no. 2022/WK/14, and the National Science Center, contracts Opus 2021/41/B/ST2/01369 and 2021/43/B/ST2/01552 (Poland); the Funda\c{c}\~ao para a Ci\^encia e a Tecnologia, grant CEECIND/01334/2018 (Portugal); the National Priorities Research Program by Qatar National Research Fund; MCIN/AEI/10.13039/501100011033, ERDF ``a way of making Europe", and the Programa Estatal de Fomento de la Investigaci{\'o}n Cient{\'i}fica y T{\'e}cnica de Excelencia Mar\'{\i}a de Maeztu, grant MDM-2017-0765 and Programa Severo Ochoa del Principado de Asturias (Spain); the Chulalongkorn Academic into Its 2nd Century Project Advancement Project, and the National Science, Research and Innovation Fund via the Program Management Unit for Human Resources \& Institutional Development, Research and Innovation, grant B37G660013 (Thailand); the Kavli Foundation; the Nvidia Corporation; the SuperMicro Corporation; the Welch Foundation, contract C-1845; and the Weston Havens Foundation (USA).
\end{acknowledgments}

\clearpage

\bibliography{auto_generated}
\cleardoublepage \appendix\section{The CMS Collaboration \label{app:collab}}\begin{sloppypar}\hyphenpenalty=5000\widowpenalty=500\clubpenalty=5000
\cmsinstitute{Yerevan Physics Institute, Yerevan, Armenia}
{\tolerance=6000
A.~Hayrapetyan, A.~Tumasyan\cmsAuthorMark{1}\cmsorcid{0009-0000-0684-6742}
\par}
\cmsinstitute{Institut f\"{u}r Hochenergiephysik, Vienna, Austria}
{\tolerance=6000
W.~Adam\cmsorcid{0000-0001-9099-4341}, J.W.~Andrejkovic, T.~Bergauer\cmsorcid{0000-0002-5786-0293}, S.~Chatterjee\cmsorcid{0000-0003-2660-0349}, K.~Damanakis\cmsorcid{0000-0001-5389-2872}, M.~Dragicevic\cmsorcid{0000-0003-1967-6783}, P.S.~Hussain\cmsorcid{0000-0002-4825-5278}, M.~Jeitler\cmsAuthorMark{2}\cmsorcid{0000-0002-5141-9560}, N.~Krammer\cmsorcid{0000-0002-0548-0985}, A.~Li\cmsorcid{0000-0002-4547-116X}, D.~Liko\cmsorcid{0000-0002-3380-473X}, I.~Mikulec\cmsorcid{0000-0003-0385-2746}, J.~Schieck\cmsAuthorMark{2}\cmsorcid{0000-0002-1058-8093}, R.~Sch\"{o}fbeck\cmsorcid{0000-0002-2332-8784}, D.~Schwarz\cmsorcid{0000-0002-3821-7331}, M.~Sonawane\cmsorcid{0000-0003-0510-7010}, S.~Templ\cmsorcid{0000-0003-3137-5692}, W.~Waltenberger\cmsorcid{0000-0002-6215-7228}, C.-E.~Wulz\cmsAuthorMark{2}\cmsorcid{0000-0001-9226-5812}
% ...
\par}
\vskip\cmsinstskip
\dag:~Deceased\\
$^{1}$Also at Yerevan State University, Yerevan, Armenia\\
$^{2}$Also at TU Wien, Vienna, Austria\\
% ...
\section{The TOTEM Collaboration\label{app:totem}}
\newcommand{\AddAuthor}[2]{#1$^{#2}$,\ }
\newcommand\AddAuthorLast[2]{#1$^{#2}$}
\noindent
\AddAuthor{G.~Antchev\cmsorcid{0000-0003-3210-5037}}{a}
\AddAuthor{P.~Aspell}{8}
\AddAuthor{I.~Atanassov\cmsorcid{0000-0002-5728-9103}}{a}
\AddAuthor{V.~Avati}{7,8}
\AddAuthor{J.~Baechler}{8}
\AddAuthor{C.~Baldenegro~Barrera\cmsorcid{0000-0002-6033-8885}}{11}
\AddAuthor{V.~Berardi\cmsorcid{0000-0002-8387-4568}}{4a,4b}
\AddAuthor{M.~Berretti\cmsorcid{0000-0003-4122-8282}}{2a}
\AddAuthor{V.~Borshch\cmsorcid{0000-0002-5479-1982}}{12}
\AddAuthor{E.~Bossini\cmsorcid{0000-0002-2303-2588}}{6a}
\AddAuthor{U.~Bottigli\cmsorcid{0000-0002-0666-3433}}{6b}
\AddAuthor{M.~Bozzo\cmsorcid{0000-0002-1715-0457}}{5a,5b}
\AddAuthor{H.~Burkhardt}{8}
\AddAuthor{F.S.~Cafagna\cmsorcid{0000-0002-7450-4784}}{4a}
\AddAuthor{M.G.~Catanesi\cmsorcid{0000-0002-2987-7688}}{4a}
\AddAuthor{M.~Deile\cmsorcid{0000-0001-5085-7270}}{8}
\AddAuthor{F.~De~Leonardis}{4a,4c}
\AddAuthor{M.~Doubek\cmsorcid{0000-0002-0026-9558}}{1c}
\AddAuthor{D.~Druzhkin}{3a,8}
\AddAuthor{K.~Eggert}{10}
\AddAuthor{V.~Eremin}{c}
\AddAuthor{A.~Fiergolski}{8}
\AddAuthor{F.~Garcia\cmsorcid{0000-0002-4023-7964}}{2a}
\AddAuthor{V.~Georgiev}{1a}
\AddAuthor{S.~Giani}{8}
\AddAuthor{L.~Grzanka\cmsorcid{0000-0002-3599-854X}}{7}
\AddAuthor{J.~Hammerbauer}{1a}
\AddAuthor{T.~Isidori\cmsorcid{0000-0002-7934-4038}}{11}
\AddAuthor{V.~Ivanchenko\cmsorcid{0000-0002-1844-5433}}{12}
\AddAuthor{M.~Janda\cmsorcid{0000-0001-5736-6183}}{1c}
\AddAuthor{A.~Karev}{8}
\AddAuthor{J.~Ka\v{s}par\cmsorcid{0000-0001-5639-2267}}{1b,8}
\AddAuthor{B.~Kaynak\cmsorcid{0000-0003-3857-2496}}{9}
\AddAuthor{J.~Kopal\cmsorcid{0000-0001-9751-7409}}{8}
\AddAuthor{V.~Kundr\'{a}t\cmsorcid{0000-0003-2868-5550}}{1b}
\AddAuthor{S.~Lami\cmsorcid{0000-0001-9492-0147}}{6a}
\AddAuthor{R.~Linhart}{1a}
\AddAuthor{C.~Lindsey}{11}
\AddAuthor{M.V.~Lokaj\'{i}\v{c}ek\cmsorcid{0000-0002-2052-1220}$^{\textrm{\dag,}}$}{1b}
\AddAuthor{L.~Losurdo\cmsorcid{0000-0002-4964-7951}}{6b}
\AddAuthor{F.~Lucas~Rodr\'{i}guez}{8}
\AddAuthor{M.~Macr\'{i}}{\dag,5a}
\AddAuthor{M.~Malawski\cmsorcid{0000-0001-6005-0243}}{7}
\AddAuthor{N.~Minafra\cmsorcid{0000-0003-4002-1888}}{11}
\AddAuthor{S.~Minutoli}{5a}
\AddAuthor{K.~Misan\cmsorcid{0009-0007-2416-042X}}{7}
\AddAuthor{T.~Naaranoja\cmsorcid{0000-0001-5797-7929}}{2a,2b}
\AddAuthor{F.~Nemes\cmsorcid{0000-0002-1451-6484}}{3a,3b,8}
\AddAuthor{H.~Niewiadomski}{10}
\AddAuthor{E.~Oliveri\cmsorcid{0000-0002-0832-6975}}{8}
\AddAuthor{F.~Oljemark\cmsorcid{0000-0003-0121-2761}}{2a,2b}
\AddAuthor{M.~Oriunno}{b}
\AddAuthor{K.~\"{O}sterberg\cmsorcid{0000-0003-4807-0414}}{2a,2b}
\AddAuthor{S.~Ozkorucuklu\cmsorcid{0000-0001-5153-9266}}{9}
\AddAuthor{P.~Palazzi\cmsorcid{0000-0002-4861-391X}}{8}
\AddAuthor{V.~Passaro\cmsorcid{0000-0003-0802-4464}}{4a,4c}
\AddAuthor{Z.~Peroutka}{1a}
\AddAuthor{O.~Potok\cmsorcid{0009-0005-1141-6401}}{9}
\AddAuthor{J.~Proch\'{a}zka\cmsorcid{0000-0002-2774-2245}}{1b,\dag}
\AddAuthor{M.~Quinto\cmsorcid{0000-0002-6363-6132}}{4a,4b}
\AddAuthor{E.~Radermacher}{8}
\AddAuthor{E.~Radicioni\cmsorcid{0000-0002-2231-1067}}{4a}
\AddAuthor{F.~Ravotti\cmsorcid{0000-0002-5709-6934}}{8}
\AddAuthor{C.~Royon\cmsorcid{0000-0002-7672-9709}}{11}
\AddAuthor{G.~Ruggiero}{8}
\AddAuthor{H.~Saarikko}{2a,2b}
\AddAuthor{V.D.~Samoylenko}{c}
\AddAuthor{A.~Scribano\cmsorcid{0000-0002-4338-6332}}{6a,11}
\AddAuthor{J.~\v{S}irok\'{y}}{1a}
\AddAuthor{J.~Smajek}{8}
\AddAuthor{W.~Snoeys\cmsorcid{0000-0003-3541-9066}}{8}
\AddAuthor{R.~Stefanovitch}{3a,8}
\AddAuthor{C.~Taylor\cmsorcid{0000-0001-6816-8051}}{10}
\AddAuthor{E.~Tcherniaev\cmsorcid{0000-0002-3685-0635}}{12}
\AddAuthor{N.~Turini\cmsorcid{0000-0002-9395-5230}}{6b}
\AddAuthor{O.~Urban}{1a}
\AddAuthor{V.~Vacek\cmsorcid{0000-0001-9584-0392}}{1c}
\AddAuthor{O.~Vavroch}{1a}
\AddAuthor{J.~Welti}{2a,2b}
\AddAuthor{J.~Williams\cmsorcid{0000-0002-9810-7097}}{11}
\AddAuthorLast{J.~Zich}{1a}
\vskip 4pt plus 4pt
\let\thefootnote\relax
\newcommand{\AddInstitute}[2]{${}^{#1}$#2\\}
\newcommand{\AddExternalInstitute}[2]{\footnote{${}^{#1}$ #2}}
\noindent
\dag Deceased\\
\AddInstitute{1a}{University of West Bohemia, Pilsen, Czech Republic}
\AddInstitute{1b}{Institute of Physics of the Academy of Sciences of the Czech Republic, Prague, Czech Republic}
\AddInstitute{1c}{Czech Technical University, Prague, Czech Republic}
\AddInstitute{2a}{Helsinki Institute of Physics, University of Helsinki, Helsinki, Finland}
\AddInstitute{2b}{Department of Physics, University of Helsinki, Helsinki, Finland}
\AddInstitute{3a}{Wigner Research Centre for Physics, RMKI, Budapest, Hungary}
\AddInstitute{3b}{MATE Institute of Technology KRC, Gy\"{o}ngy\"{o}s, Hungary}
\AddInstitute{4a}{INFN Sezione di Bari, Bari, Italy}
\AddInstitute{4b}{Dipartimento Interateneo di Fisica di Bari, University of Bari, Bari, Italy}
\AddInstitute{4c}{Dipartimento di Ingegneria Elettrica e dell'Informazione --- Politecnico di Bari, Bari, Italy}
\AddInstitute{5a}{INFN Sezione di Genova, Genova, Italy}
\AddInstitute{5b}{Universit\`{a} degli Studi di Genova, Genova, Italy}
\AddInstitute{6a}{INFN Sezione di Pisa, Pisa, Italy}
\AddInstitute{6b}{Universit\`{a} degli Studi di Siena and Gruppo Collegato INFN di Siena, Siena, Italy}
\AddInstitute{7}{Akademia G\'{o}rniczo-Hutnicza (AGH) University of Krak\'ow, Faculty of Computer Science, Krak\'ow, Poland}
\AddInstitute{8}{CERN, Geneva, Switzerland}
\AddInstitute{9}{Istanbul University, Istanbul, Turkey}
\AddInstitute{10}{Case Western Reserve University, Department of Physics, Cleveland, Ohio, USA}
\AddInstitute{11}{The University of Kansas, Lawrence, Kansas, USA}
\AddInstitute{12}{Authors affiliated with an institute or an international laboratory covered by a cooperation agreement with CERN}
\AddExternalInstitute{a}{INRNE-BAS, Institute for Nuclear Research and Nuclear Energy, Bulgarian Academy of Sciences, Sofia, Bulgaria}
\AddExternalInstitute{b}{SLAC, Stanford University, California, USA}
\AddExternalInstitute{c}{Authors affiliated with an institute or an international laboratory covered by a cooperation agreement with CERN}
\end{sloppypar}
\end{document}